\documentclass{article}

\usepackage[a4paper, 
        left=0.9in,
        right=0.9in,
        top=1in,
        bottom=1in,
]{geometry}

\usepackage{graphics}
\usepackage{amsmath}
\usepackage{stmaryrd}
\usepackage{bm}
\allowdisplaybreaks
\usepackage{fancyhdr}
\usepackage{amssymb}
\usepackage{latexsym}
\usepackage{color}
\usepackage{graphicx}
\usepackage{hyperref}
\usepackage{url}
\hypersetup{
colorlinks = true,
citecolor  = blue,
linkcolor = blue
}
\usepackage[T1]{fontenc}
\usepackage{epsfig}
\usepackage{setspace}
\usepackage{graphicx}
\usepackage{epstopdf}
\usepackage{subfigure}
\usepackage{xcolor}
\usepackage{float}
\usepackage[titletoc,toc,title]{appendix}

\usepackage{fontawesome5}
\numberwithin{equation}{section}
\numberwithin{figure}{section}

\colorlet{Changes@Color}{red}
\usepackage{authblk}
\usepackage{citesort}
\usepackage{cite}

\newcommand{\rout}[1]

\begin{document}

\title{A tutorial on simulating nonlinear behaviors of flexible structures with the discrete differential geometry (DDG) method}



\author{Weicheng Huang$^{1,\dagger,*}$, Zhuonan Hao$^{2,\dagger}$, Jiahao Li$^{3,\dagger}$, Dezhong Tong$^{4,\dagger}$, Kexin Guo$^{5}$, \\ Yingchao Zhang$^{6}$, Huajian Gao$^{6,7,*}$, K. Jimmy Hsia$^{5,8,*}$, Mingchao Liu$^{9,*}$}
\date{%
{\small \it   
$^1$ School of Engineering, Newcastle University, Newcastle upon Tyne NE1 7RU, UK \\
$^2$ Department of Mechanical and Aerospace Engineering, University of California, Los Angeles,  
\\ Los Angeles, California 90095, United States \\
$^3$ CAS Key Laboratory of Mechanical Behavior and Design of Materials, Department of Modern Mechanics, University of Science and Technology of China, Hefei 230027, People’s Republic of China \\
$^4$   Department of Material Science and Engineering, University of Michigan, Ann Arbor, \\ Ann Arbor, Michigan, 48105, USA \\
$^5$ School of Mechanical and Aerospace Engineering, Nanyang Technological University, \\ Singapore 639798, Republic of Singapore \\
$^6$ AML, Department of Engineering Mechanics, Tsinghua University, \\ Beijing 100084, People's Republic of China \\
$^7$ Mechano-X Institute, Tsinghua University, Beijing 100084, People's Republic of China \\
$^8$ School of Chemistry, Chemical Engineering and Biotechnology, Nanyang Technological University, \\ Singapore 639798, Republic of Singapore \\
$^9$ Department of Mechanical Engineering, University of Birmingham, Birmingham B15 2TT, UK \\
\vspace{8pt}
$^\dagger$ These authors contributed equally to this work. \\
$^*$E-mail: weicheng.huang@newcastle.ac.uk (W.H.) \\ gao.huajian@tsinghua.edu.cn (H.G.) \\ kjhsia@ntu.edu.sg (K.J.H.) \\ m.liu.2@bham.ac.uk (M.L.)
}}

\maketitle

\newpage

\begin{abstract}    
Flexible elastic structures, such as beams, rods, ribbons, plates, and shells, exhibit complex nonlinear dynamical behaviors that are central to a wide range of engineering and scientific applications, including soft robotics, deployable structures, and biomedical devices. While various numerical methods have been developed to simulate these behaviors, many conventional approaches struggle to simultaneously capture geometric and material nonlinearities, as well as nonlinear external interactions, particularly in highly deformable and dynamically evolving systems. The Discrete Differential Geometry (DDG) method has emerged as a robust and efficient numerical framework that intrinsically preserves geometric properties, accommodates material nonlinearity, and accurately models interactions with external environments and fields. By directly discretizing geometric and mechanical quantities, DDG provides an accurate, stable, and efficient approach to modeling flexible structures, addressing key limitations of traditional numerical methods. This tutorial provides a systematic introduction to the DDG method for simulating nonlinear behaviors in flexible structures. It covers DDG theory, simulation frameworks, and MATLAB implementation, with examples spanning dynamic systems, geometric and material nonlinearities, and external interactions like magnetics and fluids, culminating in practical insights and future directions. By offering a comprehensive and practical guide—together with open-source MATLAB code—this tutorial aims to facilitate the broader adoption of DDG-based numerical tools among researchers and engineers in computational mechanics, applied mathematics, and structural design. We seek to enhance the accessibility and applicability of DDG methods, fostering further advancements in the simulation and analysis of highly flexible structures across diverse scientific and engineering domains.
\end{abstract}

\newpage

\tableofcontents

\newpage
\section{Introduction}

Flexible structures, such as beams, rods, ribbons, plates, and shells, are fundamental components in both natural and engineered systems \cite{reis2018mechanics,marder2007crumpling}. Their ability to undergo large deformations and exhibit highly nonlinear mechanical responses under various loading conditions makes them essential in numerous scientific and engineering applications. These structures are prevalent in biological systems, where they enable functionalities such as plant morphogenesis \cite{moulton2013morphoelastic, lessinnes2017morphoelastic, moulton2020morphoelastic, zhang2018finite}, soft tissue mechanics \cite{holzapfel2001biomechanics, fu2019modeling, xu2020water, xu2022chiral}, and insect wing kinematics \cite{bos2008influence, liu2008wing, young2009details, seshadri2012novel}.
In engineering, they serve as critical elements in soft robotics \cite{rafsanjani2019programming, yang2023morphing}, micro robotics \cite{graule2016perching, jafferis2019untethered}, flexible electronics \cite{zhang2017printing,jiang2022flexible}, mechanical metamaterials \cite{walia2015flexible,bertoldi2017flexible}, medical devices \cite{khan2016monitoring,la2022flexible}, and deployable structures for civil and aerospace applications \cite{pellegrino2001deployable,flores2014review,zhang2021deployable}. The ability of these structures to adapt, morph, and sustain large deformations under external stimuli is central to their functionality. Consequently, accurate and efficient modeling of their nonlinear behaviors is crucial for performance prediction, optimal design, and control \cite{junkins1993introduction,yang2003recent,yang2007rigid,touze2021model, choi2024learning, tong2024sim2real}.

\subsection{Challenges in modeling nonlinear behaviors of flexible structures}

The nonlinear behavior of flexible structures arises due to three primary factors: (i) geometric nonlinearity, (ii) material nonlinearity, and (iii) nonlinear interactions with external fields and/or environments. These complexities make their numerical modeling inherently challenging, requiring advanced computational techniques to capture their full nonlinear mechanical response.

\subsubsection*{\textit{i}). \textit{Geometric Nonlinearity}}

\begin{itemize}

\item{Large deformations lead to nonlinear strain-displacement relationships, making linearized small-strain assumptions inapplicable. In slender and thin structures, the coupling between stretching, bending, and twisting further complicates deformation analysis \cite{audoly2000elasticity}.}

\item{Instability-driven behaviors such as buckling, wrinkling, and bistability commonly emerge due to geometric constraints and external loading conditions. These behaviors lead to bifurcations and sudden shape transformations, which require specialized numerical frameworks to track \cite{reis2018mechanics, hu2015buckling}.}

\end{itemize}

\subsubsection*{\textit{ii}). \textit{Material Nonlinearity}}

\begin{itemize}

\item{Many flexible structures are composed of soft, hyperelastic, or viscoelastic materials, whose stress-strain response is inherently nonlinear. Unlike stiff materials governed by Hookean elasticity, these materials exhibit large recoverable deformations and often require hyperelastic constitutive models (e.g., Neo-Hookean, Mooney-Rivlin, Ogden models) to characterize their behavior \cite{kim2012comparison, melly2021review, khaniki2022review}.}

\item{Hysteretic and rate-dependent responses occur in biological tissues, soft polymers, and elastomers, where time-dependent deformation (e.g., viscoelasticity, creep, and relaxation) significantly affects their structural performance. Modeling these effects demands constitutive models that incorporate strain history and time-dependent properties \cite{xiang2019physically, gomez2019dynamics, chen2023pseudo}.}

\end{itemize}

\subsubsection*{\textit{iii}). \textit{Nonlinear Interactions with External Fields}}

\begin{itemize}

\item{Multi-physical actuation: Many flexible structures respond to external fields such as magnetic, electrical, and optical stimuli, enabling programmable deformations. Magneto-mechanical coupling is widely used in magnetically actuated soft robots and shape-morphing materials, where deformation is governed by magnetic torques and forces \cite{zhao2019mechanics, wang2020hard, huang2023discrete, huang2023modeling}. Additionally, electrostatic actuation in dielectric elastomers (DEs) \cite{tian2021dynamics, luo2020dynamic, o2009finite, wissler2005modeling} and light-responsive deformations in liquid crystal elastomers (LCEs) \cite{goriely2023rod, barnes2023surface, brighenti2023multiphysics, zhou2025modified, wei2023rate} provide new opportunities for reconfigurable and adaptive structures.}

\item{Fluid- and solid-structure interactions: Flexible structures frequently interact with surrounding fluids, surfaces, and substrates, leading to complex mechanical responses. Fluid-structure interactions (FSI) are critical in applications such as underwater soft robotics, aerodynamic morphing structures, and biological locomotion (e.g., swimming and flying organisms), where hydrodynamic and aerodynamic forces strongly influence deformation dynamics \cite{jawed2015propulsion, huang2020numerical, jawed2017dynamics, guttag2017active, pezzulla2020deformation, huang2021swimming, tong2023fully}. Similarly, frictional and adhesive interactions are essential in wearable electronics, bio-integrated devices, and deployable mechanisms, where mechanical compliance and contact mechanics determine functional performance \cite{huang2020dynamic, huang2023modeling, choi2021implicit, tong2023fully}.}

\end{itemize}


Due to these intricate nonlinearities, the computational modeling of flexible structures faces several challenges:

\begin{itemize}

\item{Tracking large deformations and topological changes: Many flexible structures undergo extreme shape changes, including self-contact, snapping, and folding, requiring numerical methods that can dynamically update geometric configurations without numerical instabilities \cite{audoly2000elasticity, reis2018mechanics, reis2015perspective, marder2007crumpling}.}

\item{Capturing bifurcation and multi-stability phenomena: Many shape-changing structures exhibit bistability and multistability, where small perturbations can induce sudden, non-reversible transformations. Traditional methods often fail to trace these bifurcating solutions efficiently \cite{ikeda2002imperfect, luongo2023stability, yang2023tutorial, huang2023bifurcations}.}

\item{Handling multi-physics coupling: Incorporating fluid-structure interactions, magneto-mechanical effects, and contact mechanics requires integrating different numerical solvers, often leading to high computational costs and convergence issues \cite{wang2020hard, goriely2023rod, brighenti2023multiphysics, tian2021dynamics, jawed2014coiling, jawed2015propulsion, huang2020dynamic}.}

\end{itemize}

The complexity of these nonlinear behaviors necessitates advanced numerical methods that can naturally handle large deformations, material constitutive laws, and multi-physics interactions while ensuring numerical stability and computational efficiency.

\subsection{Traditional numerical methods and their limitations}

In addition to theoretical models such as the Elastica, Cosserat rod, and Kirchhoff-Love shell theories, computational mechanics has made significant progress in the numerical modeling of flexible structures. Various numerical methods have been developed to simulate large deformations, complex instabilities, and multi-physical interactions in slender and thin structures. However, traditional numerical approaches often struggle with efficiency, accuracy, and robustness when dealing with nonlinear behaviors, geometric complexity, and multi-field interactions. The most widely used methods include:

\subsubsection*{\textit{i}). \textit{Finite Element Method (FEM)}}

The finite element method (FEM) is one of the most established and widely used numerical techniques for structural analysis \cite{hughes2003finite, barbero2023finite, madenci2015finite}. By discretizing a continuous structure into small elements and applying variational principles, some modified FEM can provide a general framework for solving elasticity problems, including those of highly flexible structures, e.g., absolute nodal coordinate formulation \cite{shabana2001three, shabana1997definition, gerstmayr2013review}, and isogeometric analysis \cite{nagy2010isogeometric, weeger2019isogeometric, weeger2017isogeometric, echter2013isogeometric, kiendl2009isogeometric, benson2010isogeometric}

\begin{itemize}
    \item \textbf{\textit{Advantages:}} FEM is highly versatile and can handle complex geometries, heterogeneous materials, and sophisticated boundary conditions. Its extensive implementation in commercial software (e.g., Abaqus, ANSYS, COMSOL) makes it accessible to engineers and researchers across different disciplines.
    
    \item \textbf{\textit{Limitations:}} Despite its strengths, FEM suffers from challenges when applied to highly deformable flexible structures. The computational demands escalate when dealing with time-dependent problems, fluid-structure interactions, and contact mechanics. Additionally, remeshing is often necessary in extreme deformations or topology-changing scenarios, significantly increasing computational cost. It can also exhibit locking phenomena (e.g., shear locking, membrane locking), requiring specialized elements or fine meshing to maintain accuracy.

\end{itemize}

\subsubsection*{\textit{ii}). \textit{Finite Difference Method (FDM)}}

The finite difference method (FDM) is another type of traditional numerical approach based on structured grid discretization and finite difference approximations of differential equations \cite{perrone1975general, liszka1980finite, mazumder2015numerical}.

\begin{itemize}
    \item \textbf{\textit{Advantages:}} FDM is computationally efficient for regular, structured domains and offers a straightforward implementation for problems governed by ordinary differential equations (ODEs) and partial differential equations (PDEs). It is widely used in wave propagation, heat conduction, and fluid flow simulations.
    
    \item \textbf{\textit{Limitations:}} FDM is not well-suited for complex boundary conditions or highly irregular geometries, as it relies on structured grids that are difficult to conform to slender, curved, or evolving shapes. Furthermore, large deformations introduce challenges in maintaining numerical stability, and multi-physics coupling (e.g., FSI) often requires specialized modifications that reduce its efficiency compared to FEM-based formulations.

\end{itemize}

\subsubsection*{\textit{ii}). \textit{Continuation Methods}}

Continuation methods (e.g., AUTO and COCO) are specialized numerical techniques used to track bifurcations and stability transitions in nonlinear systems \cite{rheinboldt1981numerical, richter1983continuation, allgower2012numerical, krauskopf2007numerical, doedel1998auto, ahsan2022methods}. They are particularly useful in studying the post-buckling behavior and bifurcation branches of flexible structures.

\begin{itemize}
    \item \textbf{\textit{Advantages:}} The arc-length continuation method enables the systematic tracking of equilibrium solutions beyond critical points, making it a valuable tool for analyzing bistable and multistable structures. It is widely applied in nonlinear elasticity, morphing structures, and mechanical metamaterials.
    
    \item \textbf{\textit{Limitations:}} Continuation methods are primarily designed for static equilibrium problems and cannot directly handle dynamic responses, transient effects, or time-dependent external interactions. Moreover, they often struggle with isolated solutions, making them less effective for highly nonlinear systems with complex potential energy landscapes.

\end{itemize}

Despite their significant contributions to computational mechanics, these traditional numerical methods face inherent limitations when modeling the full complexity of flexible structures:

\begin{itemize}

\item{ \textbf{\textit{Capturing large deformations and geometric nonlinearity:}} Many flexible structures exhibit extreme shape changes, including snapping, wrinkling, and self-contact, which often require specialized numerical techniques beyond classical FEM and FDM.}

\item{ \textbf{\textit{Handling material and multi-physics coupling:}} Integrating hyperelasticity, viscoelasticity, fluid-structure interactions, and active actuation (e.g., magnetic, electrostatic, thermal) within a single robust framework remains a computational challenge.}

\item{ \textbf{\textit{Efficiency and stability in high-fidelity simulations:}} Balancing computational efficiency with numerical accuracy is critical, especially in scenarios requiring real-time control, optimization, and large-scale simulations.}

\end{itemize}

Given these limitations, there is a strong demand for alternative numerical frameworks that can naturally accommodate large deformations, geometric complexity, and external interactions while maintaining computational efficiency, accuracy, and robustness.

\subsection{Discrete differential geometry (DDG) as an alternative approach}

\textit{Discrete differential geometry (DDG)} has emerged as a promising numerical framework for simulating flexible structures, particularly in scenarios involving large deformations, geometric complexity, and nonlinear interactions \cite{grinspun2006discrete, bobenko2008discrete, crane2018discrete, choi2024dismech}. Initially developed in the computational graphics and computer geometry communities, DDG has gained increasing recognition in structural mechanics for its ability to preserve geometric properties while efficiently modeling thin elastic structures, such as rods, ribbons, plates, and shells. Unlike conventional numerical methods, which focus on discretizing differential equations, DDG instead discretizes the underlying geometry itself, enabling more natural and stable simulations of complex deformations \cite{vouga2013discrete}.

\begin{figure}[ht]
\centering
\includegraphics[width=0.8\columnwidth]{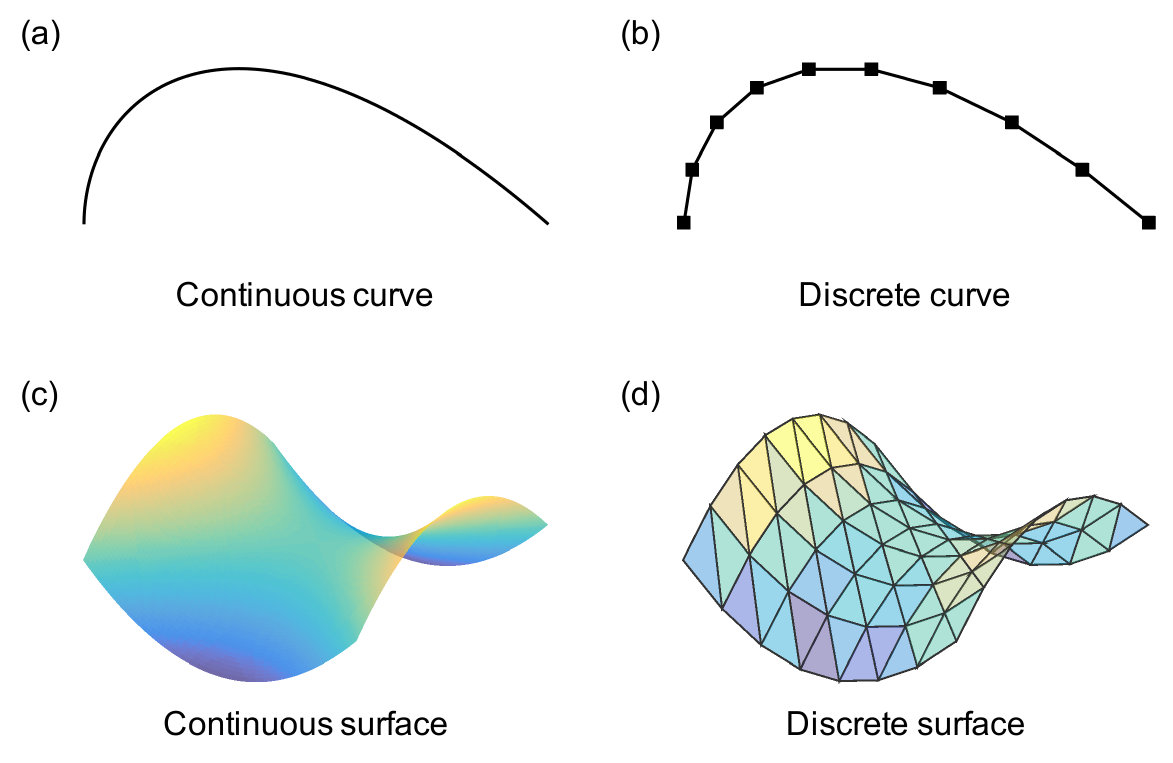}
\caption{\textbf{Geometry discretizations in DDG simulations.}
(a-b) A continuous curve is approximated using discrete linear segments. The curve is now represented as a collection of nodes and connecting edges. (c-d). A continuous surface is approximated using discrete triangular elements.}
\label{fig:ddg_demo_plot}
\end{figure}

A key feature of DDG is its intrinsic geometric preservation, which allows it to directly encode curvature, torsion, and bending mechanics. This is particularly advantageous for modeling flexible and highly deformable structures, where traditional methods often introduce numerical artifacts. The DDG approach discretizes continuous structures into meshes: nodes and edges for structures deforming from one-dimensional (1D) to two- or three-dimensional (2D/3D) forms, such as rods (Fig. \ref{fig:ddg_demo_plot}(a,b)); and vertices, edges, and faces for structures deforming from 2D to 3D, such as shells (Fig. \ref{fig:ddg_demo_plot}(c,d)). This representation naturally captures the geometric invariants of the original system, ensuring accuracy even under extreme deformations \cite{grinspun2006discrete}.

Another distinguishing aspect of DDG is its energy-based formulation. In contrast to FEM, which relies on high-order shape functions to approximate deformation fields, DDG directly expresses elastic energy in terms of nodal positions. This eliminates the need for complex interpolation functions and numerical integration, significantly improving computational efficiency. Since DDG does not suffer from locking phenomena, it provides a more stable and accurate way to model bending, twisting, and snap-through instabilities in slender structures \cite{huang2020shear, huang2021snap, huang2024integration, huang2024snap, huang2024exploiting, wang2024transient}.

In addition to its geometric and computational advantages, DDG offers a natural way to incorporate external interactions, such as frictional contact, fluid-structure coupling, and magneto-mechanical actuation. Many real-world applications, such as soft robotics, deployable structures, and morphing metamaterials, require accurate simulation of structures that not only deform significantly but also interact with external forces and constraints. DDG’s formulation allows these interactions to be seamlessly integrated while maintaining numerical stability.

Several key advantages make DDG an attractive alternative to traditional numerical methods:

\begin{itemize}

\item{\textbf{\textit{Intrinsic Geometric Preservation}} – DDG maintains curvature and topological constraints, making it particularly well-suited for thin and highly deformable structures.}

\item{\textbf{\textit{Robustness to Large Deformations}} – Unlike FEM, DDG avoids locking effects and accurately captures bending, twisting, and instability-driven deformations.}

\item{\textbf{\textit{Computational Efficiency}} – DDG simplifies numerical integration, leading to faster and more stable simulations compared to traditional methods.}

\item{\textbf{\textit{Seamless Multi-Physics Integration}} – Interaction with external fields, such as magneto-elastic coupling, fluid-structure interactions, and contact with solids can be incorporated without compromising geometric consistency.}

\end{itemize}

With its ability to efficiently and accurately model complex nonlinear behaviors, DDG presents a powerful alternative to traditional computational approaches. The following sections introduce a structured framework for applying DDG to simulate beams, rods, ribbons, plates, and shells, demonstrating its effectiveness in capturing geometric, material, and interaction-driven nonlinearities.

\subsection{Applications of DDG in flexible systems}

\begin{figure}[ht]
\centering
\includegraphics[width=\columnwidth]{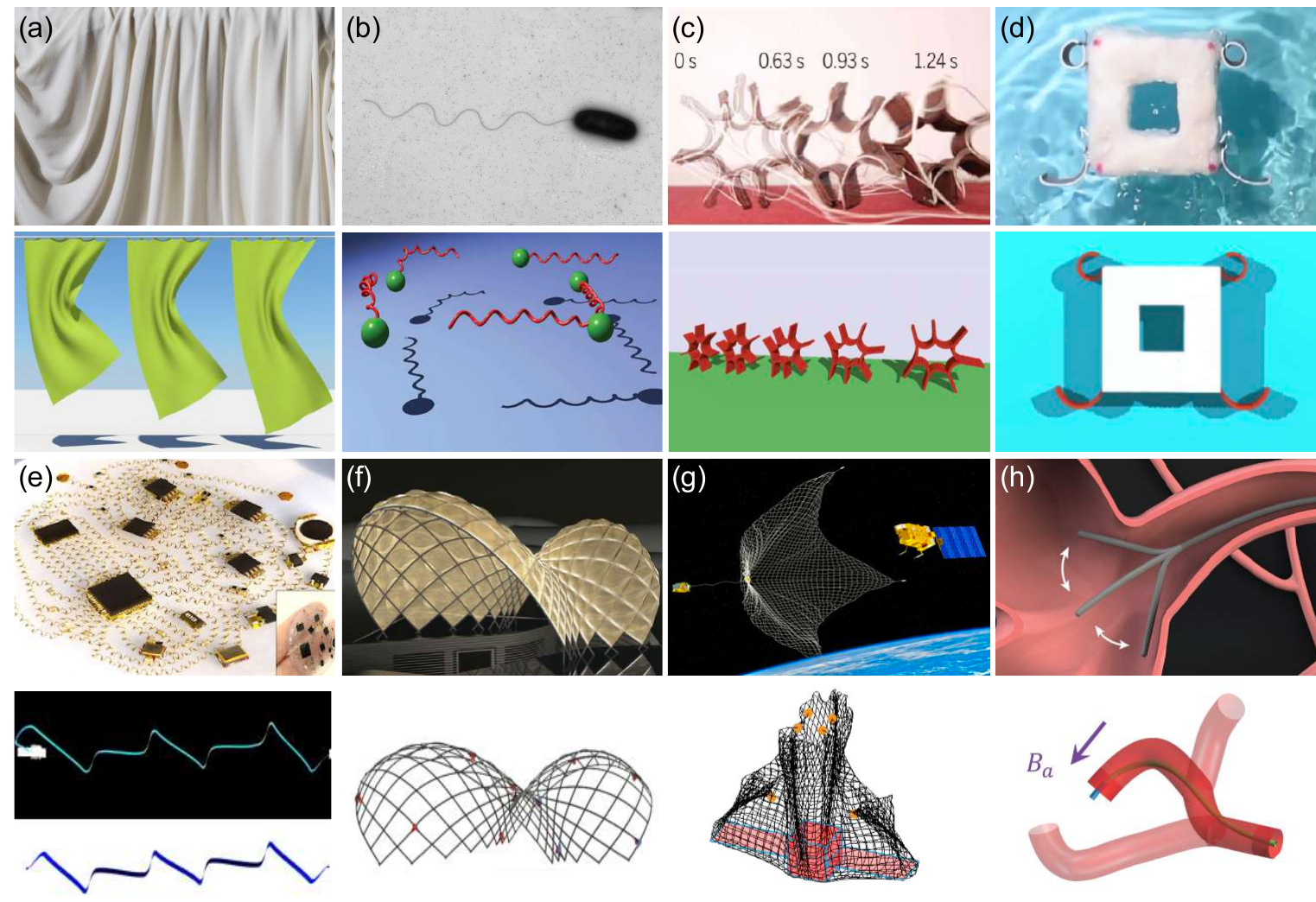}
\caption{\textbf{Applications of DDG-based simulations.} (a) Computer graphics \cite{ward2007survey, bertails2006super, kaufman2014adaptive, bouaziz2023projective}. (b) Biophysics \cite{seymour2017zooming, huang2020numerical}. (c) Soft robots \cite{huang2020dynamic}. (d) Underwater robots \cite{huang2021numerical, huang2022design}. (e) Flexible electronics \cite{nan2018compliant, jang2017self, shi2025double}. (f) Civil engineering \cite{panetta2019x, becker2023c}. (g) Aerospace engineering \cite{biesbroek2017deorbit, huang2022nonlinear, huang2023contact}. (h) Biomedical engineering \cite{kim2019ferromagnetic, tong2025real}.}
\label{fig:ddg_apply_plot}
\end{figure}

The Discrete Differential Geometry (DDG) method provides a powerful numerical framework for simulating flexible structures across a broad range of biological and artificial systems (Fig. \ref{fig:ddg_apply_plot}). Many natural systems, such as flagella, cilia, and soft tissues, rely on highly deformable structures to perform essential functions, while engineered systems, such as soft robots, deployable structures, and flexible electronics, exploit similar principles to achieve adaptability and functionality. DDG’s ability to preserve geometric invariants, efficiently model bending-dominated mechanics, and seamlessly integrate external interactions makes it particularly well-suited for simulating these systems with high accuracy and computational efficiency.

\paragraph{\textit{i}). \textit{Biological Systems and Bio-Locomotion}}

Many biological organisms rely on highly deformable structures for locomotion, fluid transport, and mechanical sensing. Flagella and cilia generate propulsion and flow regulation through coordinated bending motions, where nonlinear elastic responses and fluid-structure interactions play a crucial role~\cite{lim2022fabrication}. DDG provides an effective framework for modeling flagellar propulsion, cilia-driven flows, and undulatory swimming, capturing nonlinear bending mechanics and hydrodynamic forces while preserving geometric consistency. This capability extends to the study of microorganism motility, mucus transport in respiratory systems, and bio-inspired swimming robots, where interactions between thin, flexible filaments and surrounding fluid environments must be accurately simulated \cite{jawed2015propulsion, jawed2017dynamics, huang2020numerical, huang2023modeling}.

\paragraph{\textit{ii}). \textit{Soft Robots and Soft Actuators}}

Soft robotic systems leverage highly flexible structures to achieve dexterous and adaptive movement through shape morphing, multi-physical actuation, and dynamic environmental interactions. Unlike traditional rigid robots, soft robots require computational models that can accurately simulate large bending deformations, instability-driven motion, and frictional or fluidic constraints. DDG enables stable and efficient simulations of continuum manipulators, pneumatic and hydraulic actuators, and bio-inspired crawling and swimming robots, ensuring precise modeling of external interactions such as contact mechanics, fluid-structure coupling, and magnetically induced deformations \cite{huang2020dynamic, huang2021numerical, huang2022design, huang2023modeling, li2025harnessing, tong2024inverse, tong2025real, liu2025force, chen2024knotted}.

\paragraph{\textit{iii}). \textit{Deployable and Morphing Structures}}

Many engineered systems require structures that can transition between compact and expanded states, enabling applications in aerospace, space exploration, and adaptive architecture. These deployable structures often rely on bistability, snap-through instabilities, and morphing metamaterials, requiring accurate modeling of global deformations, localized folding mechanics, and dynamic energy redistribution. DDG provides an efficient approach for simulating thin-walled deployable mechanisms, reconfigurable morphing surfaces, and adaptive load-bearing structures, ensuring that large-scale deformations are captured while maintaining mechanical robustness and geometric fidelity \cite{del2013deployable, leng2015deployable, liu2020tapered, huang2022nonlinear, yang2024hierarchical, li2016dynamics, li2016dynamics2, huang2024integration, lu2024multistability, lu2023multiple, lu2023multiple2}.

\paragraph{\textit{iv}). \textit{Mechanical Metamaterials and Programmable Structures}}

Mechanical metamaterials achieve tunable and counterintuitive mechanical responses through geometrically programmed architectures rather than intrinsic material properties. Many of these structures exploit instability-driven transformations, hierarchical tessellations, and origami-inspired configurations to achieve negative Poisson’s ratios, bistable behavior, and controllable stiffness modulation. DDG’s ability to efficiently track local and global deformations makes it ideal for simulating programmable auxetic materials, kirigami-based structures, and reconfigurable lattices, which are widely applied in adaptive actuators, biomedical implants, and energy-absorbing materials \cite{meng2023encoding, meng2022deployable, meng2020multi, bertoldi2017flexible, rafsanjani2019programming, barchiesi2019mechanical}.

\paragraph{\textit{v}). \textit{Flexible Electronics and Bio-Integrated Devices}}

Flexible electronics and bio-integrated devices must conform to dynamic, stretchable, or curved surfaces while maintaining functionality under continuous mechanical deformation and strain. Devices such as electronic skins, soft sensors, and wearable health monitors experience repeated bending, stretching, and environmental loading. DDG enables efficient simulation of thin-film deformations, interfacial adhesion, and strain-dependent mechanical behavior, providing insight into the design of durable and adaptable electronic systems for medical and wearable applications \cite{jiang2022flexible, jang2017self, zhang2013buckling, zhang2017printing, bo2023mechanically, shen2024curvature}.

By offering a geometrically consistent and computationally efficient approach, DDG has become a valuable tool in simulating highly flexible structures across multiple disciplines. Its ability to handle large deformations, external interactions, and multi-stable behaviors makes it a preferred alternative to conventional numerical methods for applications in soft robotics, deployable structures, programmable metamaterials, and flexible electronics.

\subsection{Objective of this tutorial}

Despite its advantages, the adoption of DDG in computational mechanics remains limited due to the lack of accessible resources and tutorials. This paper aims to bridge that gap by providing a \textbf{comprehensive and practical guide} to the use of DDG for nonlinear numerical simulation of flexible structures.

\begin{itemize}
    \item Chapter 2 reviews the classical dynamic simulation of structural mechanics, presents the overall procedure for our DDG-based simulation framework, and provides the detailed description about the MATLAB implementation.
    
    \item Chapter 3 demonstrates the basic concepts of dynamic simulation framework by adopting single- and multiple-degree of freedom systems as examples. 
    
    \item Chapter 4 showcases representative examples demonstrating the key nonlinear behaviors of flexible structures, including \textit{planar beams}, \textit{3D rods}/\textit{ribbons}, \textit{rotational surfaces}, \textit{3D plates}/\textit{shells}, and \textit{hollow nets}/\textit{gridshells}.
    
    \item Chapter 5 illustrated the DDG method for handling material nonlinearity such \textit{hyperelastic} constitutive laws.
    \item Chapter 6 discusses how external interactions, such as \textit{magnetic actuation}, \textit{fluid-structure interaction}, and \textit{frictional contact}, can be incorporated into DDG-based simulations.
    \item Chapter 7 concludes with future research directions and potential improvements in DDG-based numerical modeling.
    \item Appendix A provides a numerical validation of the DDG-based method.
\end{itemize}

Through this work, we aim to \textit{facilitate the adoption of DDG in the broader research community}, providing a \textbf{versatile and efficient tool for simulating nonlinear behaviors of flexible structures}. This tutorial serves as a valuable resource for researchers and engineers working in \textit{applied mathematics, computational mechanics, structural design,} and \textit{engineering science}.

\section{General numerical framework}

In this chapter, we discuss the general numerical framework for the dynamical simulations of flexible structures. 
We first review the numerical method for solving a nonlinear structural dynamic problem and then discuss the overall procedure of our numerical implementation in MATLAB.

\subsection{Nonlinear structural dynamics}

In this subsection, we use a generalized degree of freedom (DOF) vector, $\mathbf{q} \in \mathcal{R}^{ \mathcal{N} \times 1}$, to formulate the structural dynamic equations based on Newton's second law \cite{hughes2003finite},
\begin{equation}
\mathbb{M} \ddot{\mathbf{q}}(t) + \mathbb{C} \dot{\mathbf{q}}(t) + \mathbb{K} \mathbf{q}(t) = \mathbf{F}^{\mathrm{ext}}(t),
\label{eq:dynamicEquation}
\end{equation}
where $t$ is the simulation time, $\mathbb{M} \in \mathcal{R}^{ \mathcal{N} \times \mathcal{N}} $ is the lumped mass matrix, $\mathbb{C} \in \mathcal{R}^{ \mathcal{N} \times \mathcal{N}} $ is the damping matrix, $\mathbb{K} \in \mathcal{R}^{ \mathcal{N} \times \mathcal{N}} $ is the tangential stiffness matrix, and $\mathbf{F}^{\mathrm{ext}} \in \mathcal{R}^{ \mathcal{N} \times 1}$ is the external force vector.
We will use $\dot{()}$ to represent the derivative with respect to time, i.e., $\dot{\mathbf{q}}$ is the generalized velocity vector, and $\ddot{\mathbf{q}}$ is the generalized acceleration vector.
The above dynamic equations of motion can be solved with given initial conditions, $\mathbf{q}(t=0)$ and $\dot{\mathbf{q}}(t=0)$, as well as the boundary conditions.
Also, it should be noted that the mass matrix, damping matrix, stiffness matrix, and external force vector, may not be a constant and could vary as a function of the DOF vector, $\mathbf{q}(t)$, which may make the dynamic equations of motion in Eq.~(\ref{eq:dynamicEquation}) nonlinear.
For example, the mass matrix would change for rocket launch, the damping matrix would change for viscoelastic material, and the tangential stiffness matrix would change for large deformations, which is one of the key challenges for the understanding of flexible structure mechanics, e.g., rods and shells.
In that case, we usually introduce the total elastic energy, $\mathcal{V}$, and find the local force vector and tangential stiffness matrix through a variational approach,
\begin{equation}
\mathbf{F}^{\text{int}} = - \frac {\partial \mathcal{V} } {\partial \mathbf{q}}, \; \mathrm{and} \;
\mathbb{K} = \frac {\partial^2 \mathcal{V}} {\partial \mathbf{q}^2},
\end{equation}
where $\mathcal{V}$ is the total elastic energy of the system and $\mathbf{F}^{\text{int}}$ the internal elastic force.
The key challenge for flexible structural dynamics is to formulate its total potential energy and find its first and second variations used in equations of motion, for which we will give detailed formulae for different types of structure.

Next, to solve the nonlinear dynamic equations of motion in Eq.~(\ref{eq:dynamicEquation}), the most straightforward way is the implicit Euler method, for its computational efficiency and numerical robustness \cite{huang2019newmark}. Specifically, the DOF vector $\mathbf q$, as well as its velocity $\dot{\mathbf q}$, can be updated from the current time step $ t_{k} $ to the next time step $ t_{k+1}$ by using the following rules,
\begin{equation}
\begin{aligned}
\mathbf{q}(t_{k+1}) & =  \mathbf{q}(t_{k}) +  \dot{\mathbf{q}}(t_{k+1}) \mathrm{d}t, \\
\dot{\mathbf{q}}(t_{k+1}) & = \dot{\mathbf{q}}(t_{k}) + \ddot{\mathbf{q}}(t_{k+1}) \mathrm{d}t, 
\end{aligned}
\end{equation}
where $\mathrm{d}t$ is the time step size.
Thus, the $\mathcal{N}$-sized continuous equation of motions in Eq.~(\ref{eq:dynamicEquation}) can be discretized and numerically solved step by step
\begin{equation}
\mathcal{E} \equiv \mathbb{M} \ddot{\mathbf{q}}(t_{k+1}) + \mathbb{C} \dot{\mathbf{q}}(t_{k+1}) - \mathbf{F}^{\text{int}}(t_{k+1}) - \mathbf{F}^{\text{ext}}(t_{k+1}) = \mathbf{0}.
\label{eq:implicitEuler}
\end{equation}
The Jacobian associated with Eq.~(\ref{eq:implicitEuler}) can be expressed as
\begin{equation}
\mathbb{J} = \frac {\mathbb{M}} {\mathrm{d}t^2} +  \frac {\mathbb{C}} {\mathrm{d}t} + \mathbb{K} - \frac{\partial \mathbf{F}^{\text{ext}}} {\partial \mathbf{q}}.
\label{eq:jacobian}
\end{equation}
To solve the nonlinear discrete equations of motion derived in Eq.~(\ref{eq:implicitEuler}), the iterative Newton-Raphson method is employed.
At the time step $t_{k+1}$, a new solution is first guessed on the basis of the previous state, i.e.,
\begin{equation}
\mathbf{q}_{n}(t_{k+1}) = \mathbf{q}(t_{k}) + \mathrm{d}t \; \dot{\mathbf{q}}(t_{k}).
\end{equation}
Then, it is optimized utilizing the gradient descent principle, such that the new solution at the $(n+1)$-th step is
\begin{equation}
\mathbf{q}_{n+1}(t_{k+1}) = \mathbf{q}_{n}(t_{k+1}) - \mathbb{J}_{n} \backslash \mathcal{E}_{n}.
\label{eq:newtonMethod}
\end{equation}
We will update the time step and move forward until the numerical error is smaller than tolerance, 
\begin{equation}
|| \mathcal{E}_{n}|| < \mathrm{tol}.
\end{equation}
It is worth noting that this framework is for dynamic simulation, and the static solution can be easily derived through the dynamic relaxation method \cite{barnes1999form, han2003study}; also, if the inertia and damping terms are ignored, the simulation can provide a static result by increasing the loading step by step, i.e., 
\begin{equation}
\mathbb{K} \mathbf{q}(t_{k+1}) = \mathbf{F}^{\text{ext}}(t_{k+1}),
\end{equation}
which is known as numerical continuation \cite{doedel1998auto, allgower2012numerical, huang2023bifurcations}.

\subsection{Simulation procedure}

\begin{figure}[ht]
\centering
\includegraphics[width=\columnwidth]{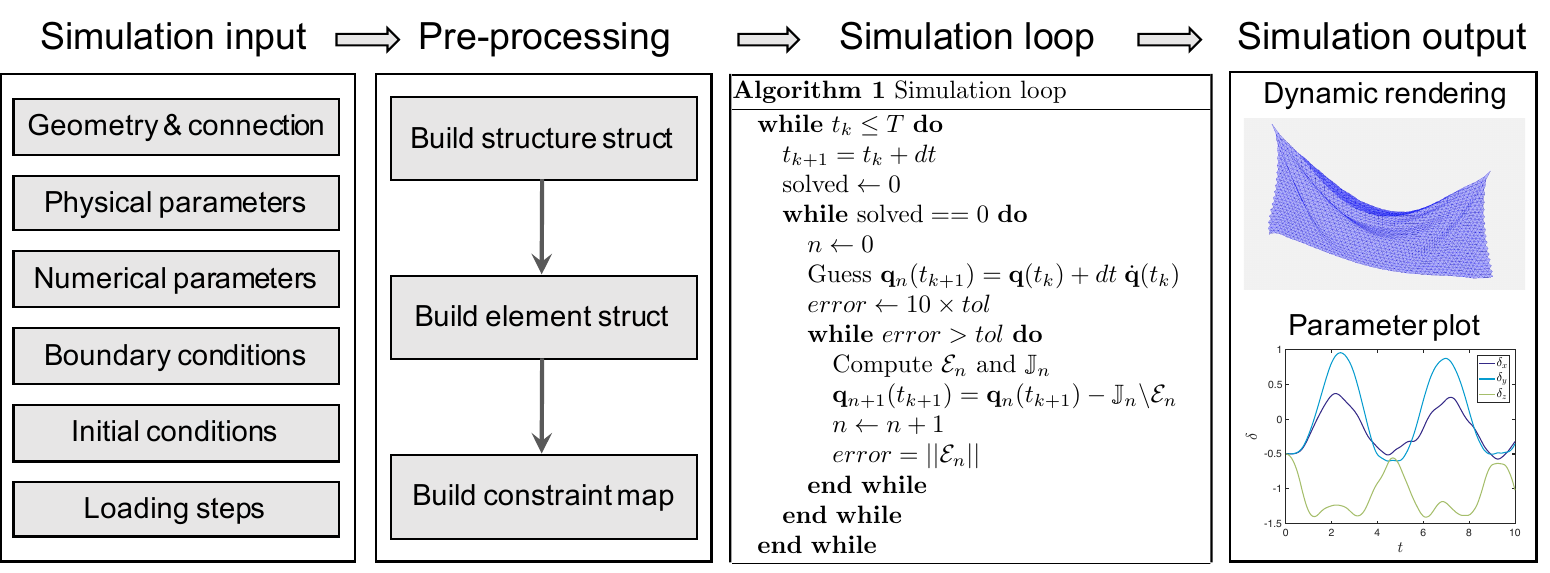}
\caption{\textbf{Simulation process flowchart.} The process begins by acquiring inputs, including geometric, material, numerical parameters, and initial conditions, followed by the construction of several fundamental data structures. The simulation loop then iteratively solves the governing equations. Finally, the results are processed and visualized through plots or animations.}
\label{fig:process_plot}
\end{figure}

In this subsection, we introduce our general numerical procedure for the nonlinear flexible structural dynamics.
The overall numerical procedure can be divided into four parts: (i) read simulation inputs, (ii) pre-processing, (iii) simulation loop, and (iv) output simulation results, referring to Fig.~\ref{fig:process_plot}.

\paragraph{Read simulation inputs.} The first step to setup a structural dynamic simulation is to select input data. There are six categories for input information: (1) The geometry and connection information for a system, e.g., nodal position, stretching element index, bending element index, triangle element index. (2) Physical parameters, e.g.,  Young's modulus, Poisson's ratio, material density, rod radius, plate thickness, gravitational acceleration, and damping viscosity. (3) Numerical parameters, such as total simulation time, time step size, numerical tolerance, and maximum iterations. (4) Boundary conditions, which are constrained index array and free index array. (5) Initial conditions, which are the positions and velocities of the DOF vector at $t = 0$. (6) Loading step, in which loadings are applied sequentially into the system and need to be defined during the time marching loop.

\paragraph{Pre-processing.} Next, the simulation needs to transfer the input data into several structs: (1) the system struct, e.g., rod struct or plate struct, which is the necessary information for a structure. (2) The element struct, which is the information for all local elements. (3) Mapping, which defines the relationship between the global DOF vector and the constrained and free DOF vectors.

\paragraph{Simulation loop.} Moving forward, we can perform the time marching scheme for the dynamic simulation, which is uniform for all structural dynamics. We use an implicit Euler method to update the system at each time step and employ Newton's method for numerical optimization within each time step. If the simulation fails to converge, check the time step size and material parameters. Reducing the time step or adjusting the stiffness values often resolves numerical instability.

\paragraph{Output simulation results.} Finally, the numerical data and the dynamic rendering will be generated during the simulation and would be ready for the result analysis.


\subsection{MATLAB implementation} 
\begin{figure}[h]
\centering
\includegraphics[width=\columnwidth]{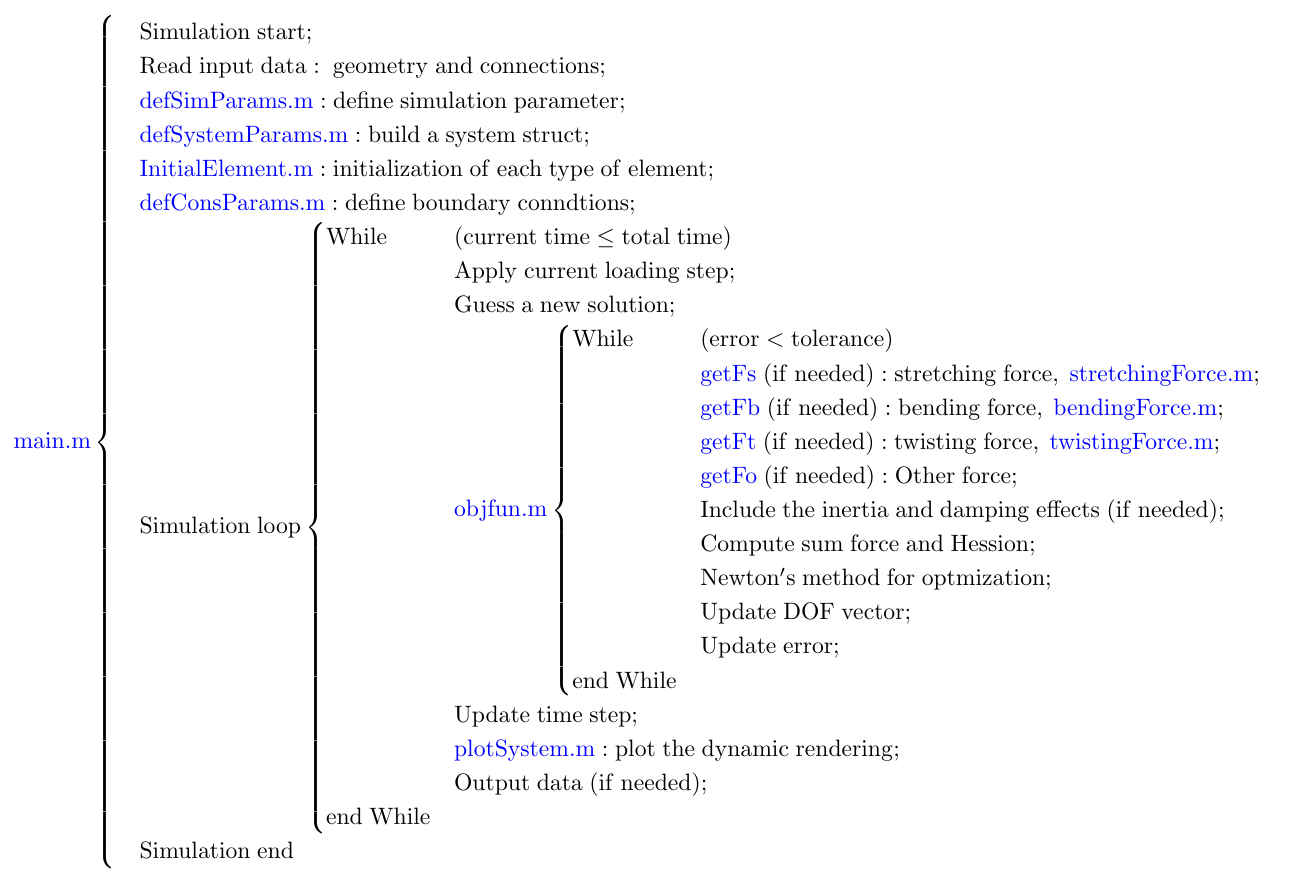}
\caption{\textbf{MATLAB implementation.} Overview of the function of each file and their interrelationships.}
\label{fig:algorithmPlot}
\end{figure}

In this subsection, we discuss the details of the MATLAB implementation.
The overall numerical algorithms can be found in Fig.~\ref{fig:algorithmPlot}, and the detailed explanations for each MATLAB file are as follows:

\begin{enumerate}

\item \textbf{main.m}: Main function, including reading inputs, building structs, and performing simulation. 

\item \textbf{defSimParams.m}:  Define the numerical parameters for the simulation.

\item \textbf{defSystemParams.m}:  Build a struct for the simulated system, e.g., beam, rod, or plate. 

\item \textbf{InitialElement.m}: The initialization of each type of element, e.g., stretching elements.

\item \textbf{defConsParams.m}: Define boundary conditions for the simulation.

\item \textbf{objfun.m}:  Update the time step by Euler method and optimize the solution by Newton's method.

\item \textbf{getForce.m}:  Loop over all elements to build the global force vector and the global stiffness matrix.

\item \textbf{elasticForce.m}: Detailed formulation for the local force vector and the local stiffness matrix.

\item \textbf{plotSystem.m}: Plot dynamic rendering for the numerical results.

\end{enumerate}
The implementation of this tutorial requires MATLAB, which should be installed prior to use.
The source code for the DDG Tutorial can be downloaded from the \href{https://github.com/weicheng-huang-mechanics/DDG_Tutorial}{GitHub Repository} \footnote{\url{https://github.com/weicheng-huang-mechanics/DDG_Tutorial}}.
For an enhanced learning experience, we have also developed a dedicated \href{http://www.ddgflexsim.com}{Website} \footnote{\url{http://www.ddgflexsim.com}}, which offers step-by-step instructions, interactive examples, and additional resources.

\section{Simulation of linear dynamic systems}

In this chapter, we establish the foundation of the overall framework by illustrating how the fundamental equations are implemented in a DDG-based simulation framework in a basic dynamic unit -- the mass-spring-damper system.

\subsection{Single-degree-of-freedom (S-DOF) system~\href{https://github.com/weicheng-huang-mechanics/DDG_Tutorial/tree/main/mass_spring_system/single_DOF}{\texorpdfstring{\faGithub}{GitHub} } } 

In this subsection, we examine the forced oscillation behavior of a simplest mass-spring-damper system -- single-degree-of-freedom (S-DOF) system.

\begin{figure}[h]
\centering
\includegraphics[width=\columnwidth]{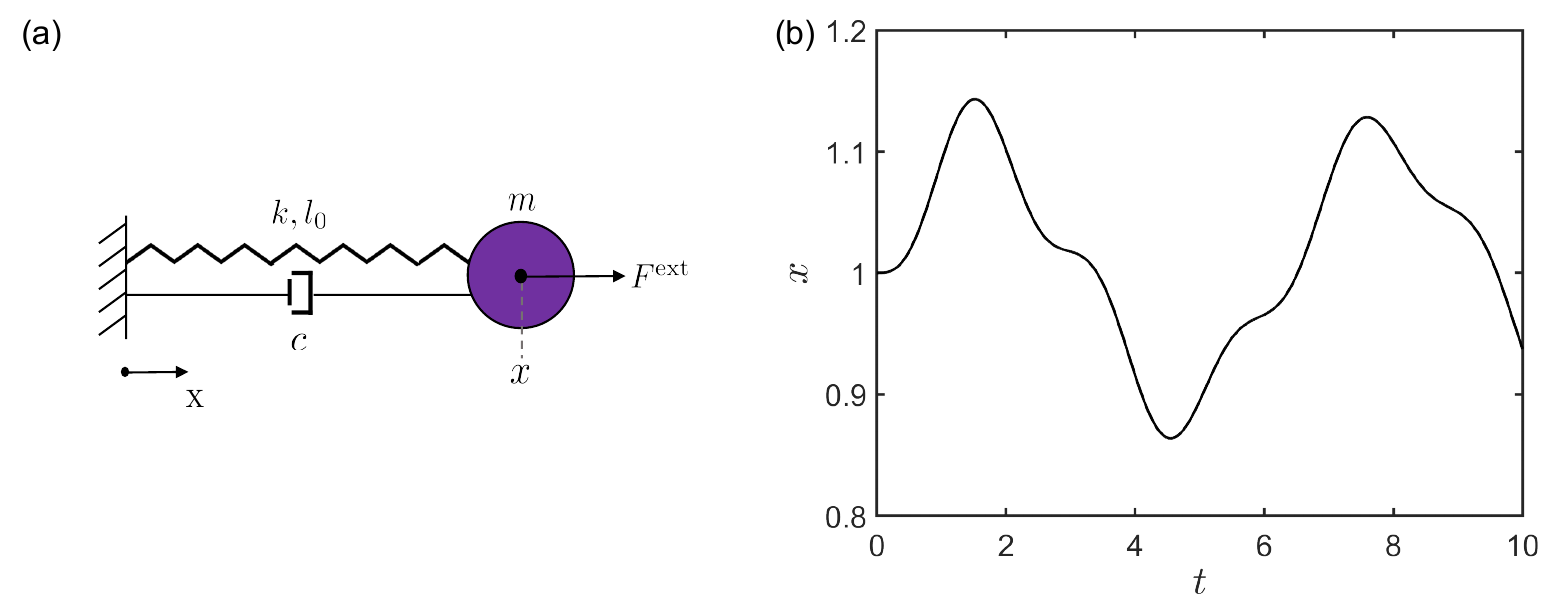}
\caption{\textbf{Oscillation of a S-DOF mass-spring-damper under a periodic external force.}
(a) Schematic of the system and the parameters.
(b) Nodal position, $x$, as a function of time, $t$.}
\label{fig:1dof_plot}
\end{figure}

\paragraph{Numerical formulation} 
As shown in Fig.~\ref{fig:1dof_plot}(a), the displacement of the mass is defined in the $x$-direction.
The total elastic energy is given by
\begin{equation}
\mathcal{V} = \frac{1} {2} k (x-l_0)^2,
\end{equation}
where $k$ is the spring constant, and $l_{0}$ is its natural length of the spring.
The elastic restoring force, derived using a variational approach, is given by $ -k(x-l_0)$. Since this is a linear S-DOF system, the stiffness matrix reduces to a scalar value $k$. 
Additionally, we introduce a time-dependent external force of the form $F^{\mathrm{ext}} = F_{0} \sin (\omega t)$, which drives the system periodically. Therefore, the equations of motion is
\begin{equation}
m \ddot{x} + c \dot{x} + k (x - l_{0}) - F_{0} \sin (\omega t) = 0.
\end{equation}

\paragraph{Simulation initialization} To initialize the simulation, the following inputs are used:

\begin{enumerate}

\item \textbf{Geometry and connection.} The spring length is $l_0=1.0\mathrm{~m}$. Since there is only a single node, we only need to input its own initial position at $x(t=0)=1.0\mathrm{~m}$, which is set to the spring's equilibrium position $x=l_0$ in this case. 

\item \textbf{Physical parameters.} (i) Mass $m = 1.0\mathrm{~kg}$. (ii) Damping viscosity $c = 0.1$. (iii) Spring stiffness $k = 10.0\mathrm{~N/m}$.

\item \textbf{Numerical parameters.} (i) Total simulation time $T=10.0\mathrm{~s}$. (ii) Time step size $\mathrm{d}t=0.01\mathrm{~s}$. (iii) Numerical force tolerance $\mathrm{tol}=1\times 10^{-6}$. 

\item \textbf{Boundary conditions.} No boundary condition is applied for this simple S-DOF system.

\item \textbf{Initial conditions.} (i) Initial position $ x(t=0) = 1.0 \mathrm{~m}$. (ii) Initial velocity $ \dot{x}(t=0) = 0.0 \mathrm{~m/s}$.

\item \textbf{Loading steps.} A periodic external force $F^{\rm ext} = F_{0}\sin(\omega t)$ is applied to the system, where the force magnitude is $F_{0}=1.0\mathrm{~N}$ and the frequency is $\omega=1.0\mathrm{~rad/s}$.

\end{enumerate}

\paragraph{Simulation results} 
The corresponding response of the system, including its displacement over time, provides a fundamental example of linear oscillatory motion. 
A schematic representation of the setup and the temporal evolution of the nodal position is illustrated in Fig.~\ref{fig:1dof_plot}(a), and the relationship between the position and time is shown in Fig.~\ref{fig:1dof_plot}(b). 
The dynamic rendering can be found~\href{https://github.com/weicheng-huang-mechanics/DDG_Tutorial/blob/main/assets/dof_single.gif}{here}. 

\subsection{Multiple-degree-of-freedom (M-DOFs) system~\href{https://github.com/weicheng-huang-mechanics/DDG_Tutorial/tree/main/mass_spring_system/multiple_DOF}{\texorpdfstring{\faGithub}{GitHub}}} 

In this subsection, we examine the forced oscillation behavior of a mass-spring-damper system with multiple-degree-of-freedom (M-DOFs).

\begin{figure}[h]
\centering
\includegraphics[width=\columnwidth]{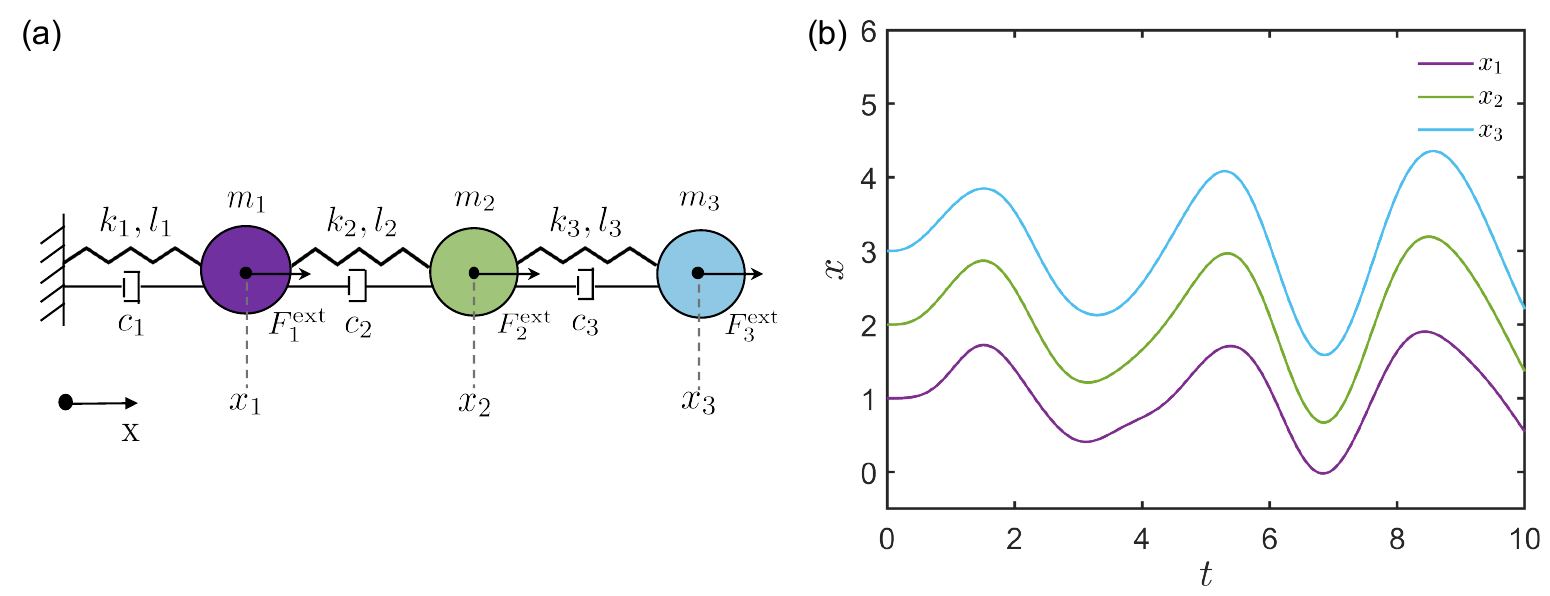}
\caption{\textbf{Oscillation of a M-DOFs mass-spring-damper system under periodic external forces.}
(a) Schematic of the system and the parameters.
(b) Node positions, $\{x_{1},x_{2},x_{3} \}$, as a function of time, $t$. }
\label{fig:3dof_plot}
\end{figure}

\paragraph{Numerical formulation} Building upon the S-DOF case, we extend the model to a M-DOFs system, as depicted in Fig.~\ref{fig:3dof_plot}(a). Here, we define a $N$-sized generalized DOF vector:
\begin{equation}
\mathbf{q} = [x_{1}; x_{2}; \ldots; x_{N}] \in \mathcal{R}^{ {N} \times 1},
\end{equation}
where each $x_i$ represents the displacement of the corresponding node in the system. 
The total elastic energy of the system is given by
\begin{equation}
\mathcal{V} =  \sum_{i=1}^{ {N}} \frac{1} {2} k_{i} (x_{i} - x_{i-1} -l_i)^2,
\end{equation}
where $k_{i} $ is the stiffness of the $i$-th spring and $l_{i}$ is its natural length.
We set $x_{0} = 0$ as the fixed reference point (the origin).
The elastic force, $\mathbf{F}^{\mathrm{int}}$, and the global stiffness matrix, $\mathbb{K}$, of the system can be derived using a variational approach, and the mass matrix and damping matrix can also be implemented straightforwardly. 
Notably, the stiffness matrix remains time-invariant, indicating that the system exhibits linear behavior.
Finally, the $N$-sized equations of motion of the M-DOFs system is
\begin{equation}
\mathbb{M} \ddot{\mathbf{q}} + \mathbb{C} \dot{\mathbf{q}} - \mathbf{F}^{\mathrm{int}} - \mathbf{F}^{\mathrm{ext}} = \mathbf{0}.
\label{eq:dynamicEquationFinal}
\end{equation}

\paragraph{Simulation initialization} To initialize the simulation, the following inputs are used:

\begin{enumerate}

\item \textbf{Geometry and connection.} The nodal positions are, $\mathbf{x} =[1.0, 2.0, 3.0]^{T}$ m, and the connection is between the two consecutive mass points. Here, we use $x_{0} = 0\mathrm{~m}$ to build a connection between the first mass point and the initial point.

\item \textbf{Physical parameters.} (i) Mass ${m}_{1} = {m}_{2} = {m}_{3} = 1.0\mathrm{~kg}$.  (ii) Damping viscosity $ c_{1} = c_{2} = c_{3} = 0.1$. (iii) Spring stiffness $k_{1} = 10.0\mathrm{~N/m}$, $k_{2} = 20.0\mathrm{~N/m}$, $k_{3} = 30.0\mathrm{~N/m}$. (iv) Natural spring length $l_{1} = l_{2} = l_{3} = 1.0\mathrm{~m}$.

\item \textbf{Numerical parameters.} (i) Total simulation time $T=10.0\mathrm{~s}$. (ii) Time step size $\mathrm{d}t=0.01\mathrm{~s}$. (iii) Numerical force tolerance $\mathrm{tol}=1\times 10^{-6}$. 

\item \textbf{Boundary conditions.} The mass point, $x_{0}$, is fixed; all other three mass points are free to move.

\item \textbf{Initial conditions.} (i) Initial position $\mathbf{q}(t=0) = [1.0, 2.0, 3.0]^{T} \mathrm{~m}$. (ii) Initial velocity is set to zeros, $\mathbf{\dot{q}}(t=0) = [0, 0, 0]^{T} \mathrm{~m/s}$.

\item \textbf{Loading steps.} The periodic force, $F_{i} \sin (\omega_{i} t) \; \mathrm{with} \; i \in [1,2,3]$, is applied into the system, where the magnitudes of the external force are ${F}_{1}^{\rm ext} = 1.0\mathrm{~N}$, ${F}_{2}^{\rm ext} = 2.0\mathrm{~N}$, ${F}_{3}^{\rm ext} =3.0\mathrm{~N}$ and the frequencies are $\omega_{1} = 1.0\mathrm{~rad/s}$, $\omega_{2} = 2.0\mathrm{~rad/s}$, $\omega_{3} = 3.0\mathrm{~rad/s}$.

\end{enumerate}

\paragraph{Simulation results} 

The temporal evolution of the node positions under external loading is examined, highlighting how M-DOFs systems capture richer vibratory characteristics compared to their S-DOF counterparts. 
The dynamic response of this M-DOFs system is provided in Fig.~\ref{fig:3dof_plot}(b).  
The dynamic rendering can be found~\href{https://github.com/weicheng-huang-mechanics/DDG_Tutorial/blob/main/assets/dof_multiple.gif}{here}. 

This fundamental simulation framework can now be readily extended to complex systems, e.g., planar beams, 3D rods, and shells, and the only difference is to formulate their total elastic potentials and the associated internal elastic force vector as well as the tangential stiffness matrix that are required for solving the nonlinear equations of motion.

\section{Flexible structures at different dimensions} 

In this chapter, we discuss the mathematical formulation and numerical implementation of the simulation for flexible structures at different dimensions, including (i) planar beam, (ii) 3D rod and ribbon, (iii) axisymmetric surface, (iv) 3D plate and shell, and (v) hollow net and gridshell.
Detailed implementation and numerical examples are also presented for illustration.

\subsection{Beam: planar curve} 

In this subsection, we start with the simplest planar beam system.
The beam's nonlinear mechanics, which can be described geometrically as a planar curve, can exhibit large deformation behavior. 
Governed by exact geometric curvature rather than linearized approximations, this theory dates back to Euler's elastic theory in the 18th century \cite{levien2008elastica, zhang2020configurations, liu2020tapered, 2502.00714}. Modern applications span compliant robotics \cite{huang2021swimming, li2025harnessing}, morphing aerospace structures\cite{huang2022nonlinear, huang2023contact, biesbroek2017deorbit}, metamaterials \cite{mehreganian2021structural, mehreganian2023impact, meng2020multi, mei2021mechanical, mao2022modular, fang2025large}, and bio-inspired systems \cite{tong2024inverse, pawashe2009modeling, pal2023programmable, tong2024inverse1}, with contemporary extensions addressing buckling and snapping behavior \cite{gomez2017critical, liu2021delayed, gomez2017critical, sano2018snap, radisson2023dynamic, radisson2023elastic, wang2024transient}. 
We first introduce the discrete formulation and the associated numerical procedure, followed by three benchmark demonstrations: (i) beam deflection under gravity, (ii) buckling of a compressive beam, and (iii) snap-through of a pre-buckled beam.

\begin{figure}[ht]
\centering
\includegraphics[width=\columnwidth]{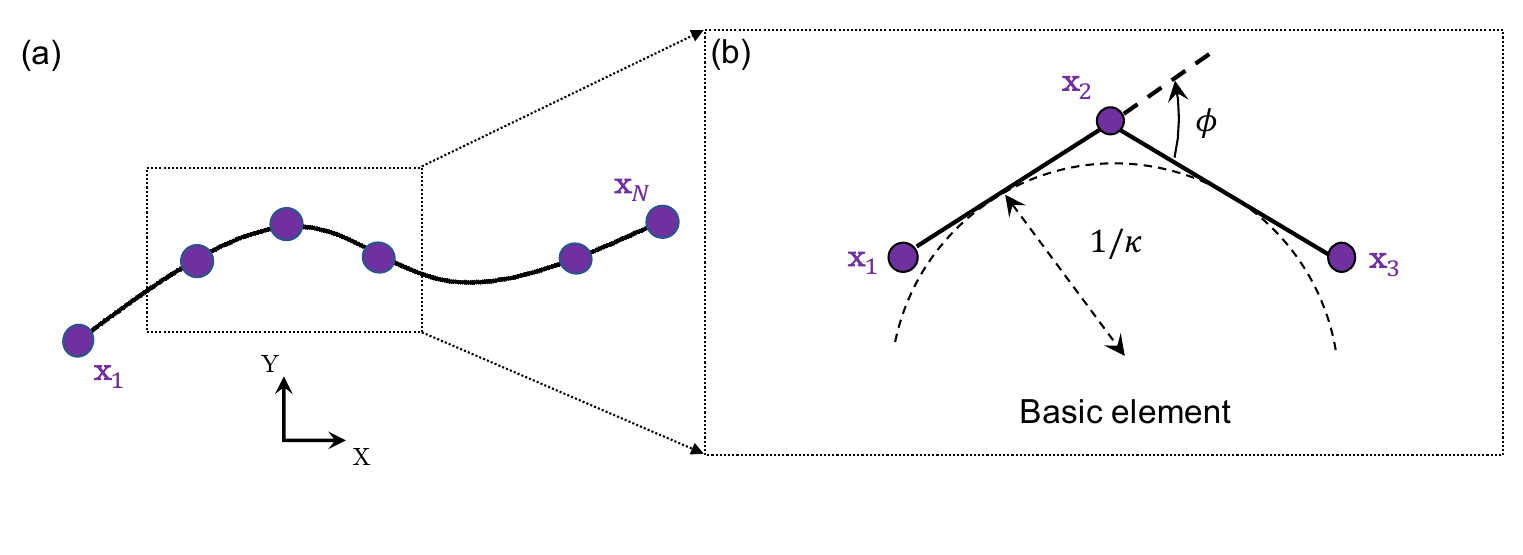}
\caption{\textbf{Planar beam model in DDG simulations.} (a) The beam is discretized into a series of nodes and segments. (b) Each pair of neighboring edges forms a bending element, characterized by a curvature $\kappa$, which is determined by the turning angle $\phi$. }
\label{fig:beam_model_plot}
\end{figure}

\subsubsection{Numerical formulation}
As shown in Fig.~\ref{fig:beam_model_plot}(a), the configuration of a planar beam is described by $N$ nodes, where each node is defined as $\mathbf{x}_{i} \equiv [x_{i}, y_{i}]^{T} \in \mathcal{R}^{2 \times 1}$. Therefore, the DOF vector can be expressed as
\begin{equation}
\mathbf{q} = [ \mathbf{x}_1; \mathbf{x}_2; \ldots; {\mathbf{x}_{N}} ] \in \mathcal{R}^{2N \times 1}.
\end{equation}
Two types of elements are used to capture the total elastic energies of a discrete planar beam: (i) stretching element and (ii) bending element, with $N_{s}$ and $N_{b}$ representing the number of each, respectively.
Note that if only the stretching element is considered, the bending-dominated beam structures would be reduced to stretching-dominated cable structures.

\paragraph{Stretching element} The stretching element is comprised of two connected nodes, defined as
\begin{equation}
\mathcal{S}: \{\mathbf{x}_{1}, \mathbf{x}_{2} \}.
\end{equation}
The local DOF vector of the stretching element is defined as 
\begin{equation}
\mathbf{q}^{s} \equiv [\mathbf{x}_{1}; \mathbf{x}_{2} ] \in \mathcal{R}^{4 \times 1}.
\end{equation}
The edge length is the $\mathcal{L}_{2}$ norm of the edge vector, defined as
\begin{equation}
l   =  || \mathbf{x}_{2}  -\mathbf{x}_{1} ||.
\end{equation}
The stretching strain is based on the uniaxial elongation of the edge, defined as 
\begin{equation}
{\varepsilon} = \frac {  l } {  \bar{l} } - 1.
\end{equation}
Hereafter, we use a bar on top to indicate the evaluation of the undeformed configuration, e.g., $\bar{l}$ is the edge length before deformation.
Using the linear elastic model, the total stretching energy is expressed as a quadratic function of the strain, following  the linear elastic constitutive law
\begin{equation}
E^s = \frac{1}{2} EA (\varepsilon)^2   \bar{l},
\end{equation}
where $EA $ is the local stretching stiffness.
The local stretching force vector, $\mathbf{F}^{s}_{\mathrm{local}} \in \mathcal{R}^{4 \times 1}$, as well as the local stretching stiffness matrix, $\mathbb{K}^{s}_{\mathrm{local}} \in \mathcal{R}^{4 \times 4}$, can be derived through a variational approach as
\begin{equation}
\mathbf{F}^{s}_{\mathrm{local}} = -\frac{\partial E^{s}}  {\partial \mathbf{q}^{s}}, \; \mathrm{and} \; \mathbb{K}^{s}_{\mathrm{local}} = \frac {\partial^2 E^{s}}  {\partial \mathbf{q}^{s} \partial \mathbf{q}^{s} }.
\end{equation}
The detailed formulation can be found in the MATLAB code.
Finally, the global stretching force vector,  $\mathbf{F}^{s}$, and the associated stiffness matrix, $\mathbb{K}^{s}$, are assembled by iterating over all stretching elements.

\paragraph{Bending element} Similarly, the bending element consists of two consecutive edges sharing a common node, defined as
\begin{equation}
\mathcal{B}: \{ \mathcal{S}_{1}, \mathcal{S}_{2}\}, \; \mathrm{with} \; \mathcal{S}_{1} : \{ \mathbf{x}_{1}, \mathbf{x}_{2} \} \; \mathrm{and} \; \mathcal{S}_{2} : \{ \mathbf{x}_{2}, \mathbf{x}_{3} \}.
\end{equation}
Here, we define $\mathbf{x}_{2}$ as the joint node, while $\mathbf{x}_{1}$ and $\mathbf{x}_{3}$ are the two adjacent nodes. Thus, the local DOF vector is defined as 
\begin{equation}
\mathbf{q}^{b} \equiv [\mathbf{x}_{1}; \mathbf{x}_{2}; \mathbf{x}_{3} ] \in \mathcal{R}^{6 \times 1}.
\end{equation}
The two edge vectors are
\begin{equation}
\begin{aligned}
\mathbf{e}_{1} &= \mathbf{x}_{2}  -\mathbf{x}_{1},\\
\mathbf{e}_{2} &= \mathbf{x}_{3}  -\mathbf{x}_{2}.
\end{aligned}
\end{equation}
The Voronoi length of the bending element is the average of the lengths of the two edges, defined as 
\begin{equation}
l = \frac{1} {2}(    || \mathbf{e}_{1}  || +  || \mathbf{e}_{2} ||  ).
\end{equation}
The bending curvature is associated with the turning angle between the two connecting edges, defined as
\begin{equation}
{\kappa} = \frac { 2 \tan ( {\phi / {2} )}} { l } ,
\end{equation}
as shown in Fig.~\ref{fig:beam_model_plot}(b).
The discrete bending energy is also expressed as a quadratic function of the curvature, given by
\begin{equation}
E^{b} = \frac{1}{2}  EI {(\kappa - \bar{\kappa})^2 }   \bar{l},
\end{equation}
where $EI$ represents the local bending stiffness.
The local bending force vector, $\mathbf{F}^{b}_{\mathrm{local}} \in \mathcal{R}^{6 \times 1}$, as well as the local bending stiffness matrix, $\mathbb{K}^{b}_{\mathrm{local}} \in \mathcal{R}^{6 \times 6}$, can be derived using a variational approach as
\begin{equation}
\mathbf{F}^{b}_{\mathrm{local}} = -\frac{\partial E^{b}}  {\partial \mathbf{q}^{b}}, \; \mathrm{and} \; \mathbb{K}^{b}_{\mathrm{local}} = \frac {\partial^2 E^{b}}  {\partial \mathbf{q}^{b} \partial \mathbf{q}^{b} }.
\end{equation}
The detailed formulation can be found in the MATLAB code.
Finally, the global bending force vector,  $\mathbf{F}^{b}$, and the associated stiffness matrix, $\mathbb{K}^{b}$, are assembled by iterating over all bending elements.

\paragraph{Equations of motion} With the formulation of the internal elastic force and the associated stiffness matrix, we can incorporate the inertial and damping effects to derive the dynamic equations of motion.
Here, the mass matrix, $\mathbb{M}$, is time-invariant and can be easily obtained based on the element size and material density. 
We then employ a simple damping matrix, which is linearly related to the mass matrix with a damping coefficient $\mu$, i.e., $\mathbb{C} = \mu \mathbb{M}$.
Finally, the equations of motion for the planar beam system are
\begin{equation}
\mathbb{M} \ddot{\mathbf{q}} + \mu \mathbb{M} \dot{\mathbf{q}} - \mathbf{F}^{s} - \mathbf{F}^{b} - \mathbf{F}^{\text{ext}} = \mathbf{0}.
\end{equation}
The implicit Euler method and Newton's method are used to solve the $(2N)$-sized nonlinear dynamic system.

\subsubsection{Case 1: Deflection of a cantilever beam under gravity~\href{https://github.com/weicheng-huang-mechanics/DDG_Tutorial/tree/main/2d_curve/case_1}{\texorpdfstring{\faGithub}{GitHub}}}


In this case study, we examine a flexible beam's free oscillation behavior under gravity's influence. 
One end of the beam is fully constrained, while the other end remains free. 
Displaced from its initial horizontal position, the beam oscillates due to the restoring elastic forces, with its motion governed by the interplay between elasticity, inertia, and external gravitational force. 
The case highlights the ability of the DDG model to accurately capture large deformations in beam structures.

\paragraph{Simulation initialization} To initialize the simulation, the following inputs are used:

\begin{enumerate}

\item \textbf{Geometry and connection.} (i) Nodal positions: the position of the nodes, $\mathbf{q}(t=0)$, with a total of $N=40$ and the beam length $L=1.0\mathrm{~m}$. (ii) Stretching elements: connection of every two consecutive nodes, with a total of $N_{s}=39$. (iii) Bending elements: connection of every two consecutive edges, with a total of $N_{b}=38$.

\item \textbf{Physical parameters.} (i) Young's modulus, $E=100$ MPa. (ii) Material density, $\rho=1000\mathrm{~kg/m^3}$. (iii) Cross-sectional radius, $r_{0}=0.01\mathrm{~m}$. (iv) Damping viscosity, $\mu = 0.1$. (v) Gravitational field, $ \mathbf{g}=[0.0, -10.0]^{T}\mathrm{~m/s^2}$. (vi) The overall simulation is dynamic, i.e., $ \mathrm{ifStatic} = 0$.

\item \textbf{Numerical parameters.} (i) Total simulation time, $T=5.0\mathrm{~s}$. (ii) Time step size, $\mathrm{d}t=0.01\mathrm{~s}$. (iii) Numerical force tolerance, $\mathrm{tol}=1\times 10^{-4}$. (iv) Maximum iterations, $N_{\mathrm{iter}}=10$.

\item \textbf{Boundary conditions.} The left tip of the beam is clamped by fixing the first two nodes, $\{\mathbf{x}_{1}, \mathbf{x}_{2} \}$, thus the constrained index array is $\mathcal{FIX} = [1,2,3,4]^T$.

\item \textbf{Initial conditions.} (i) Initial position is input from the nodal positions. (ii) Initial velocity is set to zeros for all nodes.

\item \textbf{Loading steps.} External gravitational force is applied to each node throughout the simulation.

\end{enumerate}

\begin{figure}[hbt]
    \centering
    \includegraphics[width=\linewidth]{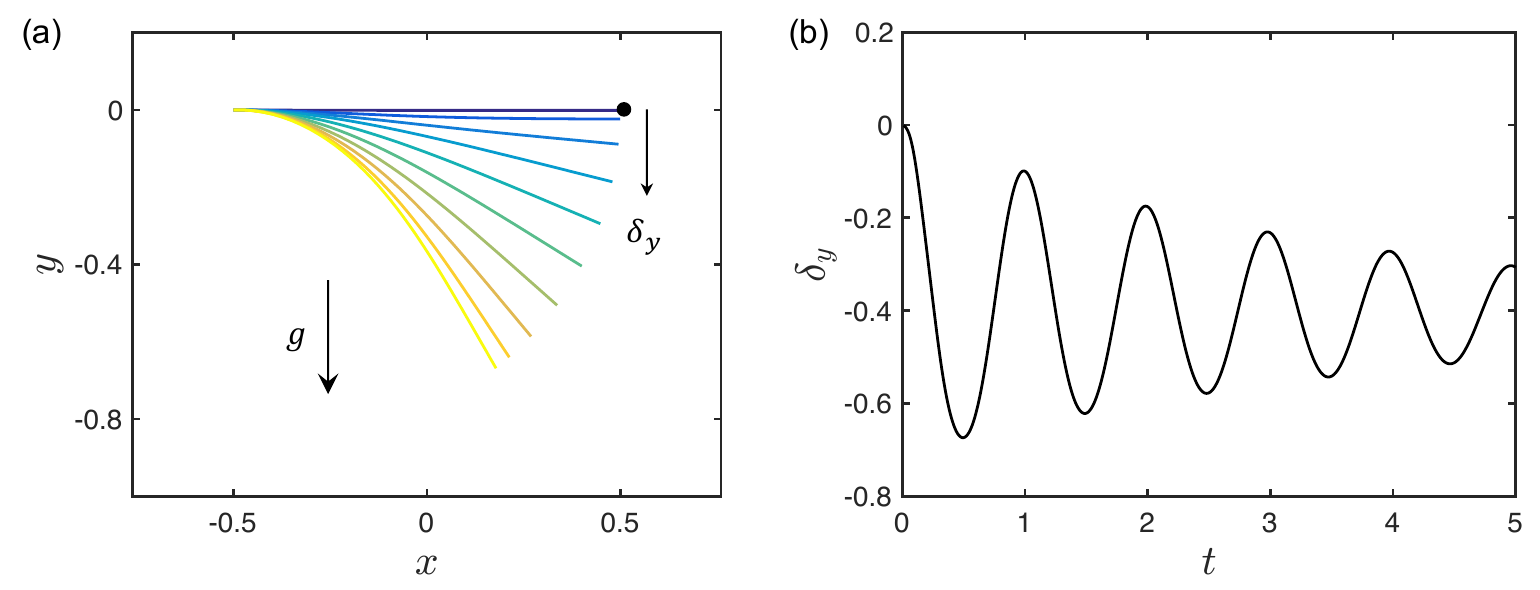}
    \caption{\textbf{Cantilever beam under gravity.} 
    {(a)} Change in the beam configuration under gravity as indicated by the arrow, where the right end freely deflects and the left is fixed. The deformation curves shown are half a wavelength of oscillation.
    {(b)} Beam tip displacement, $\delta_y$, as a function of time, $t$. }
    \label{fig:beam_case_1_plot}
\end{figure}

\paragraph{Simulation results} 
The beam undergoes deformations under gravity and oscillates along a nonlinear trajectory, as illustrated in Fig.~\ref{fig:beam_case_1_plot}(a).
The time evolution of the vertical displacement $\delta_y$ at the free end is depicted in Fig.~\ref{fig:beam_case_1_plot}(b), revealing the characteristic vibratory dynamic motion. 
Due to the presence of the damping effect, the oscillation amplitude gradually diminishes over time, leading to a progressive decay of the beam's oscillation response. 
The dynamic rendering can be found~\href{https://github.com/weicheng-huang-mechanics/DDG_Tutorial/blob/main/assets/beam_1.gif}{here}.

\subsubsection{Case 2: Buckling of a compressive beam~\href{https://github.com/weicheng-huang-mechanics/DDG_Tutorial/tree/main/2d_curve/case_2}{\texorpdfstring{\faGithub}{GitHub}}}

In this case study, we examine the classical Euler buckling phenomenon of a slender beam under axial compression.
The beam is fixed at both ends, with a gradually increasing compressive load applied at the right end.
As the load increases, the beam transitions from a straight configuration to a buckled state, illustrating the onset of structural instability.
The case highlights the ability of the DDG model to accurately capture buckling behavior in slender structures.


\paragraph{Simulation initialization} To initialize the simulation, the following inputs are used:

\begin{enumerate}

\item \textbf{Geometry and connection.} (i) Nodal positions: the position of the nodes, $\mathbf{q}(t=0)$, with a total of $N=40$ and the beam length $L=1.0\mathrm{~m}$. (ii) Stretching elements: connection of every two consecutive nodes, with a total of $N_{s}=39$. (iii) Bending elements: connection of every two consecutive edges, with a total of $N_{b}=38$.

\item \textbf{Physical parameters.} (i) Young's modulus, $E=10\mathrm{~MPa}$. (ii) Material density, $\rho=1000\mathrm{~kg/m^3}$. (iii) Cross-sectional radius, $r_{0} = 0.01\mathrm{~m}$. (iv) Damping viscosity, $\mu = 0.1$. (v) Gravitational field, $ \mathbf{g}= [0.0, 0.1]^{T}\mathrm{~m/s^2}$. (vi) The overall simulation is static, i.e., $ \mathrm{ifStatic} = 1$.

\item \textbf{Numerical parameters.} (i) Total simulation time,  $T=10.0$ s. (ii) Time step size, $\mathrm{d}t=0.01 \mathrm{~s}$. (iii) Numerical force tolerance, $\mathrm{tol} = 1\times10^{-4}$. (iv) Maximum iterations, $N_{\mathrm{iter}}=10$.

\item \textbf{Boundary conditions.} The first two nodes, $\{\mathbf{x}_{1}, \mathbf{x}_{2} \}$, and the last two nodes, $\{\mathbf{x}_{39}, \mathbf{x}_{40} \}$, are fixed to achieve clamped-clamped boundary conditions, thus $\mathcal{FIX} = [1,2,3,4,77,78,79,80]^{T}$. 

\item \textbf{Initial conditions.} (i) Initial position is input from the nodal positions. (ii) Initial velocity is set to zeros.

\item \textbf{Loading steps.} (i) Perturbation step: a small perturbation to the initial horizontal configuration is created by applying a small gravitational force (with $ \mathbf{g}=[0.0,0.1]^{T}\mathrm{~m/s^2}$) when $t \le 1.0\mathrm{~s}$. (ii) Compression step: when $t>1.0$ s, a displacement is applied to the last two nodes along the negative $X$ axis with a speed $v_{0} = 0.1\mathrm{~m/s}$, until the compression distance is larger than the target value, $\Delta X \ge 0.6$ m.

\end{enumerate}

\begin{figure}[ht]
    \centering
    \includegraphics[width=\linewidth]{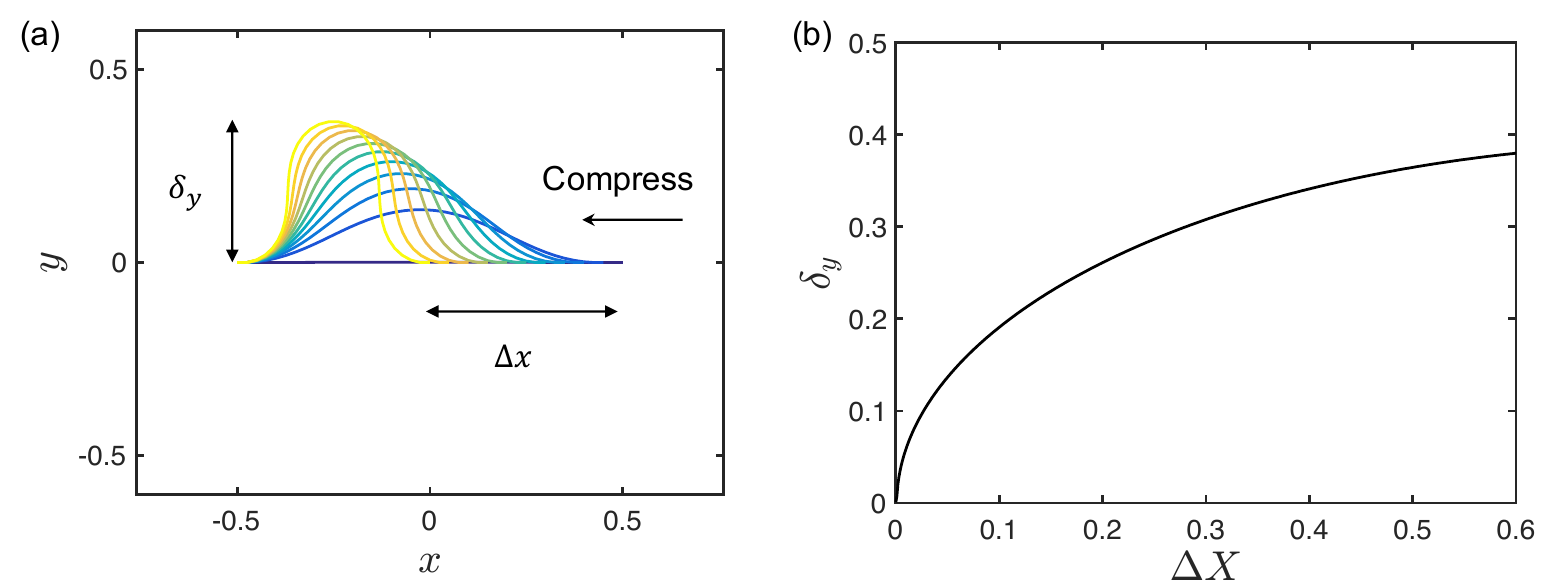}
    \caption{\textbf{Buckling instability of a compressive beam.}
    {(a)} Time evolution of the buckled beam configuration. The initially horizontal beam is compressed from the right end. 
    {(b)} Deflection of the beam at the midpoint, $\delta_y$, increases with the amount of horizontal compression to the right end, $\Delta X$. }
    \label{fig:beam_case_2_plot}
\end{figure}

\paragraph{Simulation results} 
As axial compression is gradually applied to the right end, the structure exhibits significant out-of-plane displacements, demonstrating strong geometric nonlinearity, as depicted in Fig.~\ref{fig:beam_case_2_plot}(a). 
In Fig.~\ref{fig:beam_case_2_plot}(b), we show that the transverse displacement at the midpoint of the beam, denoted by $\delta_y$, increases nonlinearly with the applied displacement loading $\Delta X$. 
A sudden increase in deflection is observed for very small values of $\Delta X$, a characteristic feature of structural instability. 
These results illustrate the effectiveness of the simulation framework in capturing the nonlinear and bifurcation behavior of slender structures under compression.
The dynamic rendering can be found~\href{https://github.com/weicheng-huang-mechanics/DDG_Tutorial/blob/main/assets/beam_2.gif}{here}.

\subsubsection{Case 3: Snapping of a pre-buckled beam~\href{https://github.com/weicheng-huang-mechanics/DDG_Tutorial/tree/main/2d_curve/case_3}{\texorpdfstring{\faGithub}{GitHub}}}

In this case study, we examine the snap-through behavior of a pre-buckled beam under rotational input. 
The beam is clamped at both ends, with one end gradually rotated to induce snap-through instability.
As the rotation increases, the beam undergoes a sudden transition, driven by the interplay between bending stiffness and applied rotational forces.
The case highlights the ability of the DDG model to effectively capture discontinuous behavior (e.g., snap-through) in slender structures.

\paragraph{Simulation initialization} To initialize the simulation, the following inputs are used:

\begin{enumerate}

\item \textbf{Geometry and connection.} (i) Nodal positions: the position of the nodes, $\mathbf{q}(t=0)$, with a total of $N=40$ and the beam length $L=1.0\mathrm{~m}$. (ii) Stretching elements: connection of every two consecutive nodes, with a total of $N_{s}=39$. (iii) Bending elements: connection of every two consecutive edges, with a total of $N_{b}=38$.

\item \textbf{Physical parameters.} (i) Young's modulus, $E=10\mathrm{~MPa}$. (ii) Material density, $\rho=1000\mathrm{~kg/m^3}$. (iii) Cross-sectional radius, $r_{0} = 0.01\mathrm{~m}$. (iv) Damping viscosity $\mu = 0.1$. (v) The overall simulation is dynamic, i.e., $ \mathrm{ifStatic} = 0$.

\item \textbf{Numerical parameters.} (i) Total simulation time,  $T=20.0$ s. (ii) Time step size, $\mathrm{d}t=0.01 \mathrm{~s}$. (iii) Numerical force tolerance, $\mathrm{tol} = 1\times10^{-4}$. (iv) Maximum iterations, $N_{\mathrm{iter}}=10$.

\item \textbf{Boundary conditions.} The first two nodes, $\{\mathbf{x}_{1}, \mathbf{x}_{2} \}$, and the last two nodes, $\{\mathbf{x}_{39}, \mathbf{x}_{40} \}$, are fixed to achieve clamped-clamped boundary conditions, thus $\mathcal{FIX} = [1,2,3,4,77,78,79,80]^{T}$. 

\item \textbf{Initial conditions.} (i) Initial position is input from the nodal positions. (ii) Initial velocity is set to zeros.

\item \textbf{Loading steps.} (i) Compression step: when $t < 3.0$ s, a displacement is applied to both the first two nodes and the last two nodes along the $X$ axis with a speed $v_{0} = 0.1\mathrm{~m/s}$ until the compressive distance reaches to the target, $\Delta X \ge 0.3$ m. Note that here we use $ \mathbf{g}=[0.0,10.0]^T\mathrm{~m/s^2}$ to induce the bifurcation direction. (ii) Delete Perturbation: when $3.0 \; \mathrm{s} \le t \le 5.0 \; \mathrm{s}$, the gravity force is removed to delete the influence of the perturbation, $ \mathbf{g}=[0.0,0.0]^T\mathrm{~m/s^2}$. (iii) Rotate left end: when $t > 5.0$ s, the left clamped edge is rotated to induce the snap-through of the bistable beam, and the rotational speed is $\omega = 1.0$ rad/s.

\end{enumerate}

\begin{figure}[ht]
    \centering
    \includegraphics[width=\linewidth]{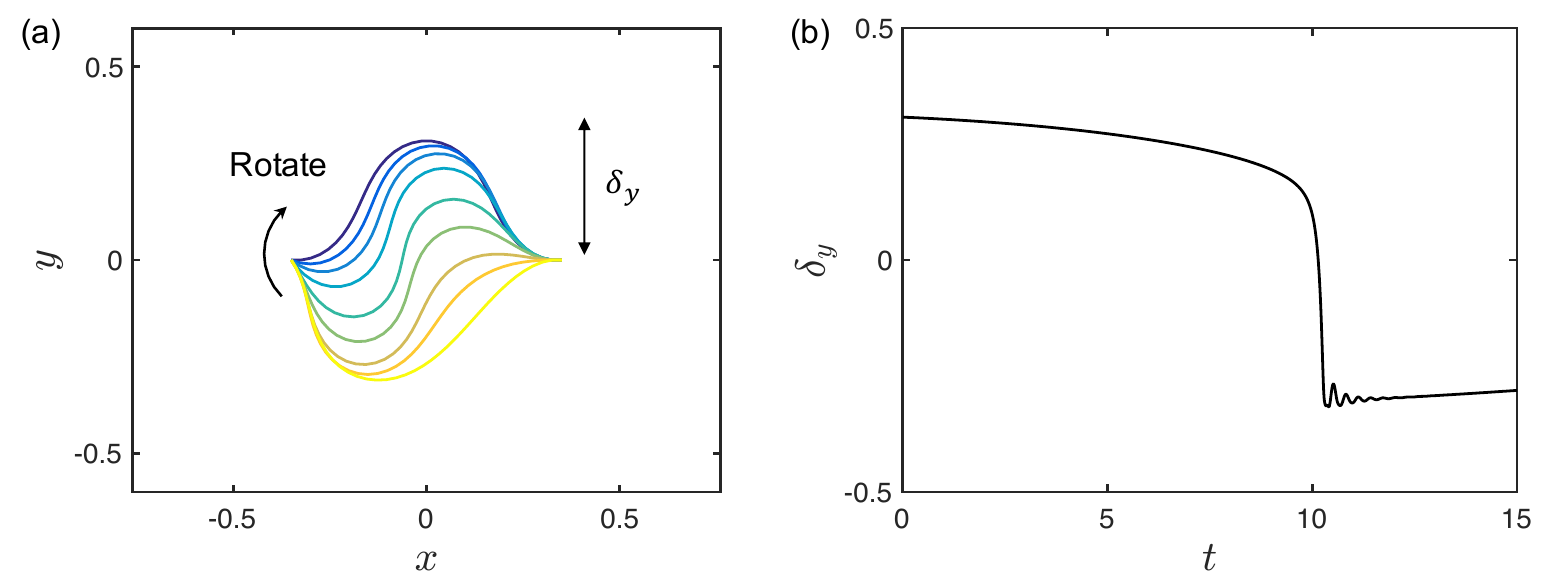}
    \caption{\textbf{Snap-through of a pre-buckled beam.}
    {(a)} Time evolution of the beam configuration during the snap-through process as the left end rotates.
    {(b)} Vertical displacement, $\delta_y$, of the midpoint with time, $t$. Snapping occurs at around $t=10.0\mathrm{~s}$. }
    \label{fig:beam_case_3_plot}
\end{figure}

\paragraph{Simulation results} 

The configuration of the beam during the snap-through process are shown in Fig.~\ref{fig:beam_case_3_plot}(a). 
In Fig.~\ref{fig:beam_case_3_plot}(b), we show the midpoint height $\delta_{y}$ as a function of loading time $t$. 
The midpoint height presents a sudden change when $t = 10.0$ s, which indicates the snap translation. 
These results illustrate the effectiveness of the simulation framework in capturing the discontinuous saddle-node bifurcation behavior of slender structures.
The dynamic rendering can be found~\href{https://github.com/weicheng-huang-mechanics/DDG_Tutorial/blob/main/assets/beam_3.gif}{here}.

\subsection{Rod \& ribbon: 3D curve}

In this subsection, we present the discrete energy formulation for 3D slender rods and ribbons using the discrete differential geometry (DDG) method.
The basic rod equation can be derived from the Cosserat rod theory \cite{altenbach2013cosserat, gazzola2018forward}; however, due to the slenderness of the structures, the stretching and shearing deformation modes can be ignored thus a simplified Kirchhoff rod theory is more common in the mechanics community \cite{dill1992kirchhoff, audoly2000elasticity}.
Kirchhoff rod model is largely used for the understanding fundamental mechanics \cite{miller2014shapes, lazarus2013continuation, goldstein2000bistable, audoly2005fragmentation}, as well as the design of engineering structures, e.g., musculoskeletal architectures \cite{zhang2019modeling}, DNA helical structure \cite{manning1996continuum, djurivckovic2013twist}, growth of plants \cite{goriely1998spontaneous, moulton2013morphoelastic, lessinnes2017morphoelastic, moulton2020morphoelastic}, bio-inspired soft robot \cite{chang2023energy, jawed2015propulsion, kaczmarski2025ultra, kaczmarski2024minimal}, marine cable deployment \cite{jawed2014coiling, tong2024sim2real}, etc. 
As for the ribbons, we also use a similar approach, i.e., we employ a simple rod model with anisotropic bending stiffness, instead of using a complex Sadowsky or Wunderlich models \cite{dias2015wunderlich, audoly2021one, korner2021simple, audoly2023analysis, huang2022discrete}, for its efficiency and accuracy when handling the nonlinear deformation of slender systems \cite{yu2019bifurcations, huang2020shear, huang2024exploiting}. 
We first introduce the discrete formulation and the associated numerical procedure \cite{bergou2008discrete, bergou2010discrete, jawed2018primer}, next examine three challenging cases: (i) the deformation of a helix under gravity, (ii) shear-induced bifurcation in a pre-buckled ribbon, and (iii) the growth of an annular ribbon.

\subsubsection{Numerical formulation}

\begin{figure}[ht]
\centering
\includegraphics[width=\columnwidth]{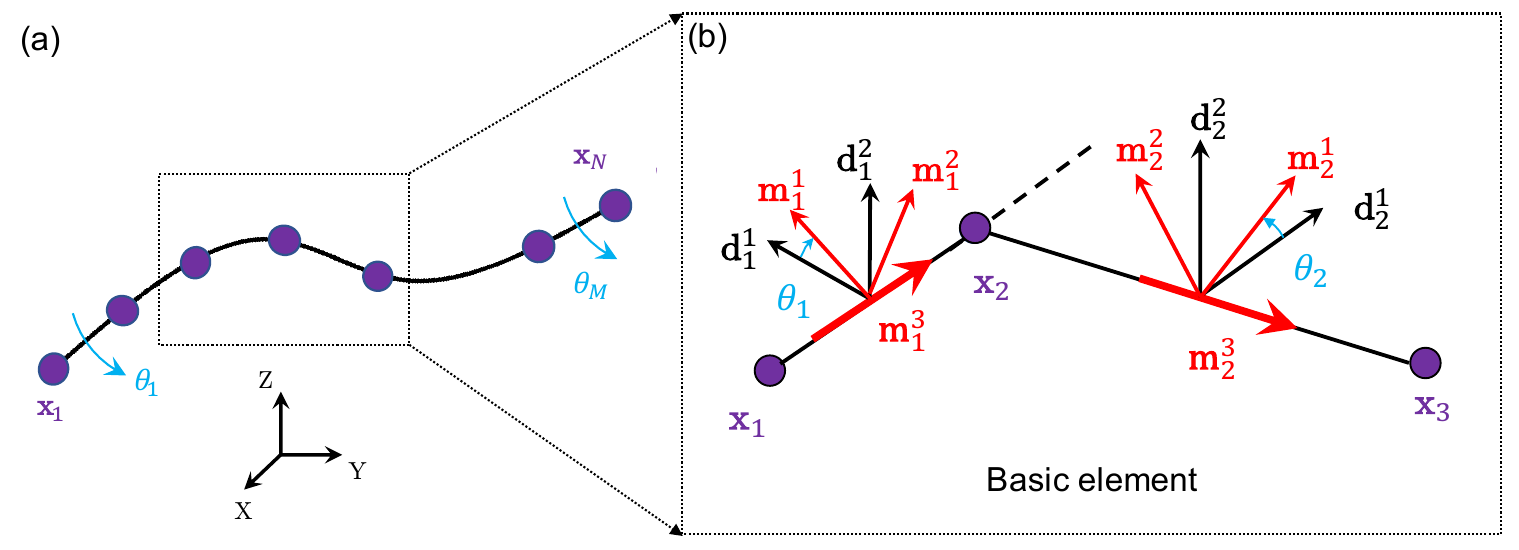}
\caption{ \textbf{3D rod model in DDG simulations.} (a) The rod is discretized into $N$ nodes and $M$ edges. (b) A basic element for bending and twisting calculations.}
\label{fig:rod_model_plot}
\end{figure}

Here, we formulate the mechanics of 3D slender structures, such as rods or ribbons, whose motion can be simplified to that of a central line, represented as a 3D curve.
As shown in Fig. \ref{fig:rod_model_plot}(a), to capture the configuration of the curve, we need $N$ nodes and $M$ edges. 
Each node is denoted as $\mathbf{x}_{i} \equiv [x_{i}, y_{i}, z_{i}]^{T} \in \mathcal{R}^{3 \times 1}$.
To define an edge, two frames are required, e.g., for the $j$-th edge, we introduce an adaptive reference frame, $\{ \mathbf{d}_{j}^{1}, \mathbf{d}_{j}^{2}, \mathbf{d}_{j}^{3}\}$, and a local material frame, $\{ \mathbf{m}_{j}^{1}, \mathbf{m}_{j}^{2}, \mathbf{m}_{j}^{3}\}$.
These frames share a common tangent direction, and the rotational angle between the two frames along the tangent direction is defined as $\theta_{j}$, i.e.,
\begin{equation}
\begin{aligned}
\mathbf{m}_{j}^{1} &= \mathbf{d}_{j}^{1} \cos(\theta_{j}) + \mathbf{d}_{j}^{2} \sin(\theta_{j}), \\
\mathbf{m}_{j}^{2} &= \mathbf{d}_{j}^{2} \cos(\theta_{j}) -\mathbf{d}_{j}^{1} \sin(\theta_{j}), \\
\mathbf{m}_{j}^{3} & \equiv  \mathbf{d}_{j}^{3}.
\end{aligned}
\end{equation}
The total DOF vector is defined as
\begin{equation}
\mathbf{q} = [ \mathbf{x}_1; \mathbf{x}_2; \ldots; {\mathbf{x}_{N}}; \theta_{1}; \theta_{2}; \ldots; \theta_{M} ] \in \mathcal{R}^{ (3N+M) \times 1}.
\end{equation}
Three types of elements are used to capture the mechanics of a discrete slender structure in 3D space (see Fig. \ref{fig:rod_model_plot}(b)): (i) stretching element, (ii) bending element, and (iii) twisting element. 
Consistent with the previous section, the number of stretching elements is denoted as $N_{s}$, while the number of bending and twisting elements are the same and denoted as $N_{b}$.
Similar to the previous planar beam case, if only the stretching element is considered, the 3D rod structures would reduce to the 3D cable structures.


\paragraph{Stretching element} The stretching element is comprised of two connected nodes, defined as
\begin{equation}
\mathcal{S}: \{\mathbf{x}_{1}, \mathbf{x}_{2}, \theta_{1} \}.
\end{equation}
As the formulation of stretching strain is independent of the twisting angle, $\theta_{1}$, the local DOF vector depends only on the nodes and is defined as 
\begin{equation}
\mathbf{q}^{s} \equiv [\mathbf{x}_{1}; \mathbf{x}_{2} ] \in \mathcal{R}^{6 \times 1}.
\end{equation}
The edge length is the $\mathcal{L}_{2}$ norm of the edge vector, defined as
\begin{equation}
l   =  || \mathbf{x}_{2}  -\mathbf{x}_{1} ||.
\end{equation}
The stretching strain is based on the uniaxial elongation of edge, defined as
\begin{equation}
{\varepsilon} = \frac {  l } {  \bar{l} } - 1.
\end{equation}
Hereafter, we use a bar on top to indicate the evaluation of the undeformed configuration, e.g., $\bar{l}$ is the edge length before deformation.
Using the linear elastic model, the total stretching energy is expressed as a quadratic function of the strain following the linear elastic constitutive law
\begin{equation}
E^s = \frac{1}{2} EA (\varepsilon)^2   \bar{l},
\end{equation}
where $EA $ is the local stretching stiffness. 
The local stretching force vector, $\mathbf{F}^{s}_{\mathrm{local}} \in \mathcal{R}^{6 \times 1}$, as well as the local stretching stiffness matrix, $\mathbb{K}^{s}_{\mathrm{local}} \in \mathcal{R}^{6 \times 6}$, can be derived through a variational approach as
\begin{equation}
\mathbf{F}^{s}_{\mathrm{local}} = -\frac{\partial E^{s}}  {\partial \mathbf{q}^{s}}, \; \mathrm{and} \; \mathbb{K}^{s}_{\mathrm{local}} = \frac {\partial^2 E^{s}}  {\partial \mathbf{q}^{s} \partial \mathbf{q}^{s}}.
\end{equation}
The detailed formulation can be found in the MATLAB code.
Finally, the global stretching force vector,  $\mathbf{F}^{s}$, and the associated stiffness matrix, $\mathbb{K}^{s}$, can be assembled by iterating over all stretching elements.

\paragraph{Bending element} Similarly, we extend the bending element to a 3D discrete slender structure as
\begin{equation}
\mathcal{B}: \{ \mathcal{S}_{1}, \mathcal{S}_{2}\}, \; \mathrm{with} \; \mathcal{S}_{1} : \{ \mathbf{x}_{1}, \mathbf{x}_{2}, \theta_{1} \} \; \mathrm{and} \; \mathcal{S}_{2} : \{ \mathbf{x}_{2}, \mathbf{x}_{3}, \theta_{2} \}.
\end{equation}
Here, we define $\mathbf{x}_{2}$ as the joint node, and $\mathbf{x}_{1}$ and $\mathbf{x}_{3}$ are the two adjacent nodes. Thus, the local DOF vector is defined as 
\begin{equation}
\mathbf{q}^{b} \equiv [\mathbf{x}_{1}; \mathbf{x}_{2}; \mathbf{x}_{3}; \theta_{1};\theta_{2} ] \in \mathcal{R}^{11 \times 1}.
\end{equation}
The two edge vectors are
\begin{equation}
\begin{aligned}
\mathbf{e}_{1} &= \mathbf{x}_{2}  -\mathbf{x}_{1},\\
\mathbf{e}_{2} &= \mathbf{x}_{3}  -\mathbf{x}_{2}.
\end{aligned}
\end{equation}
The Voronoi length of the bending element is the average of the two edges, defined as
\begin{equation}
l = \frac{1} {2}(    || \mathbf{e}_{1}  || +  || \mathbf{e}_{2} ||  ).
\end{equation}
The curvature bi-normal, which measures the misalignment between two consecutive edges at the center node, is given by
\begin{equation}
\kappa \mathbf{b}=\frac{2\;\mathbf{e}_1\times \mathbf{e}_2}{\mathbf{e}_1\cdot \mathbf{e}_2+\left\| \mathbf{e}_1 \right\| \left\| \mathbf{e}_2 \right\|}.
\end{equation}
The material curvatures are given by the inner products between the curvature bi-normal and the material frame as
\begin{equation}
\begin{aligned}
    \kappa_{1} & = \frac {\kappa \mathbf{b}} {2} \cdot \frac {\left( \mathbf m_1^{2} + \mathbf m_2^{2} \right)} { l}, \\
\kappa_{2} & = - \frac {\kappa \mathbf{b}} {2} \cdot \frac{\left( \mathbf m_1^{1} + \mathbf m_2^{1}\right) } { l}.
\end{aligned}
\end{equation}
For the linear elastic constitutive model, the discrete bending energy is a quadratic function of the curvature as 
\begin{equation}
E^{b} = \frac{1}{2}  EI_{1} {(\kappa_{1} - \bar{\kappa}_{1})^2 }   \bar{l} + \frac{1}{2}  EI_{2} {(\kappa_{2} - \bar{\kappa}_{2})^2 }   \bar{l}.
\end{equation}
where $EI_{1}$ and $EI_{2}$ are the local bending stiffness along the two bending axes of the section, respectively.
The local bending force vector, $\mathbf{F}^{b}_{\mathrm{local}} \in \mathcal{R}^{11 \times 1}$, as well as the local bending stiffness matrix, $\mathbb{K}^{b}_{\mathrm{local}} \in \mathcal{R}^{11 \times 11}$, can be derived through a variational approach as
\begin{equation}
\mathbf{F}^{b}_{\mathrm{local}} = -\frac{\partial E^{b}}  {\partial \mathbf{q}^{b}}, \; \mathrm{and} \; \mathbb{K}^{b}_{\mathrm{local}} = \frac {\partial^2 E^{b}}  {\partial \mathbf{q}^{b} \partial \mathbf{q}^{b}}.
\end{equation}
The detailed formulation can be found in the MATLAB code.
Finally, the global bending force vector,  $\mathbf{F}^{b}$, and the associated stiffness, $\mathbb{K}^{b}$, can be assembled by iterating over all bending elements. 

\begin{figure}[t]
\centering
\includegraphics[width=0.7\columnwidth]{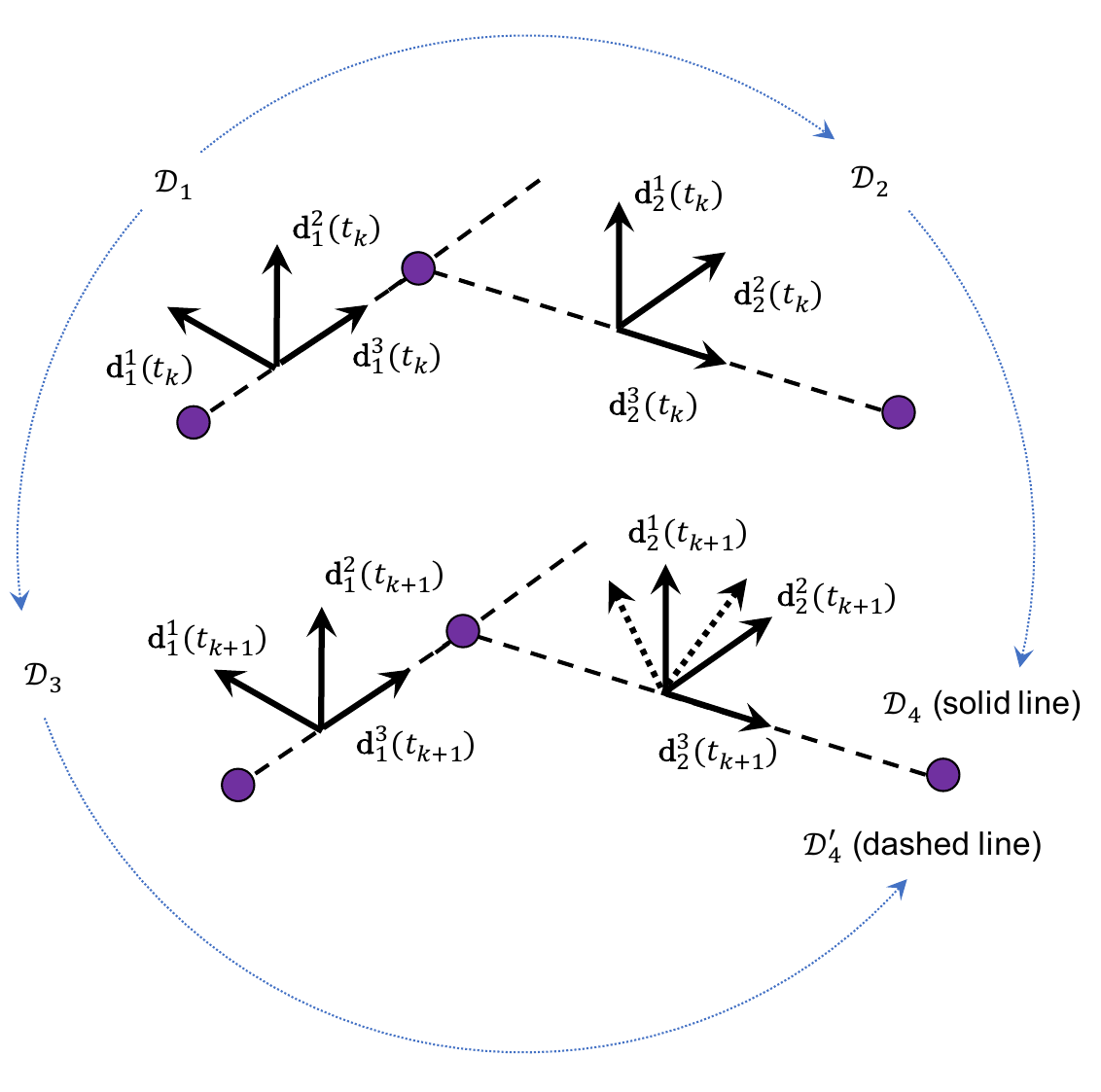}
\caption{\textbf{Space and time parallel transport.} The top configuration shows the orientations of the two reference frames at the current time step, while the bottom configuration depicts their updated orientations at the next time step.}
\label{fig:para_tran_plot}
\end{figure}

\paragraph{Twisting energy} Here, we define the twist for a 3D slender structure.
Similar to the bending element, the twisting element is defined as
\begin{equation}
\mathcal{T} \equiv \mathcal{B}: \{ \mathcal{S}_{1}, \mathcal{S}_{2}\}, \; \mathrm{with} \; \mathcal{S}_{1} : \{ \mathbf{x}_{1}, \mathbf{x}_{2}, \theta_{1} \} \; \mathrm{and} \; \mathcal{S}_{2} : \{ \mathbf{x}_{2}, \mathbf{x}_{3}, \theta_{2} \}.
\end{equation}
The local DOF vector is defined as
\begin{equation}
\mathbf{q}^{t} \equiv [\mathbf{x}_{1}; \mathbf{x}_{2}; \mathbf{x}_{3}; \theta_{1}; \theta_{2} ] \in \mathcal{R}^{11 \times 1}.
\end{equation}
The twisting curvature is the sum of the twist of the reference frame, $\tau_{r}$,  and the twist of the material frame, $\tau_{m}$.
To derive the twisting curvature, we need to introduce a parallel transport algorithm.
In order to obtain a twist-free transformation from a given frame, $\mathcal{D}_{i} \equiv \{ \mathbf{d}_{i}^{1}, \mathbf{d}_{i}^{2}, \mathbf{d}_{i}^{3}\}$, to a new frame with a specific tangential direction, $\mathcal{D}_{j} \equiv \{ \mathbf{d}_{j}^{1}, \mathbf{d}_{j}^{2}, \mathbf{d}_{j}^{3}\}$, we will use a parallel transport algorithm
\begin{equation}
\mathcal{D}_{j} = \mathcal{P}_{ij} (\mathcal{D}_{i}), 
\end{equation}
with the details as follows
\begin{equation}
\begin{aligned}
\mathbf{d}_{j}^{1} &= \left[ \mathbf{d}_{i}^{1} \cdot \left( \mathbf{d}_{i}^3 \times  \mathbf{n}_{ij} \right)  \right] \cdot \left( \mathbf{d}_{j}^3 \times \mathbf{n}_{ij} \right) +   \left( \mathbf{d}_{i}^{1} \cdot \mathbf{n}_{ij}  \right) \cdot \mathbf{n}_{ij}, \\
\mathbf{d}_{j}^{2} &= \mathbf{d}_{j}^{3} \times \mathbf{d}_{j}^{1},
\end{aligned}
\end{equation}
where $\mathbf{n}_{ij}$ is the bi-normal vector between the two tangential directions
\begin{equation}
\mathbf{n}_{ij} = \frac { \mathbf{d}_{i}^3 \times \mathbf{d}_{j}^3} { || \mathbf{d}_{i}^3 \times \mathbf{d}_{j}^3 ||}.
\end{equation}
Then, we discuss how to compute the twist associated with the reference frame during the time marching scheme.
Referring to Fig.~\ref{fig:para_tran_plot}, at time $t=t_{k}$, the reference frame for the first edge is defined as $\mathcal{D}_{1}$, and the reference frame for the second edge is defined as $\mathcal{D}_{2}$, and we assume the initial twisting between the two frames are zero, i.e., 
\begin{equation}
\mathcal{D}_{2} = \mathcal{P}_{12} (\mathcal{D}_{1}).
\end{equation}
Next, during the computation of the next time step, $t = t_{k+1}$, the two frames need to be updated accordingly to satisfy the twisting-free condition. Specifically, $\mathcal{D}_{1}$ will parallel transport to $\mathcal{D}_{3}$, while $\mathcal{D}_{2}$ will parallel transport to $\mathcal{D}_{4}$,
\begin{equation}
\begin{aligned}
\mathcal{D}_{3} &= \mathcal{P}_{13} (\mathcal{D}_{1}), \\
\mathcal{D}_{4} &= \mathcal{P}_{24} (\mathcal{D}_{2}).
\end{aligned}
\end{equation}
However, if we transport $\mathcal{D}_{3}$ to $\mathcal{D}_{4}'$,  it would differ from $\mathcal{D}_{4}$, indicating that the parallel transport algorithm is path-dependent, i.e.,
\begin{equation}
\mathcal{P}_{24}\left( \mathcal{P}_{12}( \mathcal{D}_{1} ) \right) \neq \mathcal{P}_{34}\left( \mathcal{P}_{13}(\mathcal{D}_{1}) \right).
\end{equation}
Thus, the angle difference between $\mathcal{D}_{4}$ and $\mathcal{D}_{4}'$ is denoted as reference twist, which is the twist of the reference frame during the time marching scheme,
\begin{equation}
\tau_{\mathrm{r}} = \frac { \measuredangle (\mathcal{D}_{4}, \mathcal{D}_{4}')} {l}.
\end{equation}
The twisting curvature associated with the material frame is straightly given by 
\begin{equation}
\tau_{\mathrm{m}} = \frac { \theta_{2} - \theta_{1} } {l}.
\end{equation}
The overall twisting curvature is the sum of the two as
\begin{equation}
\tau = \tau_{\mathrm{r}} + \tau_{\mathrm{m}}.
\end{equation}
Finally, for a linear elastic constitutive, the discrete twisting energy is given by
\begin{equation}
E^{t} = \frac{1}{2}  GJ {(\tau - \bar{\tau})^2 }   \bar{l} .
\end{equation}
where $GJ$ is the local twisting stiffness.
The local twisting force vector, $\mathbf{F}^{t}_{\mathrm{local}} \in \mathcal{R}^{11 \times 1}$, as well as the local twisting stiffness matrix, $\mathbb{K}^{t}_{\mathrm{local}} \in \mathcal{R}^{11 \times 11}$, can be derived through a variational approach as
\begin{equation}
\mathbf{F}^{t}_{\mathrm{local}} = -\frac{\partial E^{t}}  {\partial \mathbf{q}^{t}}, \; \mathrm{and} \; \mathbb{K}^{t}_{\mathrm{local}} = \frac {\partial^2 E^{t}}  {\partial \mathbf{q}^{t} \partial \mathbf{q}^{t} }.
\end{equation}
The detailed formulation can be found in the MATLAB code.
Finally, the global twisting force vector,  $\mathbf{F}^{t}$, and the associated stiffness, $\mathbb{K}^{t}$, can be assembled by iterating over all twisting elements. 

\paragraph{Equations of motion} With the derivation of the internal elastic force and the associated stiffness matrix, we include the inertial and damping effects to formulate the dynamic equations of motion.
Here, the mass matrix, $\mathbb{M}$, is time-invariant and can be readily determined based on the element size and material density. 
Note that the mass associated with nodes differs from that of the twisting angles.
Again, we employ a simple damping matrix, which is linearly proportional to the mass matrix with a damping coefficient $\mu$, i.e., $\mathbb{C} = \mu \mathbb{M}$.
Finally, the equations of motion for 3D rod or ribbon are
\begin{equation}
\mathbb{M} \ddot{\mathbf{q}} + \mu \mathbb{M} \dot{\mathbf{q}} - \mathbf{F}^{s} - \mathbf{F}^{b} - \mathbf{F}^{t} -  \mathbf{F}^{\text{ext}} = \mathbf{0}.
\end{equation}
The implicit Euler method and Newton's method are used to solve the $(3N+M)$-sized nonlinear dynamic system.

\subsubsection{Case 1: Deformation of a 3D helix under gravity~\href{https://github.com/weicheng-huang-mechanics/DDG_Tutorial/tree/main/3d_curve/case_1}{\texorpdfstring{\faGithub}{GitHub}}}

The deformation of a helix rod is a complex mechanical process due to the coupling between the bending mode and twisting mode, which also draws the attention of engineers, mathematicians, and computer scientists during the past decade\cite{miller2014shapes, moulton2013morphoelastic, bergou2008discrete, bergou2010discrete, bertails2006super, bertails2018inverse}. 
In this case study, we examine the free oscillation behavior of a slender rod under the influence of gravity.
The rod is clamped at one end while the other end remains free, allowing it to deform under its own weight.
As gravity acts on the structure, the helix undergoes complex bending and twisting deformations.
The case highlights the ability of the DDG model to capture intricate geometric nonlinearities in slender rod structures.

\paragraph{Simulation initialization} To initialize the simulation, the following inputs are used:

\begin{enumerate}

\item \textbf{Geometry and connection.} (i) Nodal position: the position of the nodes $\mathbf{q}(t=0)$, with a total of $N = 50$. The rod is initialized as a helix rod with a length $L = 1.54$ m. The helical radius is $ R_{h} = 0.125$ m and the helical pitch is $P_{h} = 0.234$ m. (ii) Stretching element: the connection of two nodes, with a total of $N_{s} = 49$. (iii) Bending element: the connection of two edges, with a total of $N_{b}=48$.

\item \textbf{Physical parameters.} (i) Young's modulus, $ E = 10.0$ MPa. (ii) Poisson's ratio $\nu=0.5$. (iii) Material density, $\rho=1000 $ $\mathrm{kg/m^3}$. (iv) Cross-sectional radius, $r_{0} = 0.01$ m, thus $EI_{1} = EI_{2} =  E\pi r_0^4/4$ and $GJ=G\pi r_0^4/2$. (v) Gravitational field $ \mathbf{g}=[0.0,0.0,-10.0]^T \mathrm{~m/s^2}$. (vi) Damping viscosity, $\mu = 1.0$. (vii) The overall simulation is dynamic, i.e., $ \mathrm{ifStatic} = 0$.

\item \textbf{Numerical parameters.} (i) Total simulation time, $T = 5.0$ s. (ii) Time step size, $\mathrm{d}t = 0.01$ s. (iii) Numerical tolerance, $\mathrm{tol} = 1 \times 10^{-4}$. (iv) Maximum iterations, $N_{\mathrm{iter}} = 10$.

\item \textbf{Boundary conditions.} The first two nodes, $\{ \mathbf{x}_{1}, \mathbf{x}_{2} \}$, and the first twisting angle, $\{ \theta_{1} \}$, are fixed to achieve clamped-free boundary conditions, thus the constrained array, $\mathcal{FIX} = [1,2,3,4,5,6,151]^{T}$. 

\item \textbf{Initial conditions.} (i) Initial position is input from the nodal positions. (ii) Initial velocity is set to zeros.

\item \textbf{Loading steps.} External gravitational force is applied to each node throughout the simulation.

\end{enumerate}

\begin{figure}[ht]
    \centering
    \includegraphics[width=\linewidth]{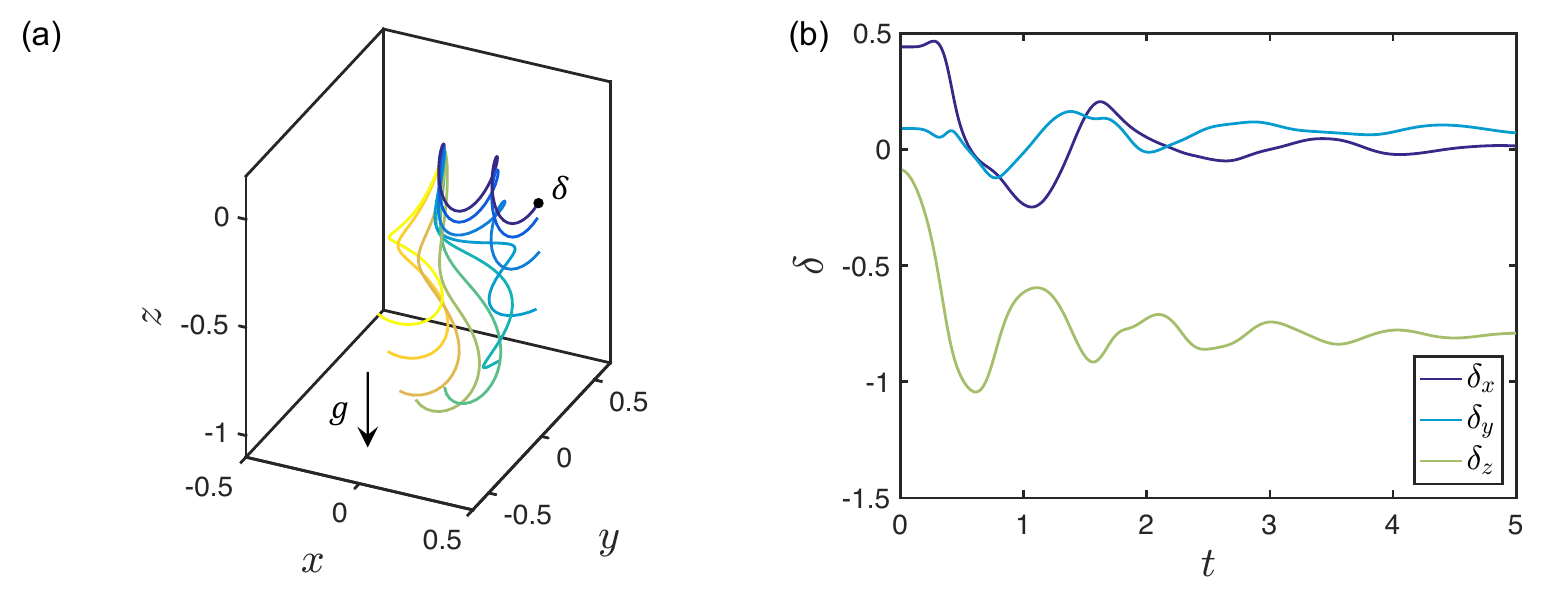}
    \caption{ \textbf{3D helix under gravity}. (a) The deformation illustration of the 3D helix rod under gravity shows the progression of the helix's shape as it falls. (b) The tip displacement in the $X$ direction, $\delta_x$, $Y$ direction, $\delta_y$, and $Z$ direction, $\delta_z$, as a function of time, $t$.}
    \label{fig:rod_case_1_plot}
\end{figure}

\paragraph{Simulation results} The helical rod deforms under gravity, as depicted in Fig.~\ref{fig:rod_case_1_plot}(a). 
In Fig.~\ref{fig:rod_case_1_plot}(b), we present the tip displacement $\{\delta_{x}, \delta_{y}, \delta_{z}\}$ over time, revealing oscillations in all three spatial directions.
The dynamic rendering can be found~\href{https://github.com/weicheng-huang-mechanics/DDG_Tutorial/blob/main/assets/rod_1.gif}{here}.

\subsubsection{Case 2: Bifurcation of a pre-buckled ribbon~\href{https://github.com/weicheng-huang-mechanics/DDG_Tutorial/tree/main/3d_curve/case_2}{\texorpdfstring{\faGithub}{GitHub}}}

The shear deformation behavior of a pre-buckled ribbon exhibits the bifurcation behavior, the continuation method is usually used to trace the deformation path, especially the static bifurcation critical point \cite{yu2019bifurcations, huang2021snap}, and the dynamics bifurcation behavior is investigated through a DDG-based approach in previous studies\cite{huang2020shear, huang2024exploiting}.
In this case study, we examine the shear deformation behavior of a pre-buckled ribbon, highlighting its bifurcation characteristics.
The ribbon is subjected to clamped-clamped boundary conditions, with specific loading conditions incorporated into the simulation.
As the ribbon undergoes shear, the simulation tracks the deformation pathway, identifying critical points between instability and stability.
This case highlights the ability of the DDG model to capture the complex bifurcation characteristics in slender structures subjected to shear deformation.

\paragraph{Simulation initialization} To initialize the simulation, the following inputs are used:

\begin{enumerate}

\item \textbf{Geometry and connection.} (i) Nodal position: the position of the nodes $\mathbf{q}(t=0)$, with a total of $N = 50$. The rod is initialized as a flat ribbon with a length $L = 1.0$ m. (ii) Stretching element: the connection of two nodes, with a total of $N_{s} = 49$. (iii) Bending element: the connection of two edges, with a total of $N_{b}=48$.

\item \textbf{Physical parameters.} (i) Young's modulus, $ E = 100.0$ MPa. (ii) Poisson's ratio $\nu=0.5$. (iii) Material density, $\rho=1000 $ $\mathrm{kg/m^3}$. (iv) Ribbon width $w=0.2$ m, ribbon thickness is $h=0.01$ m, thus $EI_{2} = Ewh^3/12$, $EI_{1} = 400EI_{2}$, and $GJ = Gwh^3/3$.  (v) Damping viscosity, $\mu = 0.1$. (vi) The overall simulation is static, i.e., $ \mathrm{ifStatic} = 1$.

\item \textbf{Numerical parameters.} (i) Total simulation time, $T = 40.0$ s. (ii) Time step size, $\mathrm{d}t = 0.01$ s. (iii) Numerical tolerance, $\mathrm{tol} = 1 \times 10^{-4}$. (iv) Maximum iterations, $N_{\mathrm{iter}} = 10$.

\item \textbf{Boundary conditions.} The first two nodes, last two nodes, $\{\mathbf{x}_{1},\mathbf{x}_{2},\mathbf{x}_{49},\mathbf{x}_{50} \}$, as well as the first twisting angle and the last twisting angle $\{\theta_{1}, \theta_{49}\}$, are fixed to achieve clamped-clamped boundary conditions, thus the constrained array, $\mathcal{FIX} = [1,2,3,4,5,6,147,148,149,150,151,199]^{T}$. 

\item \textbf{Initial conditions.} (i) Initial position is input from the nodal positions. (ii) Initial velocity is set to zeros.

\item \textbf{Loading steps.} (i) Compression step: when $t \le 3.0$ s, a displacement is applied to both ends along the $X$ axis with speed $v_{x} = 0.1\mathrm{~m/s}$, until the compressive distance reaches to the target, $\Delta X \ge 0.4$ m. Note that here we use $ \mathbf{g}=[0.0,0.0,10.0]^T\mathrm{~m/s^2}$ to induce the bifurcation direction. (ii) Change perturbation: when $3.0 \; \mathrm{s} \le t \le 5.0 \; \mathrm{s}$, the gravity force is switched to $ \mathbf{g}=[0.1,0.1,0.0]^T\mathrm{~m/s^2}$ to change the direction of the perturbation. (iii) Shear step:  when $t \ge 5.0$ s, a displacement is applied to both ends along the $Y$ axis with speed $v_{y} = 0.01\mathrm{~m/s}$, until the shear distance reaches to the target, $\Delta Y \ge 0.6$ m. 

\end{enumerate}

\begin{figure}[ht]
    \centering
    \includegraphics[width=\linewidth]{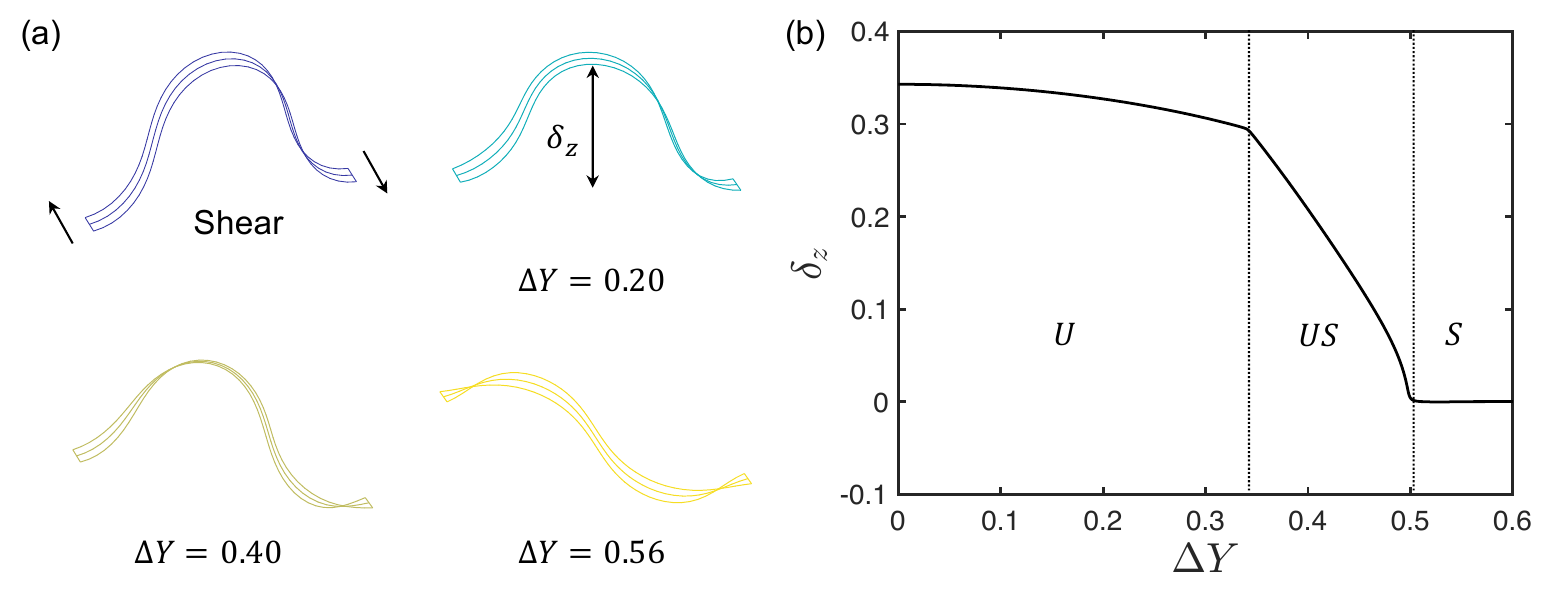}
    \caption{ \textbf{Bifurcation of pre-buckled ribbon.} (a) The shear process of the pre-buckled ribbon, illustrating the deformed shape at different shear displacements, $\Delta Y$. (b) The middle nodal displacement in the $Z$ direction, $\delta_z$, as a function of the shear displacement, $\Delta Y$, shows distinct regions of $U$ pattern, $US$ pattern, and $S$ pattern.}
    \label{fig:rod_case_2_plot}
\end{figure}

\paragraph{Simulation results} The shear process of the pre-bucked ribbon is illustrated in Fig.~\ref{fig:rod_case_2_plot}(a). 
In Fig.~\ref{fig:rod_case_2_plot}(b), midpoint height $\delta_z$ exhibits a distinct transition through the $U$ phase, $US$ phase, and $S$ phase regions as the shear displacement $\Delta Y$ increases
The boundaries between these phases are characterized as supercritical pitchfork bifurcations. 
This bifurcation behavior, captured in the displacement of the central node, highlights the sensitivity of pre-buckled structures to external shear forces.
The dynamic rendering can be found~\href{https://github.com/weicheng-huang-mechanics/DDG_Tutorial/blob/main/assets/rod_2.gif}{here}.

\subsubsection{Case 3: Growth of an annular ribbon~\href{https://github.com/weicheng-huang-mechanics/DDG_Tutorial/tree/main/3d_curve/case_3}{\texorpdfstring{\faGithub}{GitHub}}}

Growth ribbons are usually built as a model for plant tissues in nature, such as the swelling and shrinkage of pine cones\cite{dawson1997pine, zhang2022unperceivable}, pea pods\cite{ArmonShahaf2011GaMi} and leaves\cite{li2025biomimetic}, during these processes the ribbons usually have a large deformation behavior \cite{huang2024integration, dias2014non, dias2012geometric, goriely2006twisted, li2025biomimetic}.
In this case study, we examine the growth of an annular ribbon, driven by an increase in its natural curvature.
The ribbon is initially defined with a specified radius and undergoes deformation as its natural curvature increases, mimicking the behavior of plant tissues.
We track the ribbon’s deformation throughout the growth process, focusing on the resulting structural changes.
This case highlights the ability of the DDG model to accurately simulate complex growth behaviors and large deformations in soft structures.

\paragraph{Simulation initialization} To initialize the simulation, the following inputs are used:

\begin{enumerate}

\item \textbf{Geometry and connection.} (i) Nodal position: the position of the nodes $\mathbf{q}(t=0)$, with a total of $N = 40$. The system is initialized as an annular ribbon connected end to end with a radius $R = 1.0$ m. (ii) Stretching element: the connection of two nodes, with a total of $N_{s} = 40$. (iii) Bending element: the connection of two edges, with a total of $N_{b}=40$.

\item \textbf{Physical parameters.} (i) Young's modulus, $ E = 10.0$ MPa. (ii) Poisson's ratio $\nu=0.5$. (iii) Material density, $\rho=100 $ $\mathrm{kg/m^3}$. (iv) Ribbon width $w=0.2$ m, ribbon thickness is $h=0.01$ m, thus $EI_{2} = Ewh^3/12$, $EI_{1} = 400EI_{2}$, and $GJ=Gwh^3/3$.  (v) Damping viscosity, $\mu = 1.0$.  (vi) The overall simulation is static, i.e., $ \mathrm{ifStatic} = 1$.

\item \textbf{Numerical parameters.} (i) Total simulation time, $T = 25.0$ s. (ii) Time step size, $\mathrm{d}t = 0.001$ s. (iii) Numerical tolerance, $\mathrm{tol} = 1 \times 10^{-4}$. (iv) Maximum iterations, $N_{\mathrm{iter}} = 10$.

\item \textbf{Boundary conditions.} Based on the symmetry, the $Y$ and $Z$ displacements of two points where the diameter intersects the annulus ribbon are constrained, while the $X$ and $Z$ displacements of another two points where the other perpendicular diameter intersects the annulus ribbon are constrained, thus the constrained array, $\mathcal{FIX} = [2,3,31,33,62,63,91,93]^{T}$. 

\item \textbf{Initial conditions.} (i) Initial position is input from the nodal positions. (ii) Initial velocity is set to zeros.

\item \textbf{Loading steps.} (i) Perturbation step: a small perturbation to the initial horizontal configuration is created by applying a small gravitational force (with $ \mathbf{g}=[0.1,0.1,0.1]^T\mathrm{~m/s^2}$) when $t \le 1.0\mathrm{~s}$. (ii) Growth step: when $t>1.0$ s, the normalized natural curvature, $\bar{\kappa}_1 /  l$, would increase with a growth rate $\dot{\bar{\kappa}}_1 /  l=0.01 \; \mathrm{s}^{-1}$.

\end{enumerate}

\begin{figure}[ht]
    \centering
    \includegraphics[width=\linewidth]{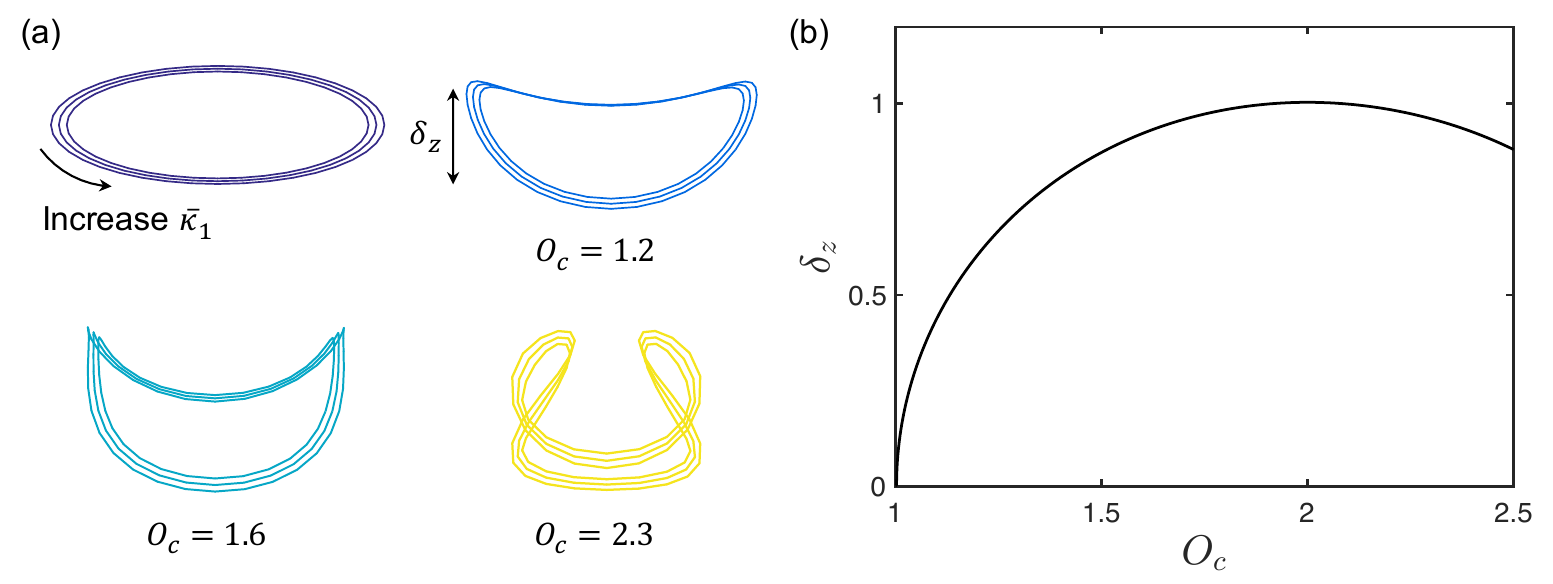}
    \caption{ \textbf{Growth of an annular ribbon}. (a) The deformation illustration of the growth annular ribbon. (b) The total height in $Z$ direction, $ \delta_z $, as a function of growing natural curvature, $O_c$.}
    \label{fig:rod_case_3_plot}
\end{figure}

\paragraph{Simulation results} 
The progressive deformation of an annular ribbon as its natural curvature increases is shown in Fig.~\ref{fig:rod_case_3_plot}(a). 
The resulting changes in the ribbon's shape, specifically the middle nodal displacement $\delta_z$, as a function of overcurvare, $O_{c} = 2 \pi \bar{\kappa}_{1} / L$, is presented in Fig.~\ref{fig:rod_case_3_plot}(b), highlight the sensitivity of annular structures to variations in natural curvature. 
The dynamic rendering can be found~\href{https://github.com/weicheng-huang-mechanics/DDG_Tutorial/blob/main/assets/rod_3.gif}{here}.

\subsection{Axisymmetric shell: rotational surface}

In this subsection, we focus on simulating the deformation of axisymmetric shell structures -- a system that exhibits rotational symmetry around a central axis \cite{nelson1978investigation}. 
Shell formations are widespread in both natural settings, such as biological vesicles, cells, and fruits, as well as in artificial designs like footballs, submarines, and space capsules \cite{hutchinson2017nonlinear, koiter1969non, turlier2014furrow, yu2023one}.
Many of these structures exhibit axisymmetric characteristics, owing to their intrinsic symmetry in geometry and loading conditions \cite{hinton2012analysis}.
By representing the shell’s middle surface as a 2D curve, we capture its complex mechanics in a simplified form by using the discrete differential geometry (DDG) method \cite{huang2024discrete}.
We first formulate the discrete energy expression of the axisymmetric shell using DDG, followed by two case studies: (i) the inflation of an axisymmetric plate \cite{liu2024simplified, sagiv1990inflation} and (ii) the snap-through behavior of a hemispherical axisymmetric shell under compressive load \cite{huang2024snap, taffetani2018static, chen2023pseudo, pezzulla2019weak}.

\begin{figure}[ht]
\centering
\includegraphics[width=\columnwidth]{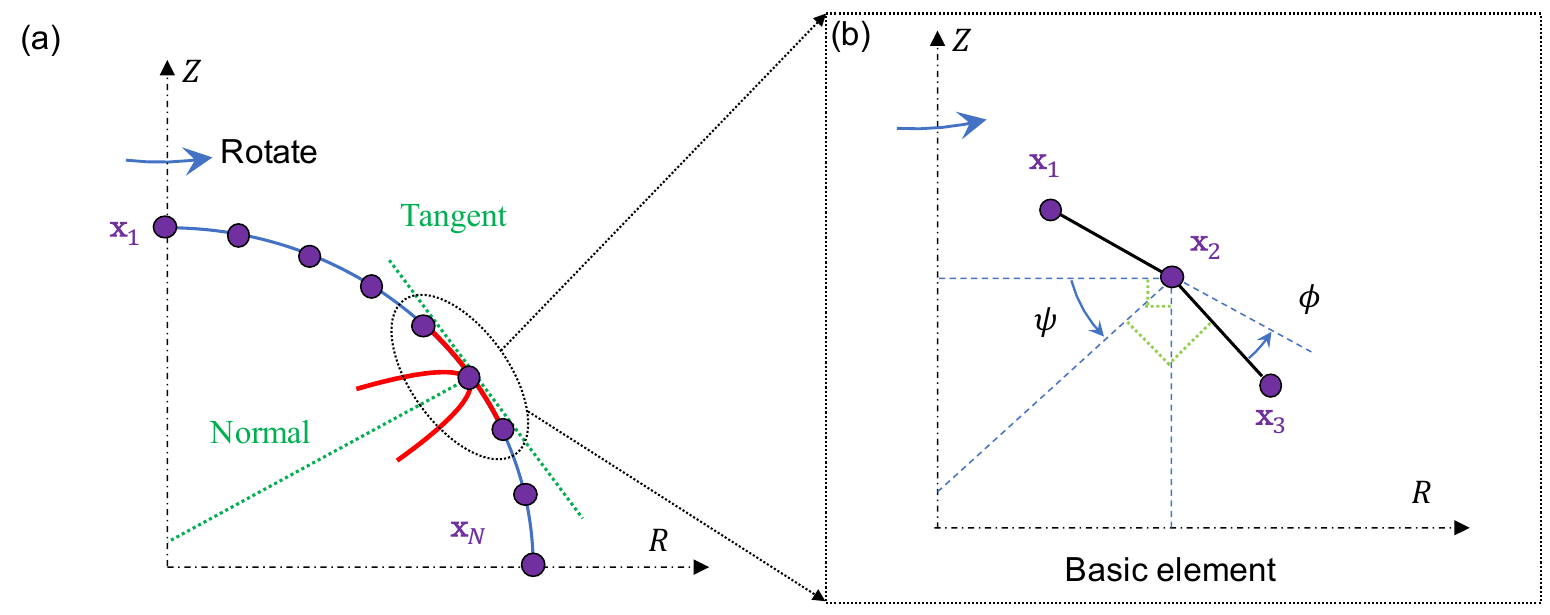}
\caption{ \textbf{The axisymmetric shell model in DDG simulations.} (a) Discretization of the axisymmetric shell, which exhibits rotational symmetry about the $Z$ axis. The deformation energy is influenced by the curvature in two orthogonal directions: the tangent and normal directions at each node. (b) The basic element of the discrete axisymmetric shell, where the curvatures in the tangent and normal directions are related to the two angles, $\phi$ and $\psi$, respectively.}
\label{fig:shell_model_plot}
\end{figure}

\subsubsection{Numerical formulation}

Here, we discuss how a simple 2D curve can represent the mechanics of an axisymmetric shell.
As shown in Fig. \ref{fig:shell_model_plot}(a), the configuration of a planar curve is presented by $N$ nodes, where each node is defined as $\mathbf{x}_{i} \equiv [r_{i}, z_{i}]^{T} \in \mathcal{R}^{2 \times 1}$. Consequently, the total DOF vector is given by
\begin{equation}
\mathbf{q} = [ \mathbf{x}_1; \mathbf{x}_2; \ldots; {\mathbf{x}_{N}} ] \in \mathcal{R}^{2N \times 1}.
\end{equation}
Two types of elements are used to capture the mechanics of a rotational surface: (i) stretching element, and (ii) bending element. 
Consistent with previous section, the number of stretching elements is denoted as $N_{s}$, while the number of bending elements is denoted as $N_{b}$.
If only the stretching element is considered, the bending-dominated plate/shell structures would reduce to the stretching-dominated membrane structures.

\paragraph{Stretching element} The stretching element is comprised of two connected nodes, defined as
\begin{equation}
\mathcal{S}: \{\mathbf{x}_{1}, \mathbf{x}_{2} \}.
\end{equation}
The local DOF vector is defined as 
\begin{equation}
\mathbf{q}^{s} \equiv [\mathbf{x}_{1}; \mathbf{x}_{2} ] \in \mathcal{R}^{4 \times 1}.
\end{equation}
The edge length is the $\mathcal{L}_{2}$ norm of the edge vector, defined as
\begin{equation}
l   =  || \mathbf{x}_{2}  -\mathbf{x}_{1} ||.
\end{equation}
The average edge radius is 
\begin{equation}
r =\frac{1}{2} (r_{1} + r_{2}).
\end{equation}
Thus, the total area of this stretching element is
\begin{equation}
s = 2 \pi r l.
\end{equation}
The element elongation along the meridional direction is straightforward, following the same formulation as the planar beam, given by
\begin{equation}
{\varepsilon}_{1} = \frac {  l} {  \bar{l} } - 1.
\end{equation}
Hereafter, we use a bar on top to indicate the evaluation of the undeformed configuration, e.g., $\bar{l}$ is the edge length before deformation.
The stretching along the circumferential direction is related to the radial expansion of the circle
\begin{equation}
{\varepsilon}_{2} = \frac { r } { \bar{r} } - 1,
\end{equation}
Due to the axisymmetric assumption, the shear coupling between the meridional direction and the circumferential direction is zero, i.e.,
\begin{equation}
{\varepsilon}_{12} ={\varepsilon}_{21} = 0.
\end{equation}
Using the linear elastic model, the total stretching energy follows a quadratic function of the strain,
\begin{equation}
E^s = \frac{1} {2} \frac {Eh} {1-\nu^2}
\left( \varepsilon_{1}^2 + 2 \nu \varepsilon_{1} \varepsilon_{2} + \varepsilon_{2}^2 \right)
 \bar{s},
\end{equation}
where $E$ is the young's modulus, $\nu$ is the Poisson ratio, and $h$ is the shell thickness.
The local stretching force vector, $\mathbf{F}^{s}_{\mathrm{local}} \in \mathcal{R}^{4 \times 1}$, as well as the local stretching stiffness matrix, $\mathbb{K}^{s}_{\mathrm{local}} \in \mathcal{R}^{4 \times 4}$, can be derived through a variational approach as
\begin{equation}
\mathbf{F}^{s}_{\mathrm{local}} = -\frac{\partial E^{s}}  {\partial \mathbf{q}^{s}}, \; \mathrm{and} \; \mathbb{K}^{s}_{\mathrm{local}} = \frac {\partial^2 E^{s}}  {\partial \mathbf{q}^{s} \partial \mathbf{q}^{s}}.
\end{equation}
The detailed formulation can be found in the MATLAB code.
Finally, the global stretching force vector,  $\mathbf{F}^{s}$, and the associated stiffness matrix, $\mathbb{K}^{s}$, can be assembled by iterating over all stretching elements. 

\paragraph{Bending element} The bending element is comprised of two consecutive edges sharing a common node, defined as
\begin{equation}
\mathcal{B}: \{ \mathcal{S}_{1}, \mathcal{S}_{2}\}, \; \mathrm{with} \; \mathcal{S}_{1} : \{ \mathbf{x}_{1}, \mathbf{x}_{2} \} \; \mathrm{and} \; \mathcal{S}_{2} : \{ \mathbf{x}_{2}, \mathbf{x}_{3} \}.
\end{equation}
Here, we define $\mathbf{x}_{2}$ as the joint node, and $\mathbf{x}_{1}$ and $\mathbf{x}_{3}$ are the two adjacent nodes. Thus, the local DOF vector is defined as 
\begin{equation}
\mathbf{q}^{b} \equiv [\mathbf{x}_{1}; \mathbf{x}_{2}; \mathbf{x}_{3} ] \in \mathcal{R}^{6 \times 1}.
\end{equation}
The two edge vectors are
\begin{equation}
\begin{aligned}
\mathbf{e}_{1} &= \mathbf{x}_{2}  -\mathbf{x}_{1},\\
\mathbf{e}_{2} &= \mathbf{x}_{3}  -\mathbf{x}_{2}.
\end{aligned}
\end{equation}
The Voronoi length of the bending element is the average of the lengths of  the two edges, defined as 
\begin{equation}
l = \frac{1} {2}(    || \mathbf{e}_{1}  || +  || \mathbf{e}_{2} ||  ).
\end{equation}
The local radius of the bending element is, 
\begin{equation}
r = \frac{1}{3}(r_{1}+r_{2}+r_{3}). 
\end{equation}
Thus, the total area is 
\begin{equation}
s = 2 \pi r l.
\end{equation}
The meridional curvature is associated with the turning angle between the two connected edges as
\begin{equation}
\kappa_{1} = \frac { 2 \tan ( {\phi / {2} )}} {   l },
\end{equation}
where $\phi$ is the turning angle between the three consecutive nodes. 
The curvature along the circumferential direction is determined by the change of the surface normal direction, as
\begin{equation}
\kappa_{2} = \frac { \cos (\psi) } { r },
\end{equation}
where $\psi$ is the orientation angle, determined by the surface normal vector relative to the $R$ axis, as shown in Fig.~\ref{fig:shell_model_plot}(b).
Similarly, the bending coupling between the meridional direction and the circumferential direction is also zero, i.e.,
\begin{equation}
\kappa_{12} = \kappa_{21} = 0.
\end{equation}
The total bending energy is
\begin{equation}
E^b = \frac{1}{2} \frac {Eh^3} {12(1-\nu^2)} \left[ (\kappa_{1} - \bar{\kappa}_{1})^2 + 2 \nu (\kappa_{1} - \bar{\kappa}_{1}) (\kappa_{2} - \bar{\kappa}_{2}) + (\kappa_{2} - \bar{\kappa}_{2})^2 \right] \bar{s} .
\end{equation}
The local bending force vector, $\mathbf{F}^{b}_{\mathrm{local}} \in \mathcal{R}^{6 \times 1}$, as well as the local bending stiffness matrix, $\mathbb{K}^{b}_{\mathrm{local}} \in \mathcal{R}^{6 \times 6}$, can be derived through a variational approach as
\begin{equation}
\mathbf{F}^{b}_{\mathrm{local}} = -\frac{\partial E^{b}}  {\partial \mathbf{q}^{b}}, \; \mathrm{and} \; \mathbb{K}^{b}_{\mathrm{local}} = \frac {\partial^2 E^{b}}  {\partial \mathbf{q}^{b} \partial \mathbf{q}^{b}}.
\end{equation}
The detailed formulation can be found in the MATLAB code.
Finally, the global bending force vector,  $\mathbf{F}^{b}$, and the associated stiffness, $\mathbb{K}^{b}$, can be assembled by iterating over all stretching elements. 

\paragraph{Equations of motion} With the formulation of the internal elastic force and the associated stiffness matrix, we include the inertial and damping effect to formulate the dynamic equations of motion.
Here, the mass matrix, $\mathbb{M}$, is time-invariant and can be easily obtained based on the element size and material density. 
We then employ a simple damping matrix, which is linearly related to the mass matrix and damping coefficient $\mu$, i.e., $\mathbb{C} = \mu \mathbb{M}$.
Finally, the equations of motion for rotational shell system are 
\begin{equation}
\mathbb{M} \ddot{\mathbf{q}} + \mu \mathbb{M} \dot{\mathbf{q}} - \mathbf{F}^{s} - \mathbf{F}^{b} -  \mathbf{F}^{\text{ext}} = \mathbf{0}.
\end{equation}
The implicit Euler method and Newton's method are used to solve the $(2N)$-sized nonlinear dynamic system.

\subsubsection{Case 1: Inflation of an axisymmetric plate~\href{https://github.com/weicheng-huang-mechanics/DDG_Tutorial/tree/main/2d_surface/case_1}{\texorpdfstring{\faGithub}{GitHub}}}

In this case study, we examine the inflation behavior of an axisymmetric plate under applied pressure \cite{liu2024simplified, sagiv1990inflation, liu2024computational}.
The $R$ displacement of the central node is constrained based on the symmetry of deformation, and the node at the right end maintains a pin boundary. 
As the pressure increases, the plate deforms into a characteristic inflated dome shape.
This case highlights the ability of the DDG model to accurately capture axisymmetric deformation and provides insights for modeling hyperelastic materials in future studies.

\paragraph{Simulation initialization} To initialize the simulation, the following inputs are used:

\begin{enumerate}

\item \textbf{Geometry and connection.} (i) Nodal position: the position of the nodes $\mathbf{q}(t=0)$, with a total of $N = 40$. The axisymmetric plate is initialized in a flat state with a radius $R = 1.0$ m. (ii) Stretching element: the connection of two nodes, with a total of $N_{s} = 39$. (iii) Bending element: the connection of two edges, with a total of $N_{b}=38$.

\item \textbf{Physical parameters.} (i) Young's modulus, $ E = 1.0$ MPa. (ii) Poisson's ratio, $\nu=0.5$. (iii) Material density, $\rho=1000 $ $\mathrm{kg/m^3}$. (iv) Plate thickness, $h=0.02$ m. (v) Damping viscosity, $\mu = 0.1$. (vi) The overall simulation is static, i.e., $ \mathrm{ifStatic} = 1$. 

\item \textbf{Numerical parameters.} (i) Total simulation time, $T = 2.0$ s. (ii) Time step size, $\mathrm{d}t = 0.01$ s. (iii) Numerical tolerance, $\mathrm{tol} = 1 \times 10^{-4}$. (iv) Maximum iterations, $N_{\mathrm{iter}} = 10$.

\item \textbf{Boundary conditions.} The $R$ displacement of the central nodal of the plate is constrained based on the symmetry of deformation, and the nodal at the right end is constrained for a pin boundary condition; thus, the constrained array, $\mathcal{FIX} = [1,79,80]^{T}$.

\item \textbf{Initial conditions.} (i) Initial position is input from the nodal positions. (ii) Initial velocity is set to zeros.

\item \textbf{Loading steps.} The external pressure is increasing with a loading rate, $\dot{p} = 10.0$ kPa/s.

\end{enumerate}

\begin{figure}[ht]
    \centering
    \includegraphics[width=\linewidth]{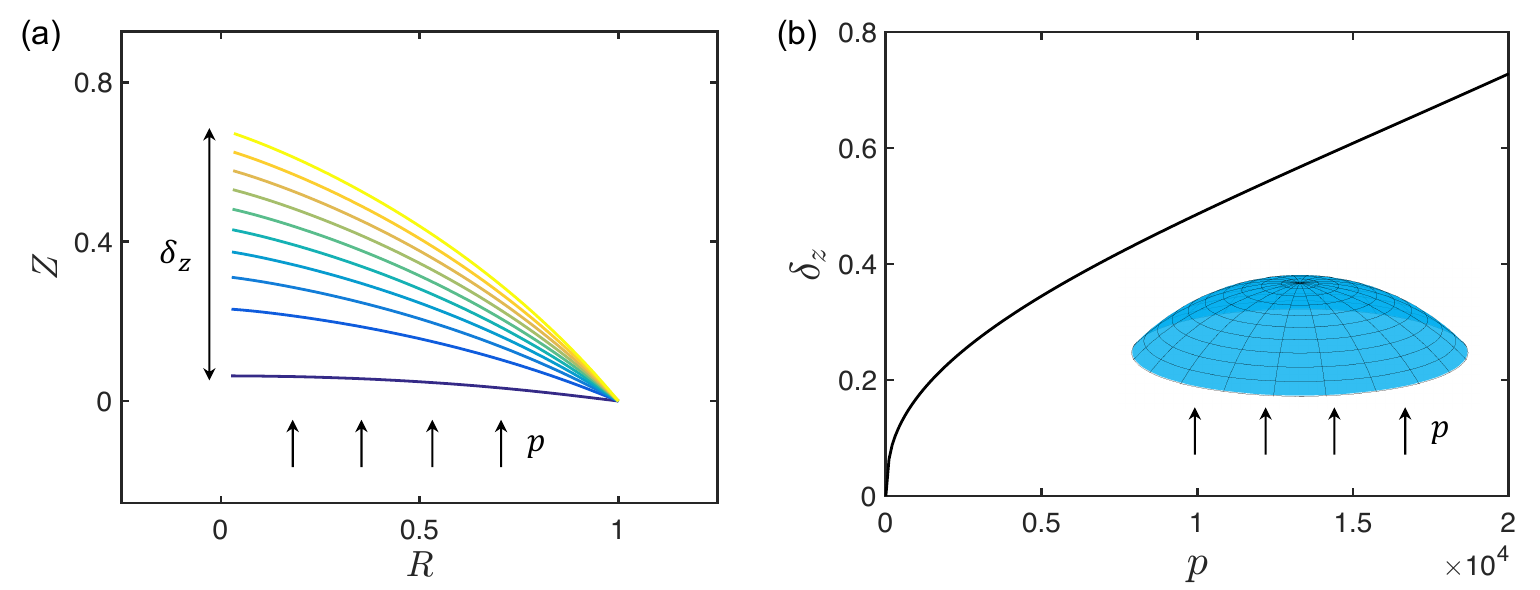}
    \caption{ \textbf{Inflation of an axisymmetric plate.} (a) The deformation illustration of the inflation axisymmetric plate. (b) The middle nodal displacement in, $\delta_{z}$, as a function of applied pressure, $p$.}
    \label{fig:shell_case_1_plot}
\end{figure}

\paragraph{Simulation results.} 
The inflation process of the axisymmetric plate is shown in Fig.~\ref{fig:shell_case_1_plot}(a). 
In Fig.~\ref{fig:shell_case_1_plot}(b), we shows the displacement $\delta_z$ of the central nodal as a function of applied pressure $p$. 
As observed, the inflated plate adopts a characteristic dome shape in response to the applied pressure.
The dynamic rendering can be found~\href{https://github.com/weicheng-huang-mechanics/DDG_Tutorial/blob/main/assets/ashell_1.gif}{here}.

\subsubsection{Case 2: Eversion of an axisymmetric shell cap~\href{https://github.com/weicheng-huang-mechanics/DDG_Tutorial/tree/main/2d_surface/case_2}{\texorpdfstring{\faGithub}{GitHub}}}

Snap-through is a fascinating nonlinear phenomenon observed in various structures, including shells, arches, and membranes. Understanding and predicting snap-through behavior is crucial in many engineering applications, ranging from micro-switches and actuators to deployable structures and biomedical devices. Understanding these characteristics is crucial for designing and using structures that exhibit snap-through behavior, particularly in applications such as switches, actuators, and energy storage devices. In this case study, we examine the snap-through behavior of an axisymmetric shell subjected to compressive loading \cite{huang2024snap, taffetani2018static, chen2023pseudo}. 
The shell is subject to boundary conditions that allow for large deformations, with the load applied uniformly along its surface. 
The snap-through phenomenon is characterized by a sudden, large-amplitude shape change as the shell transitions between two stable equilibrium states. 
This case highlights the ability of the DDG model to accurately capture the nonlinear behavior of the shell and predict the critical load at which the snap-through occurs,.

\begin{figure}[ht]
    \centering
    \includegraphics[width=\linewidth]{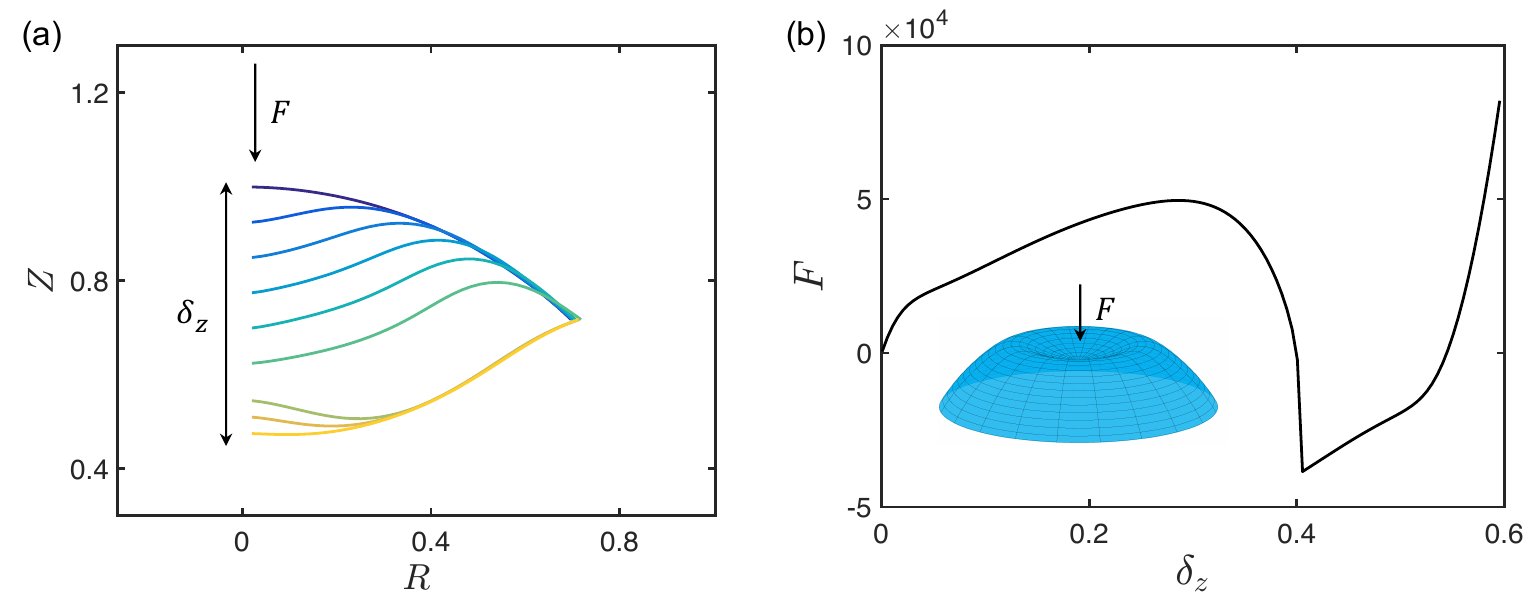}
    \caption{ \textbf{Snap-through of an axisymmetric shell cap.} (a) The deformation illustration of the compressed axisymmetric shell. (b) The relationship between the pole displacement, $\delta_z$, and the indention force, $F$.}
    \label{fig:shell_case_2_plot}
\end{figure}

\paragraph{Simulation initialization} To initialize the simulation, the following inputs are used:

\begin{enumerate}

\item \textbf{Geometry and connection.} (i) Nodal position: the position of the nodes $\mathbf{q}(t=0)$, with a total of $N = 40$. The axisymmetric shell cap is initialized as a hemisphere with a radius $R = 1.0$ m, and the size angle $0.8 \le \phi_{R} \le \pi/2$. (ii) Stretching element: the connection of two nodes, total with a total of $N_{s} = 39$. (iii) Bending element: the connection of two edges, with a total of $N_{b}=38$.

\item \textbf{Physical parameters.} (i) Young's modulus, $ E = 1.0$ GPa. (ii) Poisson's ratio, $\nu=0.5$. (iii) Material density, $\rho=1000 $ $\mathrm{kg/m^3}$. (iv) Shell thickness, $h=0.02$ m.  (v) Damping viscosity, $\mu = 0.1$. (vi) The overall simulation is static, i.e., $ \mathrm{ifStatic} = 1$. 

\item \textbf{Numerical parameters.} (i) Total simulation time, $T = 6.0$ s. (ii) Time step size, $\mathrm{d}t = 0.005$ s. (iii) Numerical tolerance, $\mathrm{tol} = 1 \times 10^{-4}$. (iv) Maximum iterations, $N_{\mathrm{iter}} = 10$.

\item \textbf{Boundary conditions.} The first two nodes, $\{\mathbf{x}_{1}, \mathbf{x}_{2} \}$, from the pole are fixed, while the last node, $\{\mathbf{x}_{40}\}$, can slide along the $R$ axis,  thus, the constrained array is $\mathcal{FIX} = [1,2,3,4,80]^{T}$.

\item \textbf{Initial conditions.} (i) Initial position is input from the nodal positions. (ii) Initial velocity is set to zeros.

\item \textbf{Loading steps.} The first two nodes are manually moved from top to bottom with a constant speed $v_{z} = 0.1$ m/s.

\end{enumerate}

\paragraph{Simulation results} 
The axisymmetric shell undergoes a snap-through process under compression, as depicted in Fig.~\ref{fig:shell_case_2_plot}(a), characterized by a sudden inversion of the shell's shape.
In Fig.~\ref{fig:shell_case_2_plot}(b), this transition leads to a sharp drop in a sudden drop in indentation force $F$ at a critical pole displacement $\delta_z=0.4$. 
Additionally, the force-displacement curve highlights the non-linear nature of this behavior, demonstrating the shell's bi-stability and the rapid release of energy during the snap-through process.
The dynamic rendering can be found~\href{https://github.com/weicheng-huang-mechanics/DDG_Tutorial/blob/main/assets/ashell_2.gif}{here}.

\subsection{Plate \& shell: 3D surface}

In this subsection, we discuss the mechanics of plates or shells, which can be simplified to the middle plane as a 3D surface based on the Kirchhoff-Love assumption \cite{schollhammer2019kirchhoff}. 
Unlike the axisymmetric shell, the deformation of general plates and shells is typically asymmetric, preventing simplification to plane deformation.
As a result, a full 3D representation is necessary to capture their complex 3D deformation behavior.  
The plate and shell models find broad applications, e.g., cloth animations \cite{baraff1998large, house2000cloth, grinspun2002charms, grinspun2003discrete, choi2005research, bridson2005simulation, wardetzky2007discrete, batty2012discrete}, the form of the guts \cite{savin2011growth}, and the growth of leaves \cite{liang2009shape}. 
We begin by introducing the discrete energy formulation for 3D surfaces within the DDG framework, which provides a principled approach to modeling bending and stretching deformations.
To illustrate the capability of this method, we present three case studies: (i) plate deflection under gravity, (ii) plate wrinkling under gravity, and (iii) indentation of a cylindrical shell.
These examples demonstrate the model’s effectiveness in capturing the intricate deformation characteristics of plates and shells under different loading conditions, highlighting its potential for structural analysis and design.

\begin{figure}[ht]
\centering
\includegraphics[width=\columnwidth]{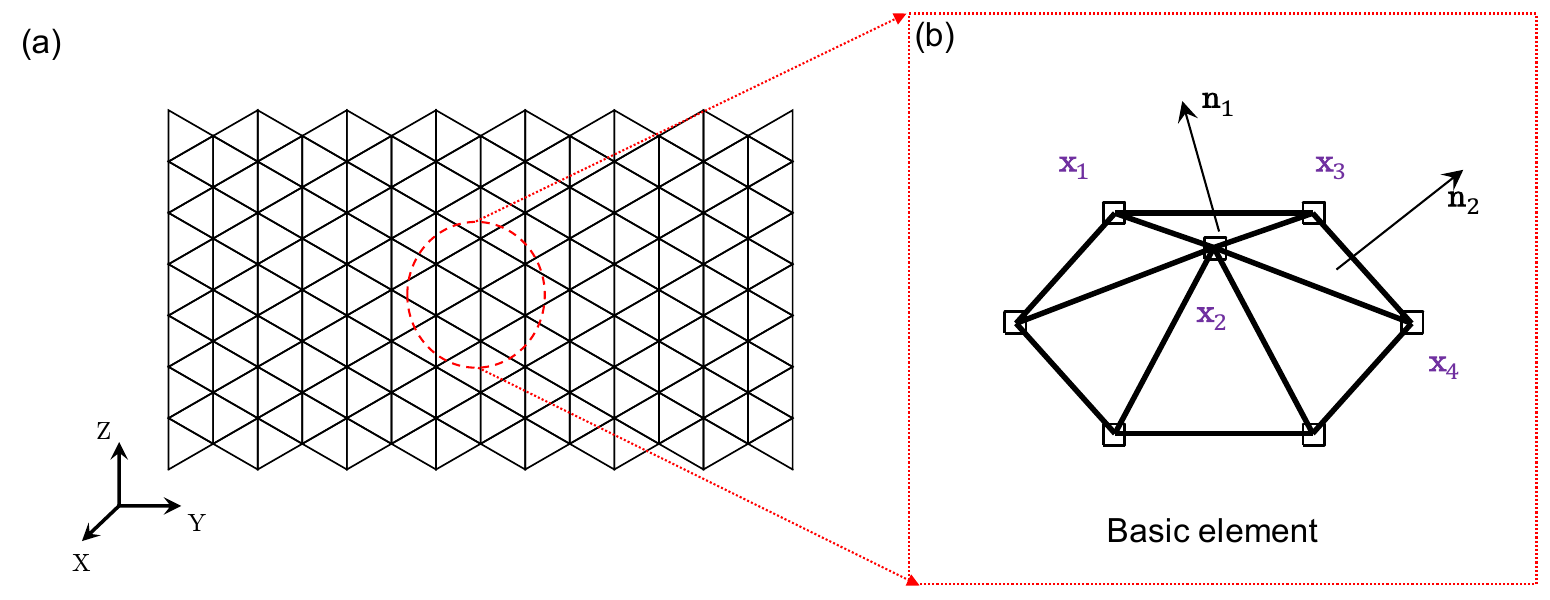}
\caption{ \textbf{The 3D plate/shell model in DDG simulations. } (a) A 3D surface discretized into triangular elements. (b) Basic element for the evaluation of surface bending curvatures.} 
\label{fig:plate_model_plot}
\end{figure}

\subsubsection{Numerical formulation}

As shown in Fig. \ref{fig:plate_model_plot}, the configuration of a 3D surface is represented by $N$ nodes, where each node is defined as $\mathbf{x}_{i} \equiv [x_{i}, y_{i}, z_{i}]^{T} \in \mathcal{R}^{3 \times 1}$. Thus, the total DOF vector is given by
\begin{equation}
\mathbf{q} = [ \mathbf{x}_1; \mathbf{x}_2; \ldots; {\mathbf{x}_{N}} ] \in \mathcal{R}^{3N \times 1}.
\end{equation}
Two types of elements are used to capture the mechanics of a 3D surface: (i) stretching element, and (ii) bending element. 
Consistent with the previous section, the number of stretching elements is denoted as $N_{s}$, while the number of bending elements is denoted as $N_{b}$.
If only the stretching elements are considered, bending-dominated plate/shell structures reduce to stretching-dominated membrane structures. 

\paragraph{Stretching element} The stretching element is comprised of two connected nodes, defined as
\begin{equation}
\mathcal{S}: \{\mathbf{x}_{1}, \mathbf{x}_{2} \}.
\end{equation}
The local DOF vector is defined as 
\begin{equation}
\mathbf{q}^{s} \equiv [\mathbf{x}_{1}; \mathbf{x}_{2} ] \in \mathcal{R}^{6 \times 1}.
\end{equation}
The edge length is the $\mathcal{L}_{2}$ norm of the edge vector, defined as
\begin{equation}
l   =  || \mathbf{x}_{2}  -\mathbf{x}_{1} ||.
\end{equation}
The stretching strain is based on the uniaxial elongation of edge, defined as
\begin{equation}
{\varepsilon} = {  l } - \bar{l}.
\end{equation}
Hereafter, we use a bar on top to indicate the evaluation of the undeformed configuration, e.g., $\bar{l}$ is the edge length before deformation.
Using the linear elastic model, the total stretching energy is in a quadratic form of the strain following the linear elastic constitutive law
\begin{equation}
E^s = \frac{\sqrt{3}}{4} Eh ( \varepsilon )^2,
\end{equation}
where $E $ is the Young's modulus and $h$ is the thickness.
Note that the Poisson's ratio is intrinsically determined in this model and equals $\nu=1/3$ when the mesh is an equilateral triangle.
The local stretching force vector, $\mathbf{F}^{s}_{\mathrm{local}} \in \mathcal{R}^{6 \times 1}$, as well as the local stretching stiffness matrix, $\mathbb{K}^{s}_{\mathrm{local}} \in \mathcal{R}^{6 \times 6}$, can be derived through a variational approach as
\begin{equation}
\mathbf{F}^{s}_{\mathrm{local}} = -\frac{\partial E^{s}}  {\partial \mathbf{q}^{s}}, \; \mathrm{and} \; \mathbb{K}^{s}_{\mathrm{local}} = \frac {\partial^2 E^{s}}  {\partial \mathbf{q}^{s} \partial \mathbf{q}^{s}}.
\end{equation}
The detailed formulation can be found in the MATLAB code.
Finally, the global stretching force vector,  $\mathbf{F}^{s}$, and the associated stiffness matrix, $\mathbb{K}^{s}$, can be assembled by iterating over all stretching elements.

\paragraph{Bending element} Before constructing the bending element, we first introduce the triangular element used in 3D surface simulation.
Each triangular element has $3$ vertices as
\begin{equation}
\mathcal{T}: \{ \mathbf{x}_{1}, \mathbf{x}_{2}, \mathbf{x}_{3} \}.
\end{equation}\
When two neighboring triangular elements share a common edge (i.e., two common nodes), a bending element is formed to capture the angular deviation between them as
\begin{equation}
\mathcal{B}: \{ \mathcal{T}_{1}, \mathcal{T}_{2} \}, \; \mathrm{with} \; \mathcal{T}_{1} : \{ \mathbf{x}_{1}, \mathbf{x}_{2}, \mathbf{x}_{3} \} \; \mathrm{and} \; \mathcal{T}_{2} : \{ \mathbf{x}_{2}, \mathbf{x}_{3}, \mathbf{x}_{4} \},
\end{equation}
where $\{ \mathbf{x}_{2} , \mathbf{x}_{3} \}$ are the joint nodes, and $\{ \mathbf{x}_{1}, \mathbf{x}_{4} \}$ are the two other nodes.
Therefore, the local DOF vector is defined as 
\begin{equation}
\mathbf{q}^{b} \equiv [\mathbf{x}_{1}; \mathbf{x}_{2};\mathbf{x}_{3};\mathbf{x}_{4} ] \in \mathcal{R}^{12 \times 1}.
\end{equation}
The edge vectors are given by
\begin{equation}
\begin{aligned}
 \mathbf{e}_{1} &= \mathbf{x}_{2}  -\mathbf{x}_{1}, \\
 \mathbf{e}_{2} &= \mathbf{x}_{3}  -\mathbf{x}_{1}, \\
 \mathbf{e}_{3} &= \mathbf{x}_{2}  -\mathbf{x}_{4}, \\
 \mathbf{e}_{4} &= \mathbf{x}_{3}  -\mathbf{x}_{4}.
\end{aligned}
\end{equation}
Next, the surface normal vectors of two triangular meshes are given by
\begin{equation}
\begin{aligned}
\mathbf{n}_1  &=  \frac {\mathbf{e}_{1} \times \mathbf{e}_{2}} { || \mathbf{e}_{1} \times \mathbf{e}_{2}||}, \\ 
\mathbf{n}_2  &=  \frac {\mathbf{e}_{3} \times \mathbf{e}_{4}} {|| \mathbf{e}_{3} \times \mathbf{e}_{4} ||}.
\end{aligned}
\end{equation}
The bending curvature is associated with the turning angle between the two connecting surfaces, which can be expressed as
\begin{equation}
{\kappa} = || \mathbf{n}_{1} - \mathbf{n}_{2} ||.
\end{equation}
The discrete bending energy is given in a similar approach
\begin{equation}
E^b = \frac{1}{12\sqrt{3}} Eh^3 (\kappa  - \bar{\kappa} )^2.
\end{equation}
The local bending force vector, $\mathbf{F}^{b}_{\mathrm{local}} \in \mathcal{R}^{12 \times 1}$, as well as the local bending stiffness matrix, $\mathbb{K}^{b}_{\mathrm{local}} \in \mathcal{R}^{12 \times 12}$, can be derived through a variational approach
\begin{equation}
\mathbf{F}^{b}_{\mathrm{local}} = -\frac{\partial E^{b}}  {\partial \mathbf{q}^{b}}, \; \mathrm{and} \; \mathbb{K}^{b}_{\mathrm{local}} = \frac {\partial^2 E^{b}}  {\partial \mathbf{q}^{b} \partial \mathbf{q}^{b}}.
\end{equation}
The detailed formulation can be found in the MATLAB code.
Finally, the global bending force vector,  $\mathbf{F}^{b}$, and the associated stiffness matrix, $\mathbb{K}^{b}$, can be assembled by iterating over all bending elements.

\paragraph{Equations of motion} With the derivation of the internal elastic force and the associated stiffness matrix, we include the inertial and damping effect to formulate the dynamic equations of motion.
Here, the mass matrix, $\mathbb{M}$, is time-invariant and can be easily determined based on the element size and material density. 
We then employ a simple damping matrix, which is linearly related to the mass matrix with a damping coefficient $\mu$, i.e., $\mathbb{C} = \mu \mathbb{M}$.
Finally, the equations of motion for 3D plate or shell system is 
\begin{equation}
\mathbb{M} \ddot{\mathbf{q}} + \mu \mathbb{M} \dot{\mathbf{q}} - \mathbf{F}^{s} - \mathbf{F}^{b} -  \mathbf{F}^{\text{ext}} = \mathbf{0}.
\end{equation}
The implicit Euler method and Newton's method are used to solve the $(3N)$-sized nonlinear dynamic system.
It is worth noting that this model is geometrically exact and can converge to the continuum limit of F{\"o}ppl-von K{\'a}rm{\'a}n equations \cite{seung1988defects}, but may not be able to provide an accurate prediction.

\subsubsection{Case 1: Deflection of a plate under gravity~\href{https://github.com/weicheng-huang-mechanics/DDG_Tutorial/tree/main/3d_surface/case_1}{\texorpdfstring{\faGithub}{GitHub}} } 

In this case study, we examine the deflection of a flexible plate under the influence of gravity. 
The plate is modeled with clamped-free boundary conditions, with the gravitational load applied uniformly across its surface.
The case highlights the ability of DDG model to accurately simulate the deflection and bending behavior of the plate, providing a basis for more complex simulations involving intricate geometries and varying loading conditions.

\begin{figure*}[!ht]
\centering
\includegraphics[width=\textwidth]{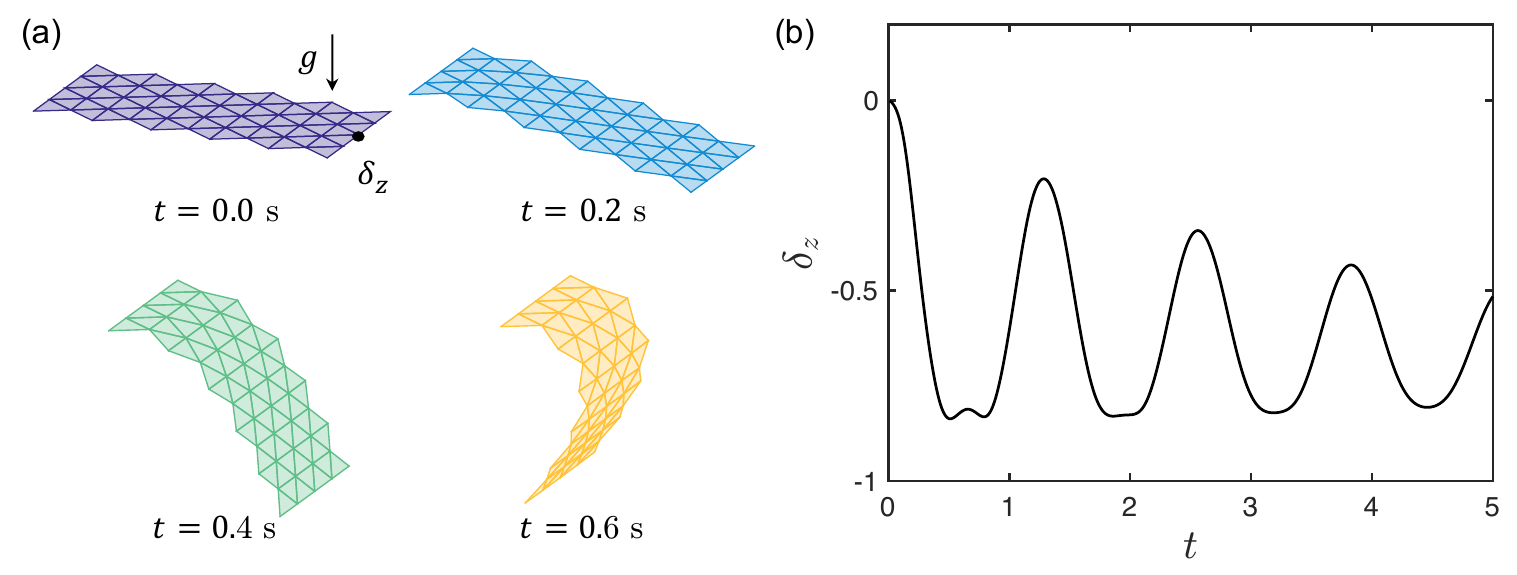}
\caption{ \textbf{A plate deformation under gravity}. (a) Deformation process of the plate at $t$ = 0.0 $s$, 0.2 $s$, 0.4 $s$, and 0.6 $s$. (b) The vertical displacement, $\delta_z$, of the plate tip, as a function of time, $t$.}
\label{fig:plate_case_1_plot}
\end{figure*}

\paragraph{Simulation initialization} To initialize the simulation, the following inputs are used:

\begin{enumerate}

\item \textbf{Geometry and connection.} (i) Nodal position: the position of the nodes $\mathbf{q}(t=0)$, with a total of $N=50$. The plate length is $L=1.0$ m, and the width is $W=0.4$ m. (ii) Stretching element: the connection of two nodes, with a total of $N_{s} = 119$. (iii) Triangular mesh, with a total of $N_{t} = 70$. (iv) Bending element: the connection of two triangular mesh, with a total of $N_{b}=91$.

\item \textbf{Physical parameters.} (i) Young's modulus, $E=100$ MPa. (ii) Material density, $\rho=1000$ $\mathrm{kg/m^3}$. (iii) Plate thickness, $h = 0.01$ m. (iv) Gravitational field $ \mathbf{g}=[0.0,0.0,-10.0]^T \mathrm{~m/s^2}$. (v) Damping viscosity, $\mu = 0.1$. (vi) The overall simulation is dynamic, i.e., $ \mathrm{ifStatic} = 0$.

\item \textbf{Numerical parameters.} (i) Total simulation time, $T=5.0$ s. (ii) Time step size, $\mathrm{d}t=0.01$ s. (iii) Numerical tolerance, $\mathrm{tol} = 1 \times 10^{-4}$. (iv) Maximum iterations, $N_{\mathrm{iter}} = 10$.

\item \textbf{Boundary conditions.} The first two rows of nodes are fixed to achieve a clamped-free boundary condition, thus, the constrained array, $\mathcal{FIX}$, can be constructed accordingly.

\item \textbf{Initial conditions.} (i) Initial position is input from the nodal positions. (ii) Initial velocity is set
to zeros for all nodes.

\item \textbf{Loading steps.} External gravitational force is applied to each node throughout the simulation.

\end{enumerate}

\paragraph{Simulation results} 
The plate undergoes deformation under gravity, as illustrated in Fig.~\ref{fig:plate_case_1_plot}(a). 
In Fig.~\ref{fig:plate_case_1_plot}(b), the displacement of the marked node along the $Z$ axis, $\delta_z$, exhibits oscillatory behavior.
The dynamic rendering can be found~\href{https://github.com/weicheng-huang-mechanics/DDG_Tutorial/blob/main/assets/plate_1.gif}{here}.

\subsubsection{Case 2: Plate wrinkling under gravity~\href{https://github.com/weicheng-huang-mechanics/DDG_Tutorial/tree/main/3d_surface/case_2}{\texorpdfstring{\faGithub}{GitHub}}}

Unlike the simple deflection case, this example introduces the phenomenon of wrinkling, a manifestation of instability in thin structures under compression \cite{seung1988defects, liang2009shape, savin2011growth, huang2024dynamic}. In this case study, we examine the wrinkling behavior of a plate subjected to gravity.
The plate is modeled with two fixed boundary corners.
As the plate deforms under the gravitational load, wrinkles develop due to instability in the structure.
This case highlights the ability of the DDG model to accurately capture the complex deformation pattern of wrinkling.

\begin{figure*}[!ht]
\centering
\includegraphics[width=\textwidth]{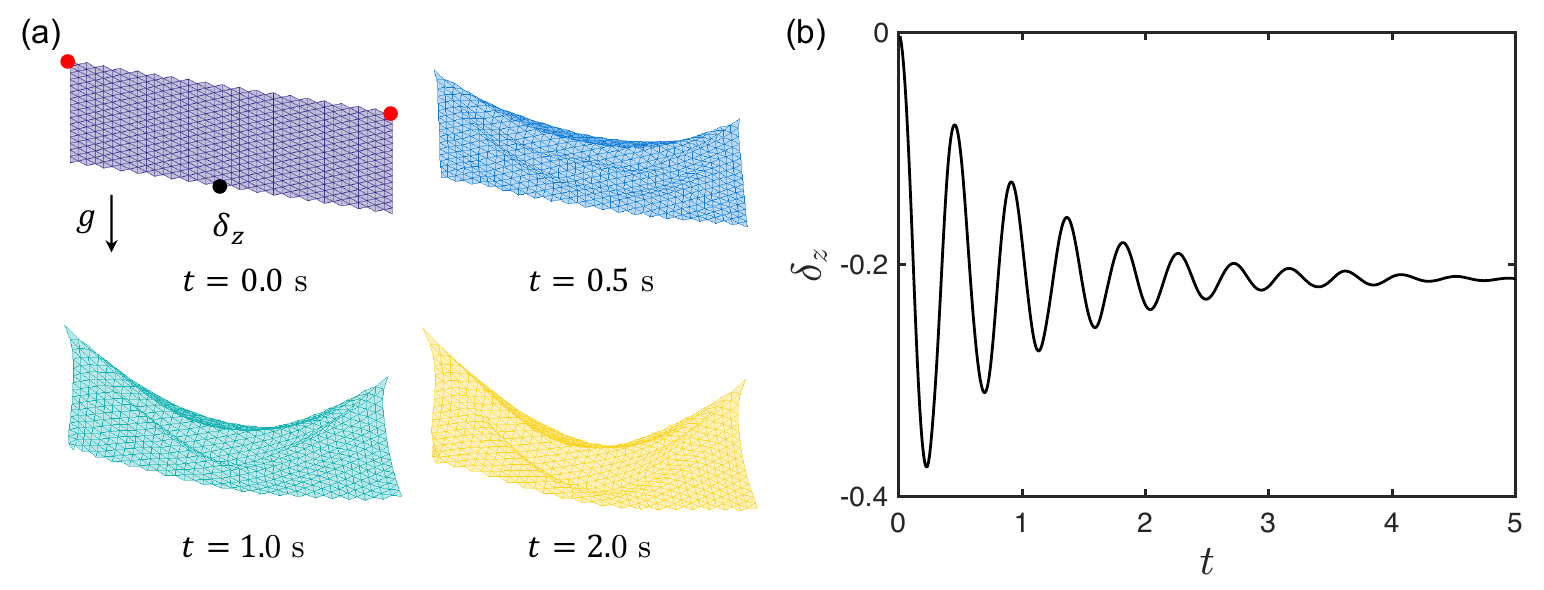}
\caption{ \textbf{Plate wrinkling under gravity.} (a) Snapshots of the plate's deformation at various time points. (b) The vertical displacement of the marked node, $\delta_z$,  as a function of time, $t$.}
\label{fig:plate_case_2_plot}
\end{figure*}

\paragraph{Simulation initialization} To initialize the simulation, the following inputs are used:

\begin{enumerate}

\item \textbf{Geometry and connection.} (i) Nodal position: the position of the nodes $\mathbf{q}(t=0)$, with a total of $N=644$. The plate length is $L=1.0$ m, and the width is $W=0.5$ m. (ii) Stretching element: the connection of two nodes, with a total of $N_{s} = 1821$. (iii) Triangular mesh, with a total of $N_{t} = 1178$. (iv) Bending element: the connection of two triangular mesh, with a total of $N_{b}=1713$.

\item \textbf{Physical parameters.} (i) Young's modulus, $E=0.1$ MPa. (ii) Material density, $\rho=1000$ $\mathrm{kg/m^3}$. (iii) Plate thickness, $h = 0.001$ m. (iv) Gravitational field $ \mathbf{g}=[1.0,1.0,-30.0]^T \mathrm{~m/s^2}$. (v) Damping viscosity, $\mu = 0.01$. (vi) The overall simulation is dynamic, i.e., $ \mathrm{ifStatic} = 0$.

\item \textbf{Numerical parameters.} (i) Total simulation time, $T=5.0$ s. (ii) Time step size, $\mathrm{d}t=0.01$ s. (iii) Numerical tolerance, $\mathrm{tol} = 1 \times 10^{-4}$. (iv) Maximum iterations, $N_{\mathrm{iter}} = 10$.

\item \textbf{Boundary conditions.} The two corner nodes, $\{\mathbf x_{17}, \mathbf x_{644} \}$, are fixed; thus, the constrained array, $\mathcal{FIX}$, can be constructed accordingly.

\item \textbf{Initial conditions.} (i) Initial position is input from the nodal positions. (ii) Initial velocity is set
to zeros for all nodes.

\item \textbf{Loading steps.} External gravitational force is applied to each node throughout the simulation.

\end{enumerate}

\paragraph{Simulation results} 
The plate exhibits wrinkling behavior under gravity, as shown in Fig.~\ref{fig:plate_case_2_plot}(a).
In Fig.~\ref{fig:plate_case_2_plot}(b), we show the displacement of the marked node along $Z$ direction, $\delta_z$, as a function of time $t$.
As time progresses, the amplitude of $\delta_z$ decreases, indicating the damping effect in this process. 
The dynamic rendering can be found~\href{https://github.com/weicheng-huang-mechanics/DDG_Tutorial/blob/main/assets/plate_2.gif}{here}.

\subsubsection{Case 3: Indentation of a cylindrical shell~\href{https://github.com/weicheng-huang-mechanics/DDG_Tutorial/tree/main/3d_surface/case_3}{\texorpdfstring{\faGithub}{GitHub}}}

Shell indention is a complicated problem and involves many nonlinear buckling, snapping, and bifurcations \cite{vella2012indentation, nasto2014localized, vaziri2008localized, lazarus2012geometry, lu2025buckling}.
In this case study, we examine the indentation of a cylindrical shell under point load. 
A half-cylindrical shell with a clamped-free boundary condition is considered, and a point indention is applied at the middle point of the free end.
This case highlights the ability of the DDG model to capture the complex deformed patterns of bifurcations in shell structures.

\begin{figure*}[!ht]
\centering
\includegraphics[width=\textwidth]{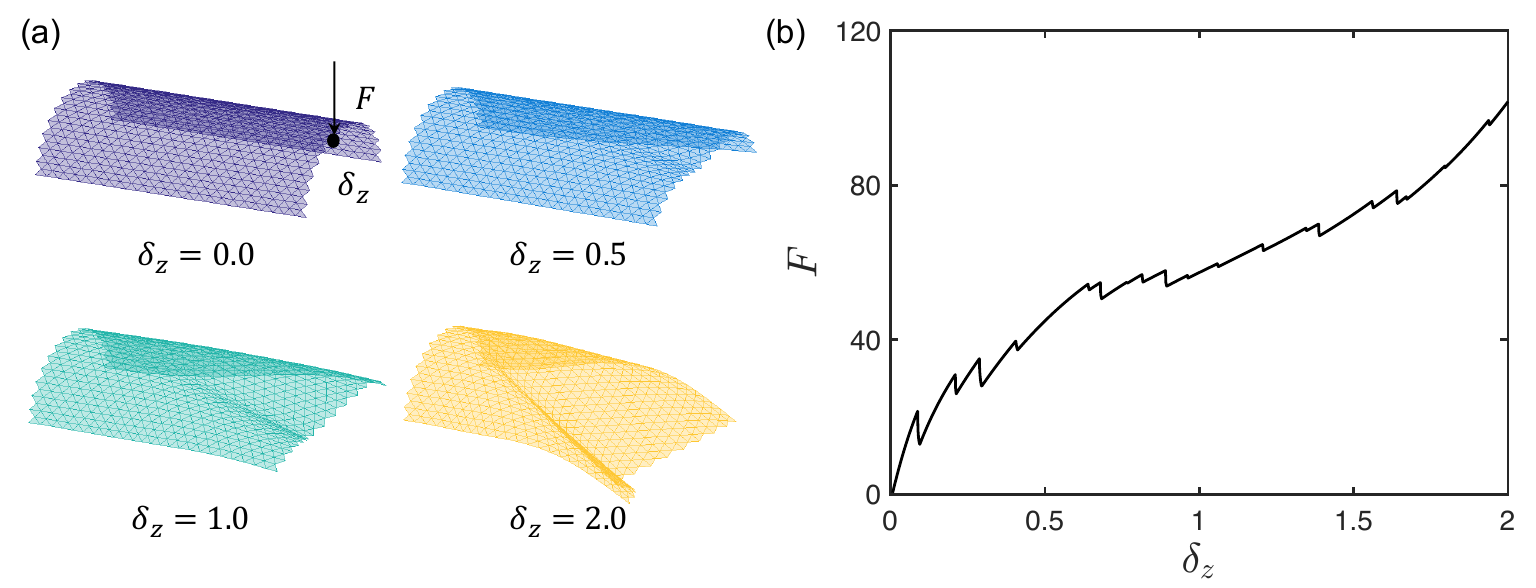}
\caption{ \textbf{Indentation of a cylindrical shell.} (a) Snapshots of the shell's deformation at various loading points. (b) The external force, $F$, as a function of vertical displacement of the marked node, $\delta_z$.}
\label{fig:plate_case_3_plot}
\end{figure*}

\paragraph{Simulation initialization} To initialize the simulation, the following inputs are used:

\begin{enumerate}

\item \textbf{Geometry and connection.} (i) Nodal position: the position of the nodes $\mathbf{q}(t=0)$, with a total of $N=760$. The half-cylindrical shell is of radius $R=1.0$ m and length $L=3.0$ m. (ii) Stretching element: the connection of two nodes, with a total of $N_{s} = 2169$. (iii) Triangular mesh, with a total of $N_{t} = 1410$. (iv) Bending element: the connection of two triangular mesh, with a total of $N_{b}=2061$.

\item \textbf{Physical parameters.} (i) Young's modulus, $E=1.0$ MPa. (ii) Material density, $\rho=1000$ $\mathrm{kg/m^3}$. (iii) Shell thickness, $h = 0.05$ m. (iv) Gravitational field $ \mathbf{g}=[0.0,0.0,0.0]^T \mathrm{~m/s^2}$. (v) Damping viscosity, $\mu = 0.01$. (vi) The overall simulation is static, i.e., $ \mathrm{ifStatic} = 1$.

\item \textbf{Numerical parameters.} (i) Total simulation time, $T=20.0$ s. (ii) Time step size, $\mathrm{d}t=0.01$ s. (iii) Numerical tolerance, $\mathrm{tol} = 1 \times 10^{-4}$. (iv) Maximum iterations, $N_{\mathrm{iter}} = 10$.

\item \textbf{Boundary conditions.} The first two columns of the nodes are fixed to achieve a clamped-free type boundary condition; the $Z$ position of the $417$-th node is fixed to apply to load; thus, the constrained array, $\mathcal{FIX}$, can be then constructed accordingly.

\item \textbf{Initial conditions.} (i) Initial position is input from the nodal positions. (ii) Initial velocity is set
to zeros for all nodes.

\item \textbf{Loading steps.} The $Z$ position of $417$-th node is manually moved along minus $Z$ axis to perform the vertical loading, the loading rate is $v_{z} = 0.1$ m/s.

\end{enumerate}

\paragraph{Simulation results} The shell exhibits complex configurations under point indention, as shown in Fig.~\ref{fig:plate_case_3_plot}(a).
In Fig.~\ref{fig:plate_case_3_plot}(b), we show the external loading force $F$ as a function of displacement of the marked node along $Z$ direction, $\delta_z$.
The complex loading curve indicates the buckling and bifurcation behaviors existing in the shell indention process.
The dynamic rendering can be found~\href{https://github.com/weicheng-huang-mechanics/DDG_Tutorial/blob/main/assets/plate_3.gif}{here}.

\subsection{Net \& gridshell: hollow surface}

In this subsection, we examine a special class of hollow structures comprised of slender rods, which lie between the 1D and 2D structures.
The hollow system can be used in many engineering systems, e.g. architectures \cite{baek2018form, baek2019rigidity, panetta2019x, becker2023c}, deployable structures \cite{huang2022nonlinear, huang2023contact, hou2021dynamic}, mechanical metamaterials \cite{mehreganian2021structural, mehreganian2023impact, meng2020multi, mei2021mechanical, mao2022modular, fang2025large}, and woven structures \cite{mahadevan2024knitting, baek2021smooth, wang2021structured}.
The numerical formulation of net and gridshell structures is straightforward, as the mechanics of the hollow rod network can be captured using rod simulations with appropriately defined input elements.
Based on their deformation characteristics, these hollow structures can be categorized into three types:
(i) flexible net -- a network of interconnected 3D rod elements where deformation is limited to stretching while bending and twisting are neglected \cite{hou2021dynamic};
(ii) elastic gridshell -- a network of 3D rod elements that undergo stretching, bending, and twisting. However, the joints between intersecting rods are free to rotate, following a pin-pin boundary condition \cite{baek2018form};
(iii) lattice structure -- similar to the elastic gridshell but with joints capable of transmitting bending and twisting moments, enforcing a clamped-clamped boundary condition.
The primary distinction between (ii) and (iii) in simulation is the inclusion of additional bending and twisting elements at the joints to account for moment and torque transmission \cite{perez2015design}.

\subsubsection{Case 1: Deflection of a flexible net under gravity~\href{https://github.com/weicheng-huang-mechanics/DDG_Tutorial/tree/main/rod_network/case_1}{\texorpdfstring{\faGithub}{GitHub}}}

Flexible nets are largely for the capture of objects with irregular shapes, e.g., the tether-net systems can be considered for space debris capture and removal mission \cite{hou2021dynamic, huang2022nonlinear, boonrath2025robustness, botta2020simulation, benvenuto2015dynamics,benvenuto2016multibody}.
In this case study, we examine the free oscillation behavior of a flexible net subjected to gravity.
The net is modeled with fixed boundary conditions at its vertices, allowing for deformation at the internal nodes.
As the gravitational load acts on the net, it undergoes oscillatory motion due to its inherent flexibility. 
This case highlights the ability of the DDG model to effectively capture the deformation behaviors of the flexible net.

\paragraph{Simulation initialization} To initialize the simulation, the following inputs are used:

\begin{enumerate}

\item \textbf{Geometry and connection.} (i) Nodal position: the position of the nodes $\mathbf{q}(t=0)$, with a total of $N = 631$. The net is initialized as a hexagon with a side length of $L = 1.0$ m. (ii) Stretching element: the connections between the nodes, with a total of $N_{s} = 720$.

\item \textbf{Physical parameters.} (i) Young's modulus, $ E = 1.0$ MPa. (ii) Material density, $\rho=1000 $ $\mathrm{kg/m^3}$. (iii) Cross-sectional radius, $r_{0} = 0.01$ m. (iv) Damping viscosity, $\mu = 0.1$. (v) Gravity, $ \mathbf{g}= [0.0, 0.0, -50]^T$ $\mathrm{m/s^2}$. (vi) The overall simulation is dynamic, i.e., $ \mathrm{ifStatic} = 0$. 

\item \textbf{Numerical parameters.} (i) Total simulation time, $T = 2.0$ s. (ii) Time step size, $\mathrm{d}t = 0.01$ s. (iii) Numerical tolerance, $\mathrm{tol} = 1 \times 10^{-4}$. (iv) Maximum iterations, $N_{\mathrm{iter}} = 10$.

\item \textbf{Boundary conditions.} The six vertices are fixed, $\{ \mathbf{x}_{2}, \mathbf{x}_{21}, \mathbf{x}_{36}, \mathbf{x}_{51}, \mathbf{x}_{66}, \mathbf{x}_{81}\}$, thus, the constrained array, $\mathcal{FIX}$, can be constructed accordingly.

\item \textbf{Initial conditions.} (i) Initial position is input from the nodal positions. (ii) Initial velocity is set
to zeros for all nodes.

\item \textbf{Loading steps.} External gravitational force is applied to each node throughout the simulation.

\end{enumerate}

\begin{figure*}[!ht]
\centering
\includegraphics[width=\textwidth]{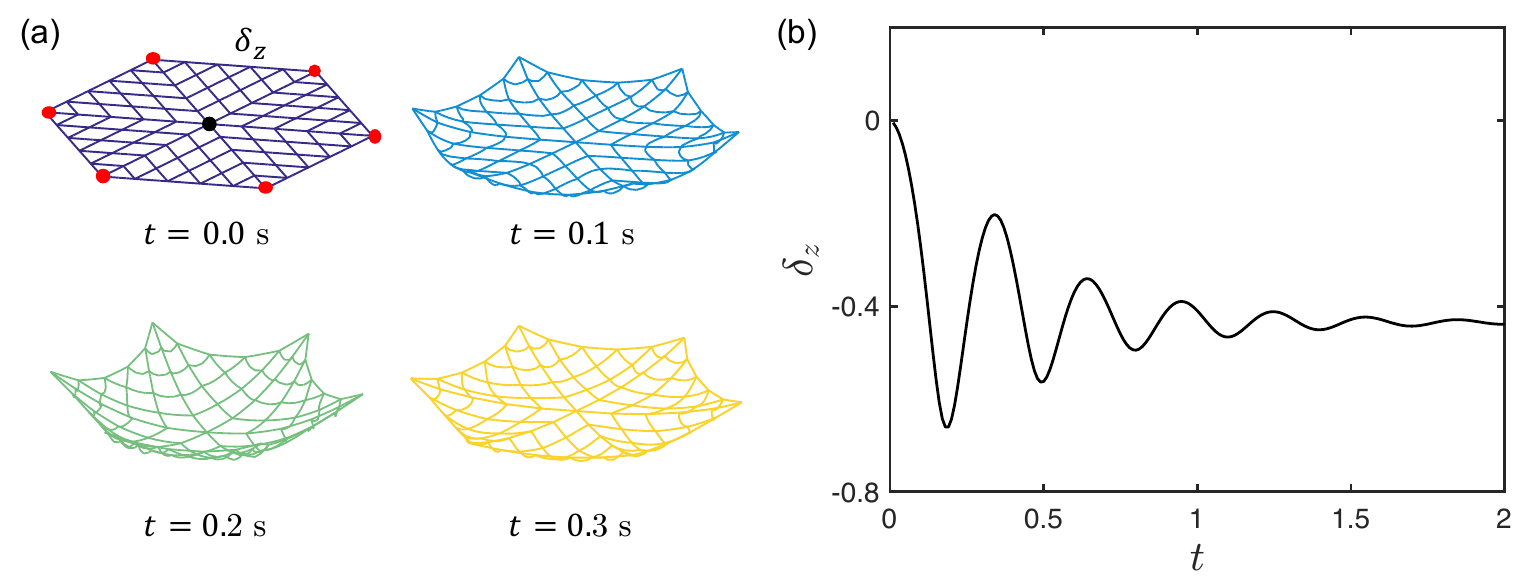}
\caption{ \textbf{Flexible net under gravity.} (a) Deformation under the gravity. (b) The central nodal displacement along $Z$-direction, $\delta_{z}$, as a function of time, $t$.}
\label{fig:net_case_1_plot}
\end{figure*}

\paragraph{Simulation results}
The net initially deforms due to its own weight, exhibiting characteristic sagging and oscillatory motion, as illustrated in Fig.~\ref{fig:net_case_1_plot}(a).
Over time, the net continues to vibrate, with these oscillations gradually damping as energy dissipates. 
In Fig.~\ref{fig:net_case_1_plot}(b), we present the time evolution of the central nodal displacement in the $Z$-direction $\delta_z$. 
The plot captures both the initial transient response and the subsequent settling phase, highlighting the interplay between gravitational loading and the net's inherent elasticity. 
The dynamic rendering can be found~\href{https://github.com/weicheng-huang-mechanics/DDG_Tutorial/blob/main/assets/net_1.gif}{here}.

\subsubsection{Case 2: Form-finding of a gridshell~\href{https://github.com/weicheng-huang-mechanics/DDG_Tutorial/tree/main/rod_network/case_2}{\texorpdfstring{\faGithub}{GitHub}}}

Elastic gridshell refers to a network of elastic rods linked by joints and finds broad applications in the civil engineering community,  e.g., Helsinki Zoo’s observatory tower and Centre Pompidou Metz.
In this case study, we examine the form-finding behavior of a hemispherical gridshell structure subjected to compressive loading \cite{baek2018form, baek2019rigidity, huang2022numerical, huang2021numericalGridshell, qin2020genetic, panetta2019x, becker2023c}. 
The grid is modeled with an initially flat shape, allowing it to deform and self-organize into a stable, curved configuration through controlled buckling. 
As the compressive forces act on the structure, the grid transitions into its equilibrium shape, illustrating the role of buckling in shape formation. 
This case highlights the ability of the DDG model to accurately capture the self-organizing behavior of slender elastic structure.

\paragraph{Simulation initialization} To initialize the simulation, the following inputs are used:

\begin{enumerate}

\item \textbf{Geometry and connection.} (i) Nodal position: the position of the nodes $\mathbf{q} (t=0)$, with a total of $N = 365$. The gridshell is constructed by the intersection of $10$ orthogonally interconnected rods, creating an overall circular structure. The length of the two longest rods is set at $L = 2.0$ m. (ii) Stretching element: the connections between the nodes, with a total of $N_{s} = 376$. (iii) Bending element: the connections between the edges, with a total of  $N_{b}=366$. (iv) Target position: the final position for the $20$ footprints.

\item \textbf{Physical parameters.} (i) Young's modulus, $ E = 100$ MPa. (ii) Material density, $\rho=1000 $ $\mathrm{kg/m^3}$. (iii) Cross-sectional radius, $r_{0} = 0.01$ m, thus $EI_{1} = EI_{2} =  E\pi r_0^4/4$ and $GJ=G\pi r_0^4/2$. (iv) Damping viscosity, $\mu = 1.0$. (v) Gravity, $ \mathbf{g}= [ 0.0, 0.0, 10.0]^{T}$ $\mathrm{m/s^2}$. (vi) The overall simulation is dynamic, i.e., $ \mathrm{ifStatic} = 0$.

\item \textbf{Numerical parameters.} (i) Total simulation time, $T = 10.0$ s. (ii) Time step size, $\mathrm{d}t = 0.01$ s. (iii) Numerical tolerance, $\mathrm{tol} = 1 \times 10^{-4}$. (iv) Maximum iterations, $N_{\mathrm{iter}} = 10$.

\item \textbf{Boundary conditions.} The $20$ boundary nodes are fixed and manually moved along a prescribed path; thus, the constrained array, $\mathcal{FIX} $, can be constructed accordingly.

\item \textbf{Initial conditions.} (i) Initial position is input from the nodal positions. (ii) Initial velocity is set
to zeros for all nodes.

\item \textbf{Loading steps.}  (i) Perturbation step: a perturbation to the initial horizontal configuration is created by applying a gravitational force (with $ \mathbf{g}= [ 0.0, 0.0, 10.0]^{T}$ $\mathrm{m/s^2}$) when $t \le 1.0\mathrm{~s}$. (ii) Compression step: the footprints are moved to induce the form-finding process during $1.0 \; \mathrm{s} < t < 5.0 \; \mathrm{s}$ . (iii) Remove Perturbation: gravitational force is removed (with $ \mathbf{g}= [ 0.0, 0.0, 0.0]^{T}$ $\mathrm{m/s^2}$) to eliminate residual perturbations when $t \ge 5.0$ s. 

\end{enumerate}

\begin{figure*}[!ht]
\centering
\includegraphics[width=\textwidth]{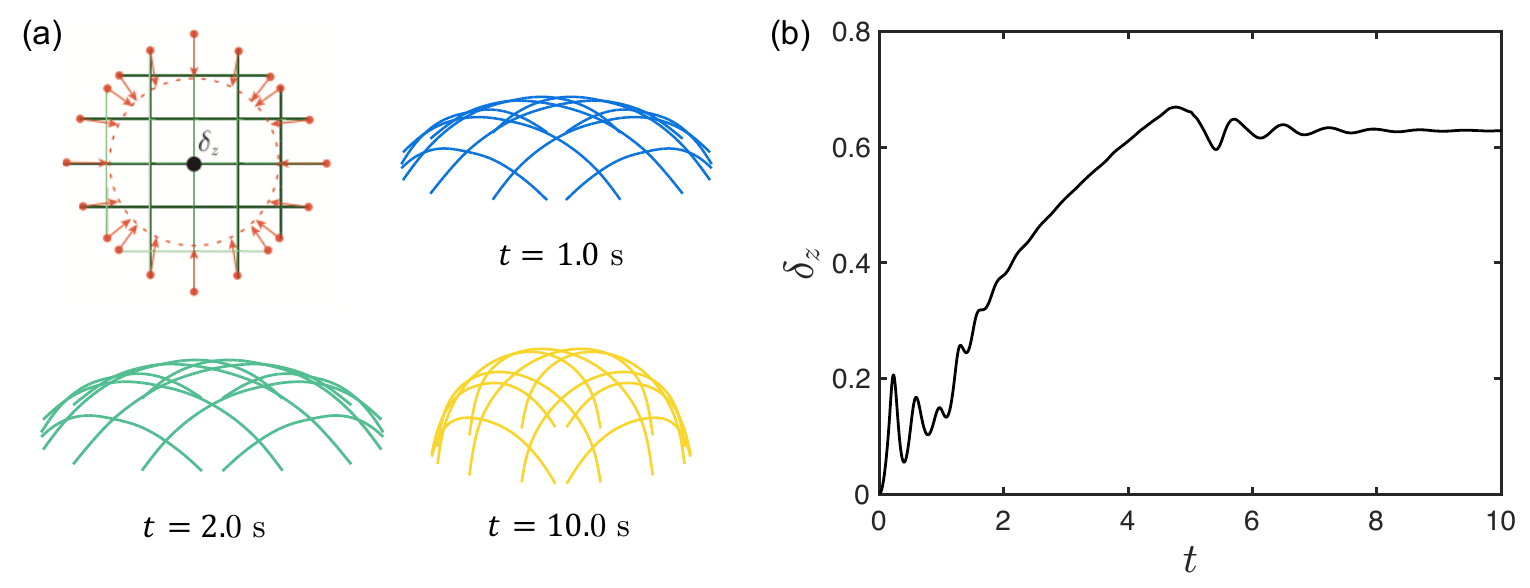}
\caption{\textbf{Buckling-induced form-finding in gridshells.} (a) Form-finding process induced by buckling. (b) Pole displacement, $\delta_{z}$, as a function of time, $t$.}
\label{fig:net_case_2_plot}
\end{figure*}

\paragraph{Simulation results} 
The hemispherical gridshell cap is generated through the controlled buckling of an initially planar gridshell structure, as illustrated in Fig.~\ref{fig:net_case_2_plot}(a). 
This transformation occurs due to the application of in-plane compressive forces, which induce an instability that leads to the formation of the curved configuration. 
The resulting gridshell cap exhibits a stable, self-supporting geometry that efficiently distributes loads through its interconnected framework. 
In Fig.~\ref{fig:net_case_2_plot}(b), we present the temporal evolution of the central nodal displacement in the $Z$-direction $\delta_z$. 
This displacement profile reveals the key stages of buckling, including the initial instability, rapid shape transition, and subsequent oscillatory adjustments as the structure settles into its final equilibrium state. 
This is a dynamic relaxation-based form-finding for an elastic gridshell, i.e., we use a dynamic approach to solve the nonlinear system and then employ the damping effect to derive its stable equilibrium configuration.
The dynamic rendering can be found~\href{https://github.com/weicheng-huang-mechanics/DDG_Tutorial/blob/main/assets/net_2.gif}{here}.

\subsubsection{Case 3: Deformation of a lattice grid under gravity~\href{https://github.com/weicheng-huang-mechanics/DDG_Tutorial/tree/main/rod_network/case_3}{\texorpdfstring{\faGithub}{GitHub}}}

Lattice grid systems are largely used in programmable metamaterial \cite{walia2015flexible, bertoldi2017flexible, mehreganian2021structural, mehreganian2023impact, meng2020multi, mei2021mechanical, mao2022modular, fang2025large}.
In this case study, we examine the free oscillation behavior of a lattice grid structure subjected to gravity.
The grid is modeled with a fixed corner, allowing it to deform under its self-weight.
As the gravitational load acts on the lattice, it undergoes oscillatory motion due to its inherent flexibility.
This case highlights the ability of the DDG model to effectively capture the structural flexibility and load-bearing characteristics of lattice frameworks.

\paragraph{Simulation initialization} To initialize the simulation, the following inputs are used:

\begin{enumerate}

\item \textbf{Geometry and connection.} (i) Nodal position: the position of the nodes $\mathbf{q}(t=0)$, with a total of $N = 104$. The lattice is constructed as a cube with a side length of $L = 1.0$ m. (ii) Stretching element: the connections between the nodes, with a total of $N_{s} = 108$. (iii) Bending element: the connections between the edges, with a total of  $N_{b}=120$. 

\item \textbf{Physical parameters.} (i) Young's modulus, $ E = 1.0$ GPa. (ii) Material density, $\rho=1000 $ $\mathrm{kg/m^3}$. (iii) Cross-sectional radius, $r_{0} = 0.01$ m, thus $EI_{1} = EI_{2} =  E\pi r_0^4/4$ and $GJ=G\pi r_0^4/2$. (iv) Damping viscosity, $\mu = 0.1$. (v) Gravity, $ \mathbf{g}= [ 0.0, 0.0, -1.0]^{T}$ $\mathrm{m/s^2}$. (vi) The overall simulation is dynamic, i.e., $ \mathrm{ifStatic} = 0$.

\item \textbf{Numerical parameters.} (i) Total simulation time, $T = 10.0$ s. (ii) Time step size, $\mathrm{d}t = 0.01$ s. (iii) Numerical tolerance, $\mathrm{tol} = 1 \times 10^{-4}$. (iv) Maximum iterations, $N_{\mathrm{iter}} = 10$.

\item \textbf{Boundary conditions.} The nodes at the one corner, $\{ \mathbf{x}_{1}, \mathbf{x}_{2}, \mathbf{x}_{11},\mathbf{x}_{20}\}$, as well as the first twisting angle, $\{\theta_{1}\}$, are fixed for a clamped like boundary condition.

\item \textbf{Initial conditions.} (i) Initial position is input from the nodal positions. (ii) Initial velocity is set
to zeros for all nodes.

\item \textbf{Loading steps.} External gravitational force is applied to each node throughout the simulation.

\end{enumerate}

\begin{figure*}[!ht]
\centering
\includegraphics[width=\textwidth]{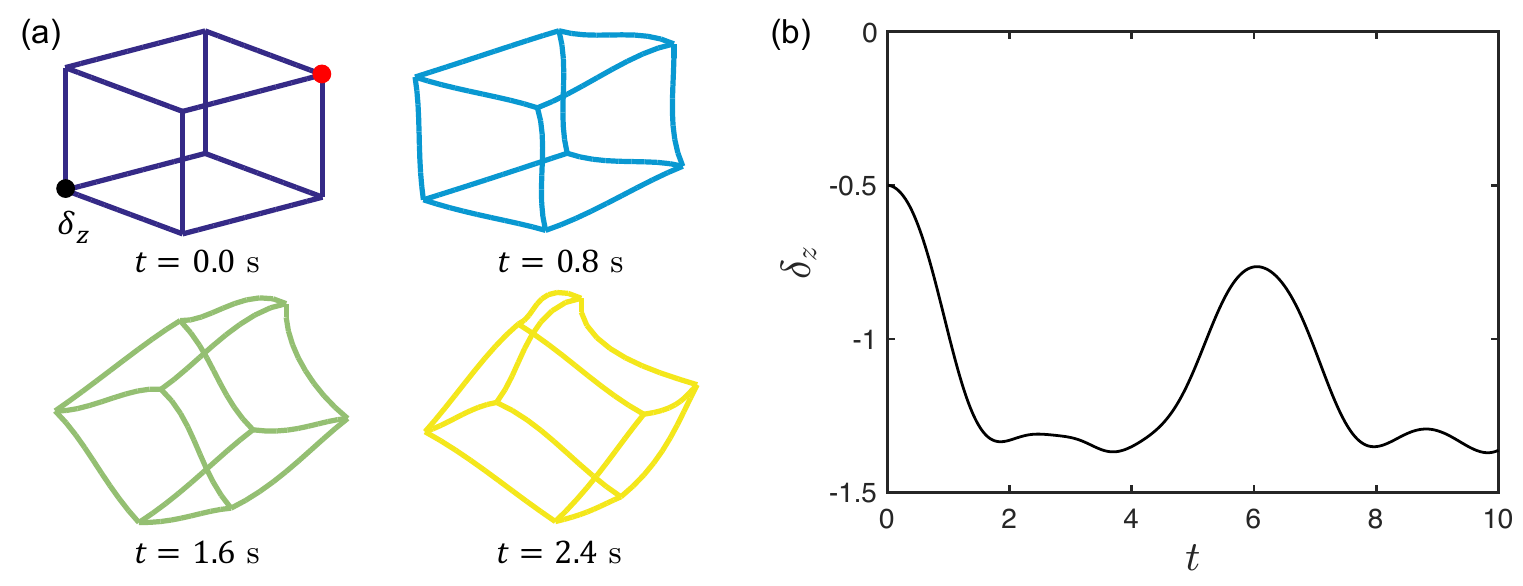}
\caption{ \textbf{Lattice grid under gravity}. (a) Deformed patterns under gravity. (b) The $Z$ displacement, $\delta_{z}$, of the black-dotted point as a function of time, $t$.}
\label{fig:net_case_3_plot}
\end{figure*}

\paragraph{Simulation results} 
The lattice undergoes deformation under gravitational loading and exhibits dynamic swinging motion around its boundary, as depicted in Fig.~\ref{fig:net_case_3_plot}(a). 
This motion results from the interplay between the lattice’s flexibility and the constraints imposed at the boundary, leading to oscillatory behavior. 
In Fig.~\ref{fig:net_case_3_plot}(b), we present the time evolution of the nodal displacement of the marked point along $Z$ direction $\delta_z$. 
This plot highlights the amplitude and frequency of oscillations. 
The dynamic rendering can be found~\href{https://github.com/weicheng-huang-mechanics/DDG_Tutorial/blob/main/assets/net_3.gif}{here}.



\section{Flexible structures with material nonlinearity}

In this chapter, we discuss how the DDG-based numerical framework handles material nonlinearity. For example, when the system is stretching-dominated (e.g., cable or membrane), the physical strain is no longer small, thus, a hyperelastic constitutive law is required \cite{khaniki2022review, liu2024computational, zhao2007method}. 
We here show two examples that employ the DDG approach for material nonlinearity.

\subsection{Case 1: Hyperelastic planar cable~\href{https://github.com/weicheng-huang-mechanics/DDG_Tutorial/tree/main/hyper_elastic/case_1}{\texorpdfstring{\faGithub}{GitHub}}} 

In this subsection, we examine a simple planar cable, where, unlike a bending-dominated beam primarily governed by bending energy, the configuration is mainly determined by stretching energy.
Mathematically, for a 1D structure of length $L$, with linear elastic stretching stiffness $EA$, bending stiffness $EI$, and subjected to an external load $F$, a beam model is appropriate when $ EA \gg F \sim EI/L^2 $, whereas a cable model should be used when $EA \sim F \gg EI/L^2 $.
In an intermediate scenario, where $EA \gg F \gg EI/L^2 $, the configuration of the structure is governed solely by its geometric characteristics and boundary conditions, rendering the problem material-independent -- such as an inextensible catenary under its self-weight, which is known as a catenary 
\cite{kacmarynski1931catenary, bernoulli2001solutions, huang2023static}.
On the other side, for sufficient large external loads, the material response may exceed the linear regime, necessitating the use of a hyperelastic model.

\paragraph{Numerical formulation}  
Similar to the planar beam system, the DOF vector for a hyperelastic cable system is defined by $N$ nodes as
\begin{equation}
\mathbf{q} = [ \mathbf{x}_1; \mathbf{x}_2; \ldots; {\mathbf{x}_{N}} ] \in \mathcal{R}^{2N \times 1},
\end{equation}
where only stretching energy is considered, as the strain induced by bending curvature is negligible compared to that from stretching.
The stretching element is comprised of two connected nodes, defined as
\begin{equation}
\mathcal{S}: \{\mathbf{x}_{1}, \mathbf{x}_{2} \}.
\end{equation}
The local DOF vector is defined as 
\begin{equation}
\mathbf{q}^{s} \equiv [\mathbf{x}_{1}; \mathbf{x}_{2} ] \in \mathcal{R}^{4 \times 1}.
\end{equation}
The edge length is the $\mathcal{L}_2$ norm of the edge vector, defined as
\begin{equation}
l   =  || \mathbf{x}_{2}  -\mathbf{x}_{1} ||.
\end{equation}
The first principal stretch is the uniaxial elongation of edge as
\begin{equation}
\lambda_{1} =  \frac {  l } {  \bar{l} }.
\end{equation}
Hereafter, we use a bar on top to indicate the evaluation of the undeformed configuration, e.g., $\bar{l}$ is the edge length before deformation.
Due to the incompressible assumption, the other two principal stretches are
\begin{equation}
\begin{aligned}
\lambda_{2} &=  \frac{1} {\sqrt{\lambda_{1}}}, \\
\lambda_{3} &=  \frac{1} {\sqrt{\lambda_{1}}}.
\end{aligned}
\end{equation}
The strain energy function for an incompressible Mooney-Rivlin material is
\begin{equation}
E^s = \left[ C_{1} (I_{1} - 3) + C_{2} (I_{2} - 3) \right] \bar{V},
\end{equation}
where $\bar{V}$ is the volume of the edge segment, $C_{1}$ and $C_{2}$ are the two material parameters, and $I_{1}$ and $I_{2}$ are the first and second principal invariant as
\begin{equation}
\begin{aligned}
I_{1} &= \lambda_{1}^2 +\lambda_{2}^2 + \lambda_{3}^2, \\
I_{2} &= \lambda_{1}^2 \lambda_{2}^2 + \lambda_{1}^2\lambda_{3}^2 + \lambda_{2}^2 \lambda_{3}^2.
\end{aligned}
\end{equation}
The local stretching force vector, $\mathbf{F}^{s}_{\mathrm{local}} \in \mathcal{R}^{4 \times 1}$, as well as the local stretching stiffness matrix, $\mathbb{K}^{s}_{\mathrm{local}} \in \mathcal{R}^{4 \times 4}$, can be derived through a variational approach as
\begin{equation}
\mathbf{F}^{s}_{\mathrm{local}} = -\frac{\partial E^{s}}  {\partial \mathbf{q}^{s}}, \; \mathrm{and} \; \mathbb{K}^{s}_{\mathrm{local}} = \frac {\partial^2 E^{s}}  {\partial \mathbf{q}^{s} \partial \mathbf{q}^{s}}.
\end{equation}
The detailed formulation can be found in the MATLAB code.
Finally, the global stretching force vector,  $\mathbf{F}^{s}$, and the associated stiffness matrix, $\mathbb{K}^{s}$, can be assembled by iterating over all stretching elements. 
With the formulation of the internal elastic force and the associated stiffness matrix, we can incorporate the inertial and damping effect to derive the dynamic equations of motion.
Here, the mass matrix, $\mathbb{M}$, is time-invariant and can be easily obtained based on the element size and material density. 
We then employ a simple damping matrix, which is linearly related to the mass matrix with a damping coefficient $\mu$, i.e., $\mathbb{C} = \mu \mathbb{M}$.
Finally, the equations of motion for hyperelastic cable system is 
\begin{equation}
\mathbb{M} \ddot{\mathbf{q}} + \mu \mathbb{M} \dot{\mathbf{q}} - \mathbf{F}^{s} -  \mathbf{F}^{\text{ext}} = \mathbf{0}.
\end{equation}
The implicit Euler method and Newton's method are used to solve the $(2N)$-sized nonlinear dynamic system.

\begin{figure*}[!ht]
\centering
\includegraphics[width=\textwidth]{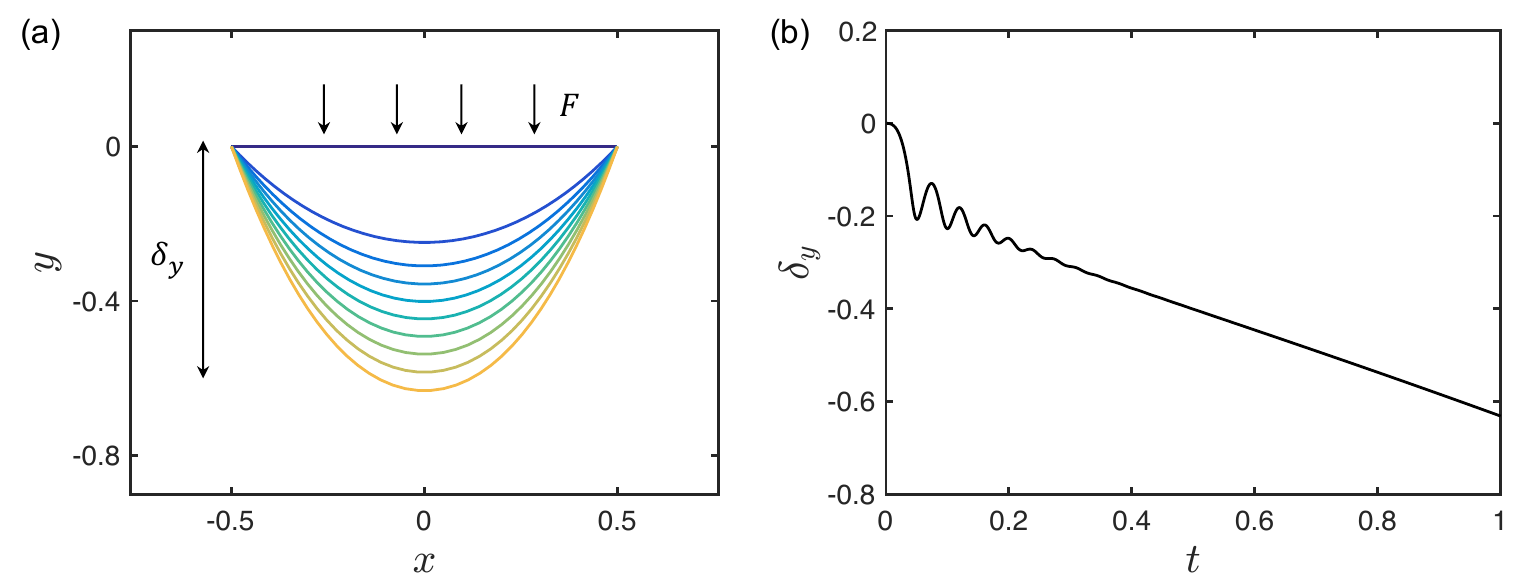}
\caption{\textbf{Hyperelastic cable under external load}. (a) Deformed process under an external vertical forces that increased at a rate $\dot{F} = 10.0$ N/s. (b) The $Y$ displacement of the center point, $\delta_{y}$, as a function of time, $t$.}
\label{fig:hyper_case_1_plot}
\end{figure*}

\paragraph{Simulation initialization} To initialize the simulation, the following inputs are used:

\begin{enumerate}

\item \textbf{Geometry and connection.} (i) Nodal position: the position of the nodes $\mathbf{q}(t=0)$, with a total of $N = 40$. The length of the cable is set at $L = 1.0$ m. (ii) Stretching element: the connections between the nodes, with a total of $N_{s} = 39$.

\item \textbf{Physical parameters.} (i) Young's modulus, $ E = 1.0$ MPa, $C_{1} = 4E/30$, and $C_{2} = E/30$. (ii) Material density, $\rho=100.0 $ $\mathrm{kg/m^3}$. (iii) Cross-sectional radius, $r_{0} = 0.01$ m. (iv) Damping viscosity, $\mu = 1.0$. (v) Gravity, $ \mathbf{g} = [0.0, 0.0]^{T}$ $\mathrm{m/s^2}$.  (vi) The overall simulation is dynamic, i.e., $ \mathrm{ifStatic} = 0$.

\item \textbf{Numerical parameters.} (i) Total simulation time, $T = 1.0$ s. (ii) Time step size, $\mathrm{d}t = 0.001$ s. (iii) Numerical tolerance, $\mathrm{tol} = 1 \times 10^{-4}$. (iv) Maximum iterations, $N_{\mathrm{iter}} = 10$.

\item \textbf{Boundary conditions.} The first node and the last node, $\{ \mathbf{x}_{1}, \mathbf{x}_{40}\}$, are fixed to achieve a pin-pin boundary condition, thus the constrained array, $\mathcal{FIX} = [1,2,79,80]^{T}$.

\item \textbf{Initial conditions.} (i) Initial position is input from the nodal positions. (ii) Initial velocity is set to zeros.

\item \textbf{Loading steps.} The external vertical force is increased at a rate $\dot{F} = 10.0$ N/s.

\end{enumerate}

\paragraph{Simulation results} 
The cable undergoes significant deformation under the influence of external load, as
depicted in Fig.~\ref{fig:hyper_case_1_plot}(a), in which the cable has been stretched a lot and went into the hyperelastic regime.
In Fig.~\ref{fig:hyper_case_1_plot}(b),  we present the displacement of the center node $\delta_y$, over time. 
The curve reveals an initial downward oscillation during the initial loading, and then gradually be stable as the enlargement of the external loads.
The dynamic rendering can be found~\href{https://github.com/weicheng-huang-mechanics/DDG_Tutorial/blob/main/assets/hyper_1.gif}{here}. 

\subsection{Case 2: Hyperelastic axisymmetric membrane~\href{https://github.com/weicheng-huang-mechanics/DDG_Tutorial/tree/main/hyper_elastic/case_2}{\texorpdfstring{\faGithub}{GitHub}}}

In this subsection, we examine an axisymmetric membrane, a commonly used hyperelastic structure for soft actuators~\cite{kumar2024deformation, zhu2025inflation, zhao2010theory, hau2018novel}, to demonstrate the effectiveness of DDG in handling nonlinear hyperelastic constitutive laws.
In contrast to the axisymmetric shell, where bending energy dominates, the axisymmetric membrane solely undergoes stretching, as seen in the inflation of a balloon.
Moreover, due to the large physical strain, the linear elastic model fails to accurately describe the membrane's inflation, making a hyperelastic model essential.~\cite{liu2021coupled, liu2024simplified, tamadapu2013finite, roychowdhury2018symmetry, kydoniefs1967finite}.

\paragraph{Numerical formulation} 
Similar to axisymmetric shell system, the DOF vector for a hyperelastic axisymmetric membrane system can be described by $N$ nodes,
\begin{equation}
\mathbf{q} = [ \mathbf{x}_1; \mathbf{x}_2; \ldots; {\mathbf{x}_{N}} ] \in \mathcal{R}^{2N \times 1},
\end{equation}
where only stretching energy is considered, as the strain induced by bending curvature is negligible compared to that from stretching.
The stretching element is comprised of two connected nodes, defined as
\begin{equation}
\mathcal{S}: \{\mathbf{x}_{1}, \mathbf{x}_{2} \}.
\end{equation}
The local DOF vector is defined as 
\begin{equation}
\mathbf{q}^{s} \equiv [\mathbf{x}_{1}; \mathbf{x}_{2} ] \in \mathcal{R}^{4 \times 1}.
\end{equation}
The edge length is the $\mathcal{L}_2$ norm of the edge vector, defined as
\begin{equation}
l   =  || \mathbf{x}_{2}  -\mathbf{x}_{1} ||.
\end{equation}
The local radius is 
\begin{equation}
r   =  \frac{1}{2} (r_{1} + r_{2}). 
\end{equation}
The first principal stretch is the elongation along the meridional direction as
\begin{equation}
{\lambda}_{1} = \frac {  l} {  \bar{l} }.
\end{equation}
Hereafter, we use a bar on top to indicate the evaluation of the undeformed configuration, e.g., $\bar{l}$ is the edge length before deformation.
The second principal stretch is along the circumferential and is related to the expansion of the circle as
\begin{equation}
{\lambda}_{2} = \frac { r } { \bar{r} }.
\end{equation}
Due to the incompressible assumption, the last principal stretch is
\begin{equation}
\lambda_{3} =  \frac{1} {{\lambda_{1}} \lambda_{2}}.
\end{equation}
The strain energy density function for an incompressible Mooney-Rivlin material is
\begin{equation}
E^s = \left[ C_{1} (I_{1} - 3) + C_{2} (I_{2} - 3) \right] \bar{V}.
\end{equation}
where $\bar{V}$ is the volume of the local membrane element, $C_{1}$ and $C_{2}$ are the two material parameters, and $I_{1}$ and $I_{2}$ are the first and second principal invariant as
\begin{equation}
\begin{aligned}
I_{1} &= \lambda_{1}^2 +\lambda_{2}^2 + \lambda_{3}^2, \\
I_{2} &= \lambda_{1}^2 \lambda_{2}^2 + \lambda_{1}^2\lambda_{3}^2 + \lambda_{2}^2 \lambda_{3}^2
\end{aligned}
\end{equation}
The local stretching force vector, $\mathbf{F}^{s}_{\mathrm{local}} \in \mathcal{R}^{4 \times 1}$, as well as the local stretching stiffness matrix, $\mathbb{K}^{s}_{\mathrm{local}} \in \mathcal{R}^{4 \times 4}$, can be derived through a variational approach as
\begin{equation}
\mathbf{F}^{s}_{\mathrm{local}} = -\frac{\partial E^{s}}  {\partial \mathbf{q}^{s}}, \; \mathrm{and} \; \mathbb{K}^{s}_{\mathrm{local}} = \frac {\partial^2 E^{s}}  {\partial \mathbf{q}^{s} \partial \mathbf{q}^{s}}.
\end{equation}
The detailed formulation can be found in the MATLAB code.
Finally, the global stretching force vector,  $\mathbf{F}^{s}$, and the associated stiffness matrix, $\mathbb{K}^{s}$, can be assembled by iterating over all stretching elements.
With the formulation of the internal elastic force and the associated stiffness matrix, we can incorporate the inertial and damping effect to derive the dynamic equations of motion.
Here, the mass matrix, $\mathbb{M}$, is time-invariant and can be easily obtained based on the element size and material density. 
We then employ a simple damping matrix, which is linearly related to the mass matrix with a damping coefficient $\mu$, i.e., $\mathbb{C} = \mu \mathbb{M}$.
Finally, the equations of motion for the hyperelastic membrane system is 
\begin{equation}
\mathbb{M} \ddot{\mathbf{q}} + \mu \mathbb{M} \dot{\mathbf{q}} - \mathbf{F}^{s} -  \mathbf{F}^{\text{ext}} = \mathbf{0}.
\end{equation}
The implicit Euler method and Newton's method are used to solve the $(2N)$-sized nonlinear dynamic system.

\begin{figure*}[!ht]
\centering
\includegraphics[width=\textwidth]{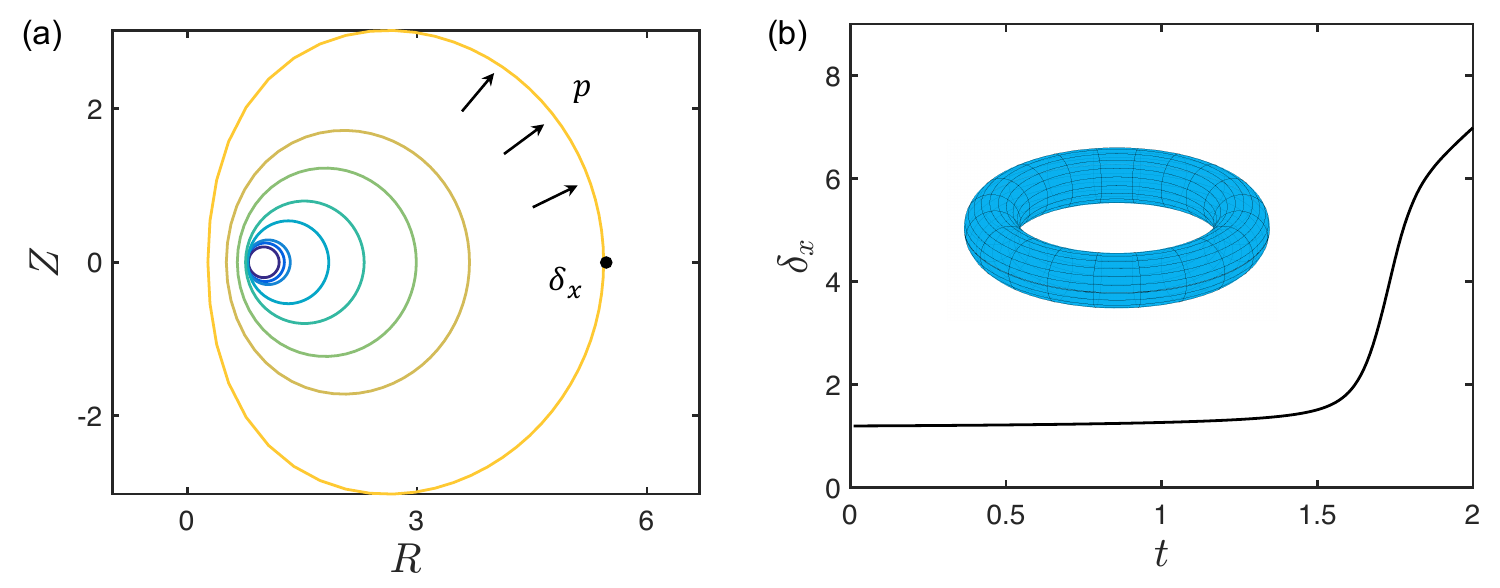}
\caption{\textbf{Hyperelastic axisymmetric membrane under internal inflation}. (a) Inflation process under an external pressure load that increased at a rate $\dot{p} = 10$ kPa/s. (b) The $x$ displacement of the black-dotted point, $\delta_{x}$, as a function of time, $t$.}
\label{fig:hyper_case_2_plot}
\end{figure*}

\paragraph{Simulation initialization} To initialize the simulation, the following inputs are used:

\begin{enumerate}

\item \textbf{Geometry and connection.} (i) Nodal position: the position of the nodes $\mathbf{q}(t=0)$, with a total of $N = 60$. The radius of the circular cross-section $R_{1} = 0.2$ m. The outer radius, which is the distance from the center of the cross-section to the center of the torus, is $R_{2} = 1.0$ m. (ii) Stretching element: the connections between the nodes, with a total of $N_{s} = 60$.

\item \textbf{Physical parameters.} (i) Shear modulus, $ G = 0.333$ MPa, thus $C_{1} = 0.4G$, and $C_{1} = 0.1G$. (ii) Material density, $\rho=100.0 $ $\mathrm{kg/m^3}$. (iii) Cross-sectional radius, $r_{0} = 0.01$ m. (iv) Damping viscosity, $\mu = 1.0$. (v) Gravity, $ \mathbf{g} = [0.0, 0.0]^{T}$ $\mathrm{m/s^2}$.  (vi) The overall simulation is dynamic, i.e., $ \mathrm{ifStatic} = 0$.

\item \textbf{Numerical parameters.} (i) Total simulation time, $T = 2.0$ s. (ii) Time step size, $\mathrm{d}t = 0.01$ s. (iii) Numerical tolerance, $\mathrm{tol} = 1 \times 10^{-4}$. (iv) Maximum iterations, $N_{\mathrm{iter}} = 10$.

\item \textbf{Boundary conditions.} The $Z$ direction of the first node is fixed to avoid rigid body motion, thus the constrained array, $\mathcal{FIX} = [2]$.

\item \textbf{Initial conditions.} (i) Initial position is input from the nodal positions. (ii) Initial velocity is set to zeros.

\item \textbf{Loading steps.} The external pressure load is increased at a rate $\dot{p} = 10.0$ kPa/s.

\end{enumerate}

\paragraph{Simulation results} 
The membrane undergoes significant inflation deformation under the influence of external pressure load, as depicted in Fig.~\ref{fig:hyper_case_2_plot}(a).
In Fig.~\ref{fig:hyper_case_2_plot}(b),  we present the displacement of the marked node $\delta_x$, over time. 
The process reveals an initially slow response followed by a rapid change after reaching a critical point at $t \approx 1.6$ s, characteristic of the material's hyperelastic properties, and is known as snap-through material instability.
The dynamic rendering can be found~\href{https://github.com/weicheng-huang-mechanics/DDG_Tutorial/blob/main/assets/hyper_2.gif}{here}. 



\section{Flexible structures interacting with external environment}

In this chapter, we discuss how the DDG-based numerical framework handles interacting with the external environment. 
In practical engineering applications, flexible structures interact with their external environment through various approaches.
For example, multi-physical actuation is essential for investigating advanced intelligent actuators~\cite{wang2020hard, zhao2019mechanics, tian2021dynamics, goriely2023rod}. 
Furthermore, the frictional contact model is crucial for the design of conformable soft electronics~\cite{wang2024design, liu2022strategies, huang2019assembly, xiao2022theoretical}, and fluid-structure interaction models are vital for developing underwater swimming robots~\cite{du2021underwater, qu2024recent, zhang2018untethered, renda2018unified}.
One of the key advantages of the DDG-based approach for flexible systems is its ability to effectively handle interactions with the external environment, enhancing its applicability in engineering solutions.
Here, we will present three examples in addressing (i) magnetic actuation, (ii) fluid-structure interaction, and (iii) frictional contact.
For clarity, we utilize a simple 2D model as a demonstration, but the framework can be easily extended to systems of varying dimensions.

\subsection{Case 1: Magnetic actuation~\href{https://github.com/weicheng-huang-mechanics/DDG_Tutorial/tree/main/soft_robot/case_1}{\texorpdfstring{\faGithub}{GitHub}}}

In this subsection, we illustrate the dynamic response of a magnetized beam subjected to external magnetic actuation~\cite{zhao2019mechanics, wang2020hard, huang2023discrete}.
Using the DDG simulation framework, we analyze how the beam deforms and oscillates in response to time-varying magnetic fields.
As magnetically actuated structures are widely explored in soft robotics, biomedical devices, and adaptive materials, this case highlights the potential of our approach in designing and optimizing magneto-mechanical systems for precise and controllable actuation~\cite{huang2023modeling, huang2023discrete, zhao2019mechanics, wang2020hard, yan2022comprehensive, sano2022kirchhoff, li2024model, tong2025real}.

\paragraph{Numerical formulation} 
Similar to the stretching element, the discrete magnetic force (or torque) is applied based on each magnetized rod segment, defined as
\begin{equation}
\mathcal{S}: \{\mathbf{x}_{1}, \mathbf{x}_{2} \}.
\end{equation}
The local DOF vector is defined as 
\begin{equation}
\mathbf{q}^{s} \equiv [\mathbf{x}_{1}; \mathbf{x}_{2} ] \in \mathcal{R}^{4 \times 1}.
\end{equation}
The local edge length is the $\mathcal{L}_2$ norm of the edge vector, defined as
\begin{equation}
l = || \mathbf{x}_{2} - \mathbf{x}_{1} ||.
\end{equation}
We use $\mathbf{t}$ and $\mathbf{n}$ to denote the local material frame for each rod segment, defined as
\begin{equation}
    \mathbf{t} = (\mathbf{x}_{2} - \mathbf{x}_{1}) / || \mathbf{x}_{2} - \mathbf{x}_{1} ||, \; \mathrm{and} \; \mathbf{n} \cdot \mathbf{t} = 0.
\end{equation}
The discrete format of the magnetic functional \cite{huang2023modeling, huang2023discrete} for this rod segment is 
\begin{equation}
E^{\mathrm{mag}} = - \bar{l} \cdot (\bm{\mathcal{M}} \cdot \mathbf{B} ),
\label{eq:discreteMagneticEnergies}
\end{equation}
where $\mathbf{B} \in \mathcal{R}^{2\times1}$ is the external magnetic field and $\bm{\mathcal{M}} \in \mathcal{R}^{2\times1}$ is the remanent magnetization density per length, i.e.,
\begin{equation}
\bm{\mathcal{M}} = A \left[ \left( \mathbf{t}\otimes \bar{\mathbf{t}} + \mathbf{n} \otimes \bar{\mathbf{n}} \right) \cdot \bar{\mathbf{B}}_{r} \right ].
\end{equation}
Here, $A$ is the local cross-sectional area, and $\bar{\mathbf{B}}_{r}$ is the remanent magnetization density per unit volume of the segment in the reference configuration.
Hereafter, we use a bar on top to indicate the evaluation of the undeformed configuration, e.g., $\bar{l}$ is the edge length before deformation.
The magnetic force vector,  $\mathbf{F}^{\mathrm{mag}}_{\mathrm{local}}$, can be derived by finding the gradient of the magnetic potential as
\begin{equation}
\mathbf{F}^{\mathrm{mag}}_{\mathrm{local}} = -\frac{\partial E^{\mathrm{mag}}}  {\partial \mathbf{q}^{s}}.
\end{equation}
The detailed formulation can be found in the MATLAB code.
Finally, the global magnetic force vector,  $\mathbf{F}^{\mathrm{mag}}$, can be assembled by iterating over all stretching elements.  
With the formulation of the internal elastic force and the associated stiffness matrix, we can incorporate the inertial and damping effect to derive the dynamic equations of motion.
Here, the mass matrix, $\mathbb{M}$, is time-invariant and can be easily obtained based on the element size and material density. 
We then employ a simple damping matrix, which is linearly related to the mass matrix with a damping coefficient $\mu$, i.e., $\mathbb{C} = \mu \mathbb{M}$.
Finally, the equations of motion for the magnetic beam system is 
\begin{equation}
\mathbb{M} \ddot{\mathbf{q}} + \mu \mathbb{M} \dot{\mathbf{q}} - \mathbf{F}^{s} - \mathbf{F}^{b} -  \mathbf{F}^{\text{mag}} = \mathbf{0},
\end{equation}
where $\mathbf{F}^{s}$ and $\mathbf{F}^{b} $ are the elastic stretching force vector and elastic bending force vector for a beam system.
The implicit Euler and Newton's methods are used to solve the $(2N)$-sized nonlinear dynamic system.

\paragraph{Simulation initialization} To initialize the simulation, the following inputs are used:

\begin{enumerate}

\item \textbf{Geometry and connection.} (i) Nodal position: the position of the nodes $\mathbf{q}(t=0)$, with a total of $N = 40$. The length of the magnetized beam is set at $L = 1.0$ m. (ii) Stretching element: the connections between the nodes, with a total of $N_{s} = 39$. (iii) Bending element: the connections between the edges, with a total of $N_{b} = 38$. 

\item \textbf{Physical parameters.} (i) Young's modulus, $ E = 100$ MPa. (ii) Material density, $\rho=10.0 $ $\mathrm{kg/m^3}$. (iii) Cross-sectional radius, $r_{0} = 0.01$ m. (iv) Damping viscosity, $\mu = 0.1$. (v) Gravitational field $ \mathbf{g}=[0.0, 0.0]^{T}\mathrm{~m/s^2}$. (vi) The beam magnetization is $\mathbf{B}_{r} = [1000.0, 0.0] \; \mathrm{T/m^3}$. (vii) The overall simulation is dynamic, i.e., $ \mathrm{ifStatic} = 0$.

\item \textbf{Numerical parameters.} (i) Total simulation time $T=20.0\mathrm{~s}$. (ii) Time step size $\mathrm{d}t=0.01\mathrm{~s}$. (iii) Numerical force tolerance $\mathrm{tol}=1\times 10^{-4}$. (iv) Maximum iterations $N_{\mathrm{iter}}=10$.

\item \textbf{Boundary conditions.} The left tip of the beam is clamped by fixing the first two nodes (starting from the left), thus, the constrained DOF-index array is $\mathcal{FIX} = [1,2,3,4]^T$.

\item \textbf{Initial conditions.} (i) Initial position is input from the nodal positions. (ii) Initial velocity is set to zeros for all nodes.

\item \textbf{Loading steps.} A magnetic field with $B_{x} = A \cos ( \omega t)$ and $B_{y} = A \sin (\omega t)$ (where $A = 3.0$ T and $\omega = 1.0$ rad/s) is applied to the entire beam.

\end{enumerate}

\begin{figure*}[!ht]
\centering
\includegraphics[width=\textwidth]{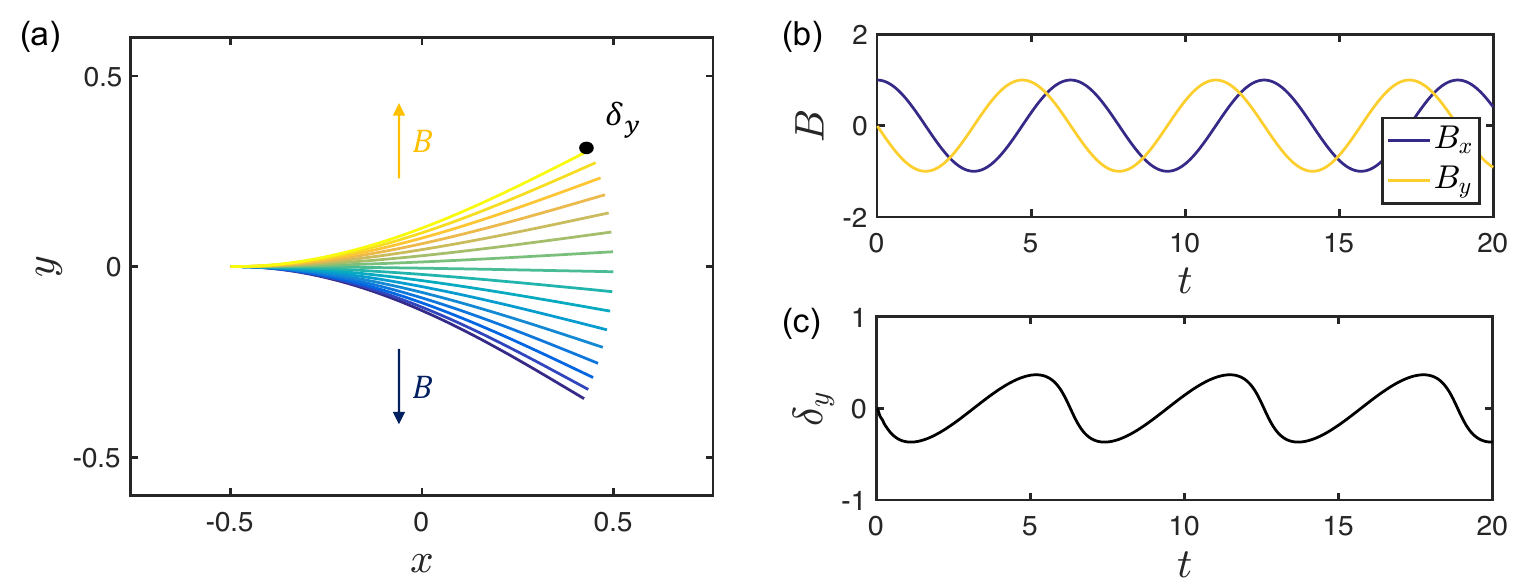}
\caption{ \textbf{Magnetized beam under periodic actuation.} (a) The dynamic response of the beam under actuation of an external magnetic field. (b) The magnetic field, $\{B_{x}, B_{y}\}$, as a function of time, $t$. (c) Tip displacement in the $Y$-direction, $\delta_{y}$, as a function of time, $t$.}
\label{fig:robot_case_1_plot}
\end{figure*}

\paragraph{Simulation results} 
The beam undergoes deformation and oscillatory motion under the influence of a periodically varying magnetic field, as shown in Fig.~\ref{fig:robot_case_1_plot}(a). 
The time-dependent changes in the magnetic field, depicted in Fig.~\ref{fig:robot_case_1_plot}(b), drive these oscillations by inducing cyclic bending and restoring forces in the beam.
Interestingly, in the absence of gravity, the tip displacement along the $Y$-direction exhibits an asymmetric rate of change during its downward and upward movements. 
Specifically, the beam’s tip moves at different speeds in these two phases, suggesting an imbalance in acceleration. 
This asymmetry arises due to the nonuniform influence of the oscillating magnetic field on the beam’s deformation dynamics. 
Consequently, the beam experiences variations in effective acceleration between the upward and downward phases, resulting in a noticeable asymmetry in its motion. 
This effect is further illustrated in Fig.~\ref{fig:robot_case_1_plot}(c), where the displacement profile of the beam’s tip in the $Y$-direction $\delta_{y}$ highlights the differing rates of change in the two phases. 
These findings provide insight into how periodic magnetic actuation influences flexible structures, leading to complex, directionally dependent dynamic behaviors. 
The dynamic rendering can be found~\href{https://github.com/weicheng-huang-mechanics/DDG_Tutorial/blob/main/assets/robot_1.gif}{here}.

\subsection{Case 2: Fluid-structure interaction~\href{https://github.com/weicheng-huang-mechanics/DDG_Tutorial/tree/main/soft_robot/case_2}{\texorpdfstring{\faGithub}{GitHub}}}

In this subsection, we demonstrate the fluid-structure interaction between a soft swimming robot and the surrounding fluid under external magnetic actuation. 
Using the DDG simulation framework, we analyze how the robotic swimmer responds to time-varying magnetic fields, generating an undulatory motion that propels it through the fluid \cite{huang2021swimming, huang2022design, liu2012numerical, mallick2014study, kelasidi2014modeling}. 
The interplay between magnetic forces, elastic deformation, and hydrodynamic resistance shapes the swimming dynamics, highlighting the model’s ability to capture complex coupled interactions. 
As magnetically actuated soft robots hold great promise for biomedical and underwater applications, this case underscores the potential of our approach in designing and optimizing efficient, adaptive locomotion strategies. 

\paragraph{Numerical formulation}
Similar to the stretching element, the discrete drag force is applied based on each rod segment, defined as
\begin{equation}
\mathcal{S}: \{\mathbf{x}_{1}, \mathbf{x}_{2} \}.
\end{equation}
We use $\mathbf{t}$ and $\mathbf{n}$ to denote the local tangential direction and the local normal direction, defined as
\begin{equation}
    \mathbf{t} = (\mathbf{x}_{2} - \mathbf{x}_{1}) / || \mathbf{x}_{2} - \mathbf{x}_{1} ||, \; \mathrm{and} \; \mathbf{n} \cdot \mathbf{t} = 0
\end{equation} 
The segment velocity components along the tangential and normal directions are given by
\begin{equation}
\begin{aligned}
\mathbf{v}_{\parallel} & = \left ( \mathbf{v} \cdot \mathbf{t} \right) \cdot \mathbf{t}, \\
\mathbf{v}_{\perp} & = \left ( \mathbf{v} \cdot \mathbf{n} \right) \cdot \mathbf{n},
\end{aligned}
\end{equation}
where 
\begin{equation}
\mathbf{v} =  \frac{1} {2} {(\mathbf{v}_{1} + \mathbf{v}_{2})}, 
\end{equation}
is the segment velocity.
For most underwater locomotion, the Reynolds number is intermediate, i.e., $ 1.0 < Re < 1000.0$, the local fluid drag force is typically quadratic with the velocity,
\begin{equation}
\begin{aligned}
(\mathbf{F}_{\mathrm{local}}^{\text{drag}})_{\parallel} &= - \frac{1} {2} \rho_{0} C_{\parallel} || \mathbf{v}_{\parallel} ||  \mathbf{v}_{\parallel} , \\
(\mathbf{F}_{\mathrm{local}}^{\text{drag}})_{\perp} &= - \frac{1} {2} \rho_{0} C_{\perp} || \mathbf{v}_{\perp} ||  \mathbf{v}_{\perp},
\end{aligned}
\end{equation}
where $\rho_{0}$ is the density of the water, $C_{\parallel}$ and $C_{\perp}$ are the drag coefficients.
Finally, the global fluid force vector, $\mathbf{F}^{\text{drag}}$, is the sum of the local force by iterating over all stretching segments.
With the formulation of the internal elastic force and the associated stiffness matrix, we can incorporate the inertial effect to derive the dynamic equations of motion.
Here, the mass matrix, $\mathbb{M}$, is time-invariant and can be easily obtained based on the element size and material density. 
We ignore the damping term since the drag force inherently acts as a damping force.
Finally, the equations of motion for a flexible beam moving in a liquid environment are
\begin{equation}
\mathbb{M} \ddot{\mathbf{q}} - \mathbf{F}^{\text{drag}} - \mathbf{F}^{s} - \mathbf{F}^{b} -  \mathbf{F}^{\text{ext}} = \mathbf{0},
\end{equation}
where $\mathbf{F}^{s}$ and $\mathbf{F}^{b} $ are the elastic stretching force vector and elastic bending force vector for a beam system.
The external force vector here, $\mathbf{F}^{\text{ext}}$, is the magnetic actuation.
The implicit Euler and Newton's methods are used to solve the $(2N)$-sized nonlinear dynamic system.

\begin{figure*}[!ht]
\centering
\includegraphics[width=\textwidth]{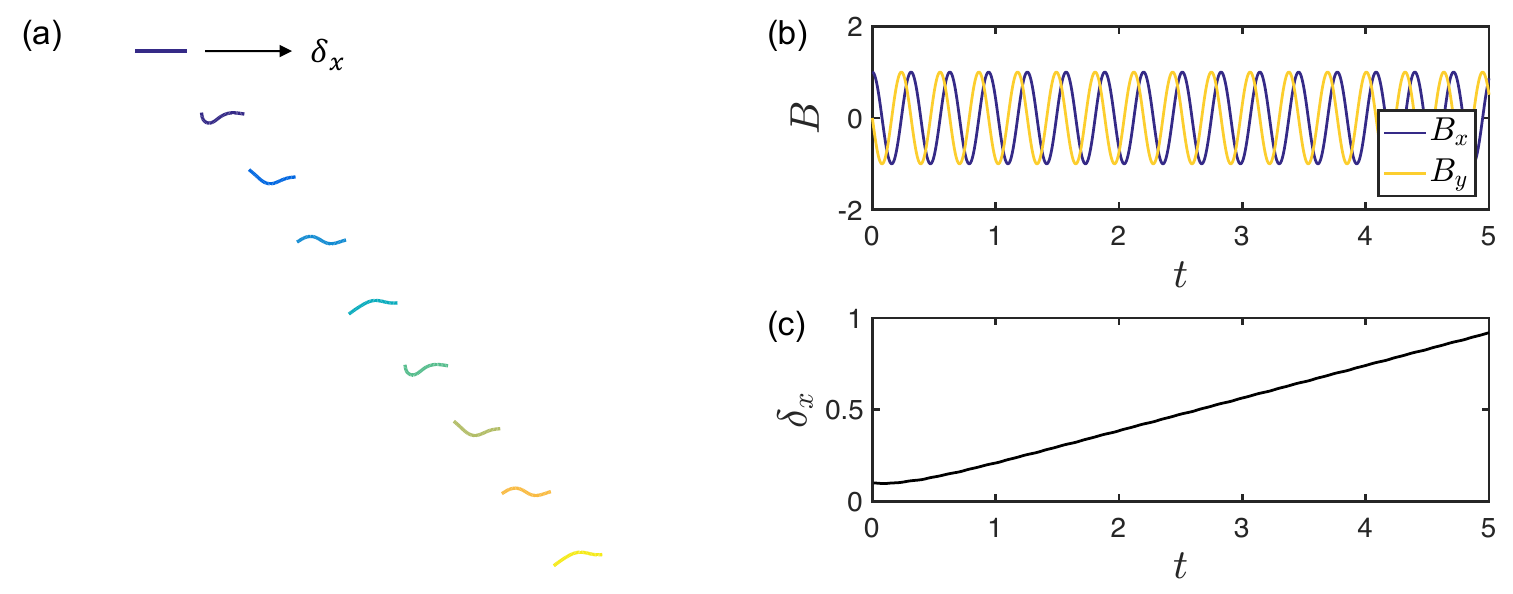}
\caption{ \textbf{Soft swimming robot.} (a) Swimming locomotion enabled by fluid-structure interaction under actuation of an external magnetic field. (b) The magnetic field, $\{B_{x}, B_{y}\}$, as a function of time, $t$. (c) Displacement of the rightmost nodes in the $X$-direction, $\delta_{x}$, as a function of time, $t$.}
\label{fig:robot_case_2_plot}
\end{figure*}

\paragraph{Simulation initialization} To initialize the simulation, the following inputs are used:

\begin{enumerate}

\item \textbf{Geometry and connection.} (i) Nodal position: the position of the nodes $\mathbf{q}(t=0)$, with a total of $N = 40$. The robot body is a magnetized beam with a length of $L = 0.1$ m. (ii) Stretching element: the connections between the nodes, with a total of $N_{s} = 39$. (iii) Bending element: the connections between the edges, with a total of $N_{b} = 38$. 

\item \textbf{Physical parameters.} (i) Young's modulus, $ E = 0.1$ MPa. (ii) Material density, $\rho=1000 $ $\mathrm{kg/m^3}$. (iii) Cross-sectional radius, $r_{0} = 0.001$ m. (iv) Gravitational field $ \mathbf{g}=[0.0, 0.0]^{T}\mathrm{~m/s^2}$. (v) Fluid parameters, including fluid density, $\rho_{0}=1000 $, normal drag coefficient $C_{\perp} = 1.0$, and shear drag coefficient $C_{\parallel} = 0.01$. (vi) The beam magnetization is ${B}_{r}^{x} = {100}  \cos ( 2 \pi s)$ $\mathrm{T/m^3}$, ${B}_{r}^{y} = {100}  \sin ( 2 \pi s)$ $\mathrm{T/m^3}$, where $s \in [0,1]$ is the arc length parameter of the beam. (vii) The overall simulation is dynamic, i.e., $ \mathrm{ifStatic} = 0$. 

\item \textbf{Numerical parameters.} (i) Total simulation time, $T = 5.0$ s. (ii) Time step size, $\mathrm{d}t = 0.01$ s. (iii) Numerical tolerance, $\mathrm{tol} = 1 \times 10^{-4}$. (iv) Maximum iterations, $N_{\mathrm{iter}} = 10$.

\item \textbf{Boundary conditions.} The $Y$ direction of the last node is fixed, thus, the constrained array is $\mathcal{FIX} = [80]$.

\item \textbf{Initial conditions.} (i) Initial position is input from the nodal positions. (ii) Initial velocity is set to zeros for all nodes.

\item \textbf{Loading steps.} A magnetic field with $B_{x} = A\cos (\omega t)$ and $B_{y} = A \sin (\omega t)$ (with $A=1.0$ T and $\omega = 20.0$ rad/s) is applied to the entire beam.

\end{enumerate}

\paragraph{Simulation results} 
The robot achieves forward propulsion through the combined effects of fluid-structure interaction and the actuation of an external magnetic field, as illustrated in Fig.~\ref{fig:robot_case_2_plot}(a). 
The interaction between the deformable structure and the surrounding fluid generates thrust, while the external magnetic field modulates the actuation forces driving the motion. 
The periodic variation of the magnetic field over time is depicted in Fig.~\ref{fig:robot_case_2_plot}(b), demonstrating the controlled oscillations that facilitate continuous propulsion.
In Fig.~\ref{fig:robot_case_2_plot}(c), we present the motion of the rightmost nodes in the $X$-direction, highlighting the effectiveness of the actuation strategy in achieving sustained forward movement. 
The dynamic rendering can be found~\href{https://github.com/weicheng-huang-mechanics/DDG_Tutorial/blob/main/assets/robot_2.gif}{here}. 

\subsection{Case 3: Frictional contact~\href{https://github.com/weicheng-huang-mechanics/DDG_Tutorial/tree/main/soft_robot/case_3}{\texorpdfstring{\faGithub}{GitHub}}}

In this subsection, we present the frictional contact interaction of a soft crawling robot with multiple legs as it moves across a surface under external magnetic actuation. 
Using the DDG simulation framework, we analyze how the legs interact with the ground, generating propulsion through controlled frictional forces \cite{huang2020dynamic, huang2023modeling, tong2023fully, choi2021implicit, li2020incremental}. 
The interplay between magnetic excitation, elastic deformation, and contact dynamics determines the robot’s locomotion efficiency and stability. 
By capturing the complex coupling between actuation and ground interaction, this model provides valuable insights into the design of magnetically controlled soft robots for applications in biomedical engineering, inspection, and search-and-rescue operations.

\paragraph{Numerical formulation} 
We use the incremental potential method to model frictional contact between flexible structures and a rigid surface. 
Here, the local nodal position is defined as $\mathbf{x} \equiv [x, y]^T$, and its velocity as $\dot{\mathbf{x}} \equiv [\dot{x}, \dot{y}]^T$.
The ground is assumed to have a surface normal $\mathbf{n}$.
The contact force is applied through each node as
\begin{equation}
(\mathbf{F}^{\mathrm{con}}_{\mathrm{local}})_{n} =
\begin{cases}
K_{c} \left[ - 2 (d - \hat{d}) \log(  {d} / {\hat{d}}) -  {(d - \hat{d})^2} /{d} \right] \cdot  \mathbf{n} \; &\mathrm{when} \; 0 < d < \hat{d}, \\
\mathbf{0} \; & \mathrm{when} \; d \geq \hat{d},
\end{cases}
\label{eq:barrierContact}
\end{equation}
where $d$ is the approaching distance between the node and ground. When the ground is flat with zero height, $d = y$. The parameter $\hat{d}$ is the barrier parameter, and $K_{c}$ denotes the contact stiffness.
The tangential frictional force is formulated based on the maximum dissipation principle as
\begin{equation}
(\mathbf{F}^{\mathrm{con}}_{\mathrm{local}})_{t} =
\begin{cases}
- \mu || (\mathbf{F}^{\mathrm{con}}_{\mathrm{local}})_{n}  || \frac {\mathbf{v}_{t}} {||\mathbf{v}_{t}||} \left(- \frac {||\mathbf{v}_{t}||^2} {\epsilon_{v}^2} +  \frac {||\mathbf{v}_{t}||} {\epsilon_{v}} \right) \; &\mathrm{when} \; 0 < ||\mathbf{v}_{t}|| < \epsilon_{v}, \\
- \mu || (\mathbf{F}^{\mathrm{con}}_{\mathrm{local}})_{n}  || \frac {\mathbf{v}_{t}}  {||\mathbf{v}_{t}||} \; & \mathrm{when} \;  ||\mathbf{v}_{t}|| \geq \epsilon_{v},
\end{cases}
\label{eq:barrierFriction}
\end{equation}
where 
\begin{equation}
\mathbf{v}_{t} = \mathbf{v} - \mathbf{v} \cdot \mathbf{n},
\end{equation}
is the relative velocity along the surface tangential direction, and $\epsilon_{v}$ is the velocity parameter.
The continuous contact defection must iterate over all nodes for a discrete beam system to build the global contact force vector, $\mathbf{F}^{\text{con}}$.
With the formulation of the internal elastic force and the associated stiffness matrix, we can incorporate the inertial and damping effect to derive the dynamic equations of motion.
Here, the mass matrix, $\mathbb{M}$, is time-invariant and can be easily obtained based on the element size and material density. 
We then employ a simple damping matrix, which is linearly related to the mass matrix with a damping coefficient $\mu$, i.e., $\mathbb{C} = \mu \mathbb{M}$.
Finally, the equations of motion for a crawling robot moving on a rigid surface is given by, 
\begin{equation}
\mathbb{M} \ddot{\mathbf{q}} + \mu \mathbb{M}  \dot{\mathbf{q}} - \mathbf{F}^{\text{con}} - \mathbf{F}^{s} - \mathbf{F}^{b} -  \mathbf{F}^{\text{ext}} = \mathbf{0}.
\end{equation}
where $\mathbf{F}^{s}$ and $\mathbf{F}^{b} $ are the elastic stretching force vector and elastic bending force vector for the system.
The external force vector, $\mathbf{F}^{\text{ext}}$, is the sum of the gravitational force and the magnetic actuation.
The implicit Euler and Newton's methods are used to solve the $(2N)$-sized nonlinear dynamic system.

\paragraph{Simulation initialization} To initialize the simulation, the following inputs are used:

\begin{enumerate}

\item \textbf{Geometry and connection.} (i) Nodal position: the position of the nodes $\mathbf{q}(t=0)$, with a total of $N = 169$. The robot's body trunk is a magnetized beam with a length of $L = 0.08$ m, and it has 15 limbs, each also a magnetized beam with a length of $L_{0} = 0.0134$ m. (ii) Stretching element: the connections between the nodes, with a total of $N_{s} = 168$. (iii) Bending element: the connections between the edges, with a total of $N_{b} = 182$. 

\item \textbf{Physical parameters.} (i) Young's modulus, $ E = 1.0$ MPa. (ii) Material density, $\rho=1000 $ $\mathrm{kg/m^3}$. (iii) Cross-sectional radius, $r_{0} = 0.001$ m. (iv) Damping viscosity, $\mu = 0.1$. (v) Gravity, $ \mathbf{g} = [0.0, -10]^{T}$ $\mathrm{m/s^2}$. (vi) The leg magnetization is ${B}_{r}^{x} = {1000}  \cos ( 4 \pi s) $  $\mathrm{T/m^{3}}$, ${B}_{r}^{y} = {1000}  \sin ( 4 \pi s)$ $\mathrm{T/m^{3}}$, where $s \in [0,1]$ is the arc length parameter of the robotic body.  (vii) Contact parameters, including $\hat{d} = 1 \times 10^{-3}$ m, $K_{c} = 1.0$ kPa, $\mu=0.3$, and $\epsilon_{v} = 1 \times 10^{-6}$ m/s

\item \textbf{Numerical parameters.} (i) Total simulation time, $T = 2.0$ s. (ii) Time step size, $\mathrm{d}t = 0.001$ s. (iii) Numerical tolerance, $\mathrm{tol} = 1 \times 10^{-4}$. (iv) Maximum iterations, $N_{\mathrm{iter}} = 10$.

\item \textbf{Boundary conditions.} No nodes are constrained in this case, thus, the constrained array is $\mathcal{FIX} = \emptyset$.

\item \textbf{Initial conditions.} (i) Initial position is input from the nodal positions. (ii) Initial velocity is set to zeros for all nodes.

\item \textbf{Loading steps.} A magnetic field with $B_{x} = A\cos (\omega t)$ and $B_{y} = A\sin ( \omega t)$ (with $A=1.0$ T and $\omega = 10\pi$ rad/s) is applied to the entire robotic system.

\end{enumerate}

\begin{figure*}[!ht]
\centering
\includegraphics[width=\textwidth]{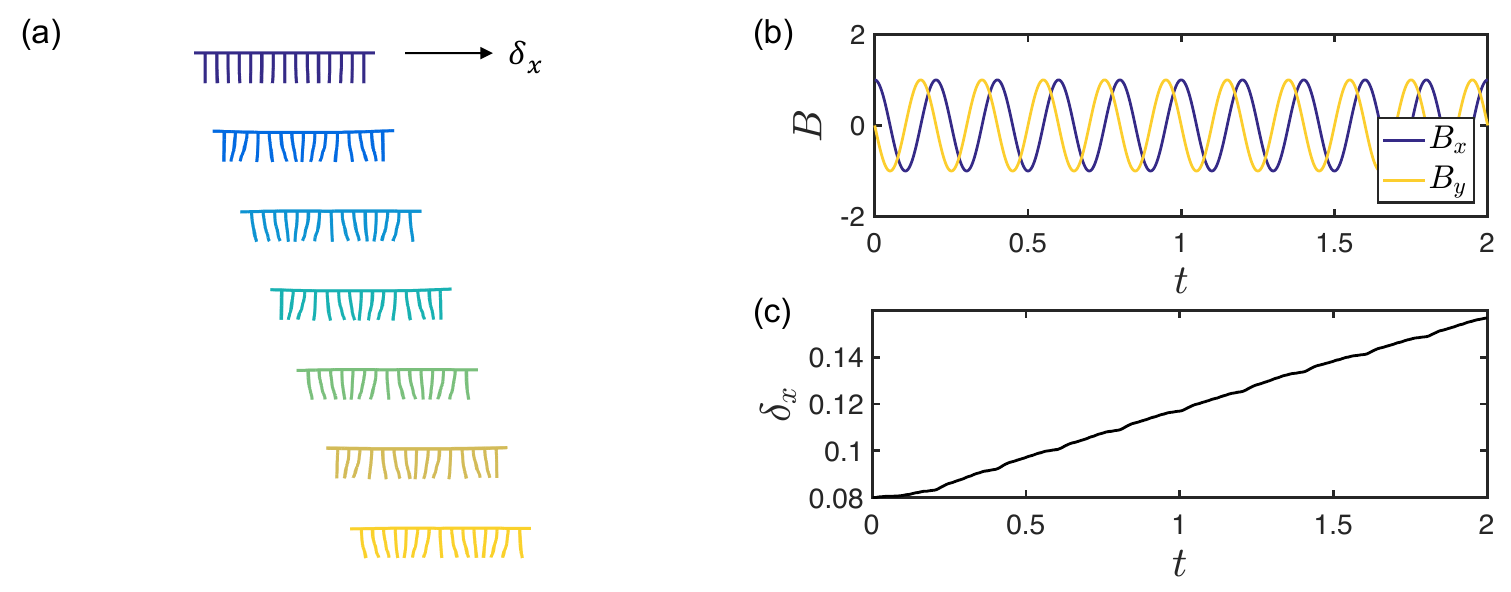}
\caption{ \textbf{Soft crawling robot.} (a) Crawling locomotion enabled by frictional contact interaction under the actuation of an external magnetic field. (b) The magnetic field, $\{B_{x}, B_{y}\}$, as a function of time, $t$. (c) Displacement of the rightmost nodes in the $X$-direction, $\delta_{x}$, as a function of time, $t$.}
\label{fig:robot_case_3_plot}
\end{figure*}

\paragraph{Simulation results} 
The robot advances forward through a crawling motion enabled by the interplay between frictional contact with the ground and the actuation of an external magnetic field, as illustrated in Fig.~\ref{fig:robot_case_3_plot}(a). 
The cyclic deformation induced by the magnetic actuation generates a coordinated sequence of contact and sliding phases, allowing the robot to incrementally propel itself. 
Fig.~\ref{fig:robot_case_3_plot}(b) depicts the periodic variation of the magnetic field over time, which governs the actuation cycle and influences the crawling efficiency.
In Fig.~\ref{fig:robot_case_3_plot}(c), we present the displacement of the rightmost nodes in the $X$-direction, highlighting the stepwise progression of the robot as it moves forward. 
The dynamic rendering can be found~\href{https://github.com/weicheng-huang-mechanics/DDG_Tutorial/blob/main/assets/robot_3.gif}{here}. 



\section{Discussions and Conclusions}

In this tutorial, we have provided a comprehensive introduction to the Discrete Differential Geometry (DDG) method for simulating the nonlinear behaviors of flexible structures. By leveraging the intrinsic geometric properties of DDG, we demonstrated its capability to accurately capture large deformations for flexible structures in different dimensions, as well as the material nonlinearities and the complex external interactions. Through a structured simulation framework and a practical MATLAB implementation guide, we aimed to bridge the gap between theoretical foundations and practical applications, making DDG more accessible to researchers and engineers.

The case studies presented in this work illustrate the versatility and robustness of the DDG method in modeling geometrically and materially nonlinear behaviors across different structural forms. By incorporating external interactions such as magnetic actuation, fluid-structure interactions, and frictional contact, we showcased the potential of DDG-based simulations in addressing real-world engineering challenges.

Despite its advantages, several areas remain open for further exploration. Future research directions include enhancing the computational efficiency of DDG algorithms for large-scale simulations, integrating machine learning techniques for data-driven modeling and optimization, and extending DDG-based formulations to multi-scale modeling, e.g., including the Van der Waals effects and hydrogen bond for the modeling and simulation of the polymer network. Additionally, real-time model-based design and control will be crucial in refining the models and broadening their applicability to emerging fields such as soft robotics, deployable structures, and biomedical engineering.

By providing this tutorial along with open-source MATLAB code, we hope to facilitate the widespread adoption of DDG in applied mathematics, computational mechanics, structural design, and engineering science.
It is worth noting that this MATLAB-based tutorial is for demonstration purposes and its computational speed may be limited. For real engineering applications, we have also released an open-source C/C++ version, which can be accessed through the \href{https://github.com/weicheng-huang-mechanics/DDG_Tutorial_Fast}{GitHub Repository} \footnote{See \url{https://github.com/weicheng-huang-mechanics/DDG_Tutorial_Fast}}.
We believe that the continued development of DDG-based numerical tools will drive significant advancements in the simulation and analysis of highly flexible structures, opening new opportunities for innovative engineering and scientific applications.

\newpage
\section*{Acknowledgments}
\addcontentsline{toc}{section}{Acknowledgements}

W.H. acknowledges the start-up funding from Newcastle University, UK. K.J.H. acknowledges the financial support from the Ministry of Education, Singapore under its Academic Research Fund Tier 3 (Grant MOE-MOET32022-0002). M.L. acknowledges the start-up funding from The University of Birmingham, UK.

\section*{Author Contributions}
\addcontentsline{toc}{section}{Author Contributions}

Weicheng Huang: Conceptualization; algorithm development; code development; data generation; graphics and video visualization; web development; manuscript organization and structuring; original draft writing.

\noindent
Zhuonan Hao: Algorithm development; code development; data generation; graphics and video visualization; web development; original draft writing.

\noindent
Jiahao Li: Algorithm development; code development; data generation; graphics and video visualization; original draft writing.

\noindent
Dezhong Tong: Algorithm development; code development; graphics and video visualization; manuscript revision.

\noindent
Kexin Guo: Code development; graphics and video visualization; manuscript revision.

\noindent
Yingchao Zhang: Graphics and video visualization; manuscript revision.

\noindent
Huajian Gao: Conceptualization; manuscript organization and structuring; manuscript revision; overall guidance.

\noindent
K. Jimmy Hsia: Conceptualization; manuscript organization and structuring; manuscript revision; overall guidance.

\noindent
Mingchao Liu: Conceptualization; graphics and video visualization; web development; manuscript organization and structuring; original draft writing; overall coordination.


\newpage
\section*{Appendix A: Model validation}
\addcontentsline{toc}{section}{Appendix A: Model validation}

\renewcommand{\thefigure}{A.\arabic{figure}} 
\setcounter{figure}{0} 

\begin{figure*}[!ht]
\centering
\includegraphics[width=\textwidth]{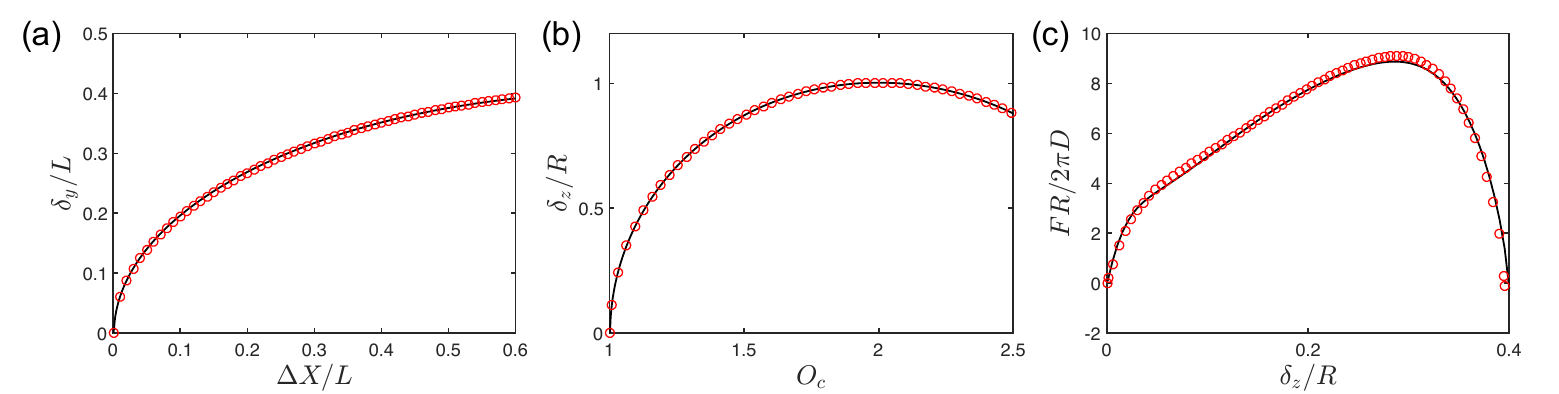}
\caption{ 
    \textbf{Numerical validation.} 
    (a) Buckling of a compressive beam, comparison between the discrete simulation (solid line) and elastica theory (red dots) \cite{zhang2020configurations}.
    (b) Growth of an annular ribbon, comparison between the discrete simulation (solid line) and AUTO continuation  (red dots) \cite{huang2024integration}. 
    (c) Snapping of an axisymmetric shell, comparison between the discrete simulation (solid line) and finite element analysis  (red dots)
    \cite{huang2024discrete}.
}
\label{fig:validate_plot_figure}
\end{figure*}

In this appendix, we provide validation for the basic DDG-based numerical method. First of all, we use the buckling of a compressive beam as an example to validate the correctness of our planar beam model, referring to Fig.~\ref{fig:validate_plot_figure}(a). The results from our DDG-based simulation show good agreement with the Elastica theory \cite{zhang2020configurations}. Next, we use the growth of an annular ribbon as an example to validate the correctness of our 3D ribbon model, referring to Fig.~\ref{fig:validate_plot_figure}(b). The results from our DDG-based simulation show good agreement with the AUTO continuation \cite{huang2024integration}. Finally, we use the snapping of an axisymmetric shell as an example to validate the correctness of our rotational surface model, referring to Fig.~\ref{fig:validate_plot_figure}(c). The results from our DDG-based simulation show good agreement with the finite element analysis \cite{zhang2020configurations}. Note that the 3D plate/shell model is only geometrically exact (which can converge to the continuum limit of F{\"o}ppl-von K{\'a}rm{\'a}n equations) and may not be able to provide an accurate prediction \cite{seung1988defects}. The validation for other extended models can be found in the existing literature \cite{baek2018form,  liu2024simplified, huang2023static, huang2020dynamic, huang2021swimming}.

\newpage
\addcontentsline{toc}{section}{References}
\bibliographystyle{ieeetr}
\bibliography{paper}

\end{document}